\documentclass[a4paper,pdftex,reqno,12pt]{amsart}
\pdfoutput=1

\allowdisplaybreaks[1]

\usepackage{geometry}
\usepackage{enumerate}
\usepackage{courier}
\usepackage{times}
\usepackage{amssymb}
\usepackage{amsmath,cite}
\usepackage{color}
\usepackage[boxed,vlined,longend]{algorithm2e}

\theoremstyle{plain}
\newtheorem{Theorem}{Theorem}[section]
\newtheorem{Proposition}[Theorem]{Proposition}

\newtheorem{Lemma}[Theorem]{Lemma}

\newenvironment{Proof}
{\begin{trivlist}\item[]{{\sc Proof.}}}{\hfill{$\square$}\noindent\end{trivlist}}

\newcommand{\F}[2]{\mathbb{F}_{#2}^{#1}}
\newcommand{\EF}[2]{\operatorname{EF}_{#2}(#1)}
\newcommand{\PG}[2]{\mathcal{P}_{#2}(#1)}
\newcommand{\G}[3]{\mathcal{G}_{#3}(#1,#2)}
\newcommand{\gaussm}[3]{\genfrac{[}{]}{0pt}{}{#1}{#2}_{#3}}
\newcommand{\rk}{\operatorname{rk}}
\newcommand{\rb}{\color{red}{\bullet}}
\newcommand{\bb}{\color{blue}{\bullet}}

\theoremstyle{definition}

\theoremstyle{remark}

\begin{document}


\title{Bounds for the multilevel construction}

\author{Tao Feng}
\address{Department of Mathematics, Beijing Jiaotong University, 100044 Beijing, P. R. China}
\email{tfeng@bjtu.edu.cn}
\footnotetext{Supported by NSFC under Grant 11871095 (T. Feng)}

\author{Sascha Kurz}
\address{Department of Mathematics, University of Bayreuth, 95440 Bayreuth, Germany}
\email{sascha.kurz@uni-bayreuth.de}

\author{Shuangqing Liu}
\address{Department of Mathematics, Suzhou University of Science and Technology, Suzhou, P. R. China}
\email{shuangqingliu@usts.edu.cn}

\date{}

\begin{abstract}
  One of the main problems in random network coding is to compute good lower and upper bounds on the achievable
  cardinality of the so-called subspace codes in the projective space $\PG{n}{q}$ for a given minimum distance. The
  determination of the exact maximum cardinality is a very tough discrete optimization problem involving a huge number of
  symmetries. Besides some explicit constructions for \textit{good} subspace codes several of the most success full constructions
  involve the solution of discrete optimization subproblems itself, which mostly have not been
  not been solved systematically. Here we consider the multilevel a.k.a.\ Echelon--Ferrers construction and given lower and
  upper bounds for the achievable cardinalities. From a more general point of view, we solve maximum clique problems
  in weighted graphs, where the weights can be polynomials in the field size $q$.

  \medskip

\end{abstract}

\maketitle

\section{Introduction}

\noindent
Let $\mathbb{F}_q$ be the finite field of order $q$, i.e., $q$ is a prime power.
Consider the $n$-dimensional vector space $\F{n}{q}$ consisting of all vectors
of length $n$ over $\mathbb{F}_q$. For $0\le k\le n$ we denote by $\G{n}{k}{q}$
the set of all $k$-dimensional subspaces of $\F{n}{q}$, which is also called
\emph{Grassmannian}, and by $\gaussm{n}{k}{q}:=\#\G{n}{k}{q}$ its cardinality.
The \emph{projective space} of order $n$ over $\mathbb{F}_q$
is given by $\PG{n}{q}=\bigcup_{0\le k\le n} \G{n}{k}{q}$. An information-theoretic
analysis of the so-called Koetter-Kschischang-Silva model \cite{silva2008rank}
motivates the subspace distance
$$
  d_S(U,U'):=\dim U+\dim U'-2\dim(U\cap U')
$$
and the injection distance
$$
  d_I(U,U'):=\max \{\dim U,\dim U'\}-\dim(U\cap U')
$$
as suitable metrics, where $U,U'\in \PG{n}{q}$. With these metrics, one can define codes on $\PG{n}{q}$ and
$\G{n}{k}{q}$, which are called \emph{subspace codes} and \emph{constant dimension
codes}, respectively. We remark $d_S(U,U')=2d_I(U,U')$ for $U,U'\in\G{n}{k}{q}$, i.e.,
the two metrics are equivalent on $\G{n}{k}{q}$, and we have $d_I(U,U')\le d_S(U,U')\le 2
d_I(U,U')$ in general.

In this paper we will restrict ourselves to constant dimension codes,
i.e., all codewords have the same dimension, and the subspace distance. We  say that $\mathcal{C}\subseteq \G{n}{k}{q}$
is an $(n,M,d;k)_q$ code if $\mathcal{C}$ has cardinality $\#\mathcal{C}=M$ and \emph{minimum
subspace distance} $D(\mathcal{C}):=\min_{U\neq U'\in\mathcal{C}} d_S(U,U')\ge d$. One main
problem is the determination of the maximum size $A_q(n,d;k)$ of an $(n,M,d;k)_q$ code in $\G{n}{k}{q}$.
In principle, the determination of $A_q(n,d;k)$ can be formulated as a maximum clique (or maximum
independent set) problem, see e.g.\ \cite{kohnert2008construction}. However, there are two
challenging facts that prevent a successful application of most of the available maximum clique
algorithms:
\begin{itemize}
  \item the order of the corresponding graphs grow very quickly, i.e.,
        $$
          \left|\G{n}{k}{q}\right|=\gaussm{n}{k}{q}:=
          \prod_{i=1}^{k} \frac{q^{n-k+i}-1}{q^i-1}
          =q^{k(n-k)}(1+o(1));
        $$
  \item the problem is very \textit{symmetric}, i.e., the automorphism group of
        $\G{n}{k}{q}$ viewed as a metric space with respect to the subspace distance is
        given by the the  projective  general semilinear group $\operatorname{P\Gamma L}(n,q)$
        having an order of
        $$
          r\cdot \frac{\prod_{i=0}^{n-1} \left(q^n-q^i\right)}{q-1}
          =r\cdot q^{n^2-1}- r\cdot O\!\left(q^{n^2-3}\right),
        $$
        where $q=p^r$.
\end{itemize}
And indeed, for $n-k\ge k\ge 2$ and $2<d<2k$\footnote{Since $A_q(n,d;k)=A_q(n,d;n-k)$ one generally assumes $n-k\ge k$.
For $d=2$, we have $A_q(n,2;k)=\gaussm{n}{k}{q}$. Clearly, we can have $A_q(n,d;k)\ge 2$ for $d\le 2k$ only. The case
$d=2k$ is known under the name \textit{partial spreads} and permits the application of farreaching analytical
tools like e.g.\ the theory of divisible codes, see e.g.~\cite{honold2018partial}.} the only three values that were determined exactly are
$A_2(6,4;3)=77$ \cite{hkk77}, $A_2(8,6;4)=257$ \cite{heinlein2019classifying}, and $A_2(13,4;3)=1,597,245$ \cite{qsteiner}. The
\textit{smallest} open case $333\le A_2(7,4;3)\le 381$ seems to be a tough nut, see e.g.~\cite{heinlein2019subspace,kiermaier2018order}.

While there are a few explicit constructions for subspace and constant dimension codes, the currently most successfully constructions are
parameterized and involve search spaces itself. Here we will consider the so-called multilevel construction \cite{etzion2009error}
a.k.a.~Echelon--Ferrers construction. The maximum possible cardinality of an $(n,\star,d;k)_q$ code within this class of constructions
is denoted by $M_q(n,d;k)$. The precise description of the multilevel method will be postphoned to Section~\ref{sec_multilevel}. Here
we only mention that it essentially consists of a constant-weight code $\mathcal{H}\subseteq \mathbb{F}_2^n$, with codewords of Hamming weight
$k$ and minimum Hamming distance $d$, and a weight function $w\colon \mathbb{F}_2^n\rightarrow \mathbb{N}$. The cardinality of the corresponding
constant dimension code then is given by $w(\mathcal{H}):=\sum_{h\in\mathcal{H}} w(h)$, where $w$ depends on the field size $q$ and also needs
to be determined. However, there is a precise conjecture on the exact value of $w(h)$, see \cite[Conjecture 1]{etzion2009error}. A lot of research
has been done in proving this conjecture in special cases, see e.g.~\cite{antrobus2019maximal,antrobus2019state,etzion2016optimal,
liu2019constructions,liu2019several,randrianarisoa2019some,zhang2019constructions}. The maximization of $w(\mathcal{H})$ can be described by a weighted maximum clique problem,
where the weights might be polynomials in $q$. For the special case where $w(h)=1$ for all $h\in\mathbb{F}_2^n$, the maximization of $w(\mathcal{H})$
equals the maximization of $\#\mathcal{H}$, which is a classical, very hard, problem, see e.g.~\cite{brouwer1990new}. Slightly abusing notation, we
denote the maximum possible value by $A_1(n,d;k)$ -- considering sets as vector spaces over the \textit{field with $q=1$ element}. A lot of papers are
studying lower bounds for $M_q(n,d;k)$, see e.g.~\cite{gorla2017subspace,liu2019parallel,he2019hierarchical,etzion2009error}. Here we present upper
bounds for $M_q(n,d;k)$ for all parameters satisfying $2\le k\le  n\le 19$ and $2\le\tfrac{d}{2}\le k\le 9$, most of which can indeed be attained
if \cite[Conjecture 1]{etzion2009error} is true.\footnote{For $d=4$ we only consider the cases for $n\le 14$. Here we have $M_q(n,d;k)=\overline{M}_q(n,d;k)$. For heuristically
obtained lower bounds we refer to \cite{lifted_codes}.} To this end we denote by $\overline{M}_q(n,d;k)$ maximum possible cardinality of an $(n, \star, d; k)_q$
code within the class of the multilevel construction assuming that \cite[Conjecture 1]{etzion2009error} is true. We also state a lot of improved lower bounds on $M_q(n,d;k)$.

We remark that the multilevel construction has been refined by using pending dots and blocks, see \cite{ST}. As mentioned before, there are also
other generic constructions for constant dimension codes, which involve combinatorial search spaces. For a recent overview we refer to
\cite{cossidente2019combining}. For the currently best known lower and upper bounds on $A_q(n,d;k)$ we refer the reader to the online database
subspacecodes.uni-bayreuth.de associated with the survey \cite{heinlein2016tables}.

The remaining part of this article is structured as follows. In Section~\ref{sec_multilevel} we describe the multilevel construction and Ferrers diagram rank metric codes.
Our algorithmic approach for weighted maximum clique problems whose weights are polynomials is presented in Section~\ref{sec_weighted_max_clique}.
The resulting upper bounds for $M_q(n,d;k)$ are summarized in Appendix~\ref{sec_upper_polynomial}. Lower bounds, i.e., constructions, are the topic of
Section~\ref{sec_constructions}. For the special case of partial spreads, i.e., minimum subspace distance $d=2k$, we analytically solve the determination
of $M_q(n,2k;k)$ in Section~\ref{sec_partial_spread}. 

\section{Multilevel construction}
\label{sec_multilevel}
The elements of a constant dimension code $\mathcal{C}\subseteq \G{n}{k}{q}$, also called codewords, are $k$-dimensional
subspaces of $\mathbb{F}_q^n$. As for linear codes we use generator matrices in order to describe them. Given a matrix
$A\in\mathbb{F}_q^{k\times n}$ of (full) rank $k$, the row-space $\langle A\rangle$ of $A$ forms
a $k$-dimensional subspace of $\mathbb{F}_q^n$, so that the matrix $A$ is called a \emph{generator matrix} of $\langle A\rangle$.
Since the  application of the Gaussian elimination algorithm onto a generator matrix $A$ does not change the row-space,
we can restrict ourselves onto generator matrices which are in \emph{reduced row echelon form} (rre), i.e., the matrix has the
shape resulting from a Gaussian elimination. It is well known that this representation is unique and does not depend on the
elimination algorithm, i.e., it gives a bijection. For later reference we denote the mapping from a $k$-dimensional subspace
$U$ of $\mathbb{F}_q^n$ to its unique generator matrix in rre by $\tau(U)$ (ignoring the parameters $n$, $k$, and $q$ for the
ease of notation). Note that $\rk(\tau(U))=k$, where $\rk$ denotes the rank of a matrix. Given a matrix $A\in \mathbb{F}_q^{k\times n}$ of full rank we denote by
$p(A)\in\mathbb{F}_2^n$ the binary vector whose $1$-entries coincide with the pivot columns of $A$. By construction,
the (Hamming) weight of $p(A)$ equals $k$. For each $v\in\mathbb{F}_2^n$ let $\EF{v}{q}$ denote the set of all $k\times n$
matrices over $\mathbb{F}_q$ that are in reduced row echelon form with pivot columns described by $v$, where $k$ is the weight of $v$.

As an example consider the pivot vector $v=(0,0,0,1,0,1,0,0,0,1,1,1,0,0)\in\mathbb{F}_2^{14}$, which has weight $5$. The corresponding set of matrices
is given by
$$
  \EF{v}{q}=\left\{\begin{pmatrix}
  0&0&0&1&\star&0&\star&\star&\star&0&0&0&\star&\star\\
  0&0&0&0&0    &1&\star&\star&\star&0&0&0&\star&\star\\
  0&0&0&0&0    &0&0    &0    &0    &1&0&0&\star&\star\\
  0&0&0&0&0    &0&0    &0    &0    &0&1&0&\star&\star\\
  0&0&0&0&0    &0&0    &0    &0    &0&0&1&\star&\star
  \end{pmatrix}\right\},
$$
where the $\star$s represent arbitrary elements of $\mathbb{F}_q$, i.e., $\#\EF{v}{q}=q^{17}$. In general we have
$$
  \#\EF{v_1,\dots,v_n}{q}=  q^{\sum\limits_{i=1}^{n} \big((1-v_i)\cdot \sum\limits_{j=1}^{i} v_j \big) }
$$
and the structure of the corresponding matrices can be read off from the corresponding
\emph{(Echelon)-Ferrers diagram}
$$
  \begin{array}{llllll}
    \bullet & \bullet & \bullet & \bullet & \bullet & \bullet \\
            & \bullet & \bullet & \bullet & \bullet & \bullet \\
            &         &         &         & \bullet & \bullet \\
            &         &         &         & \bullet & \bullet \\
            &         &         &         & \bullet & \bullet
  \end{array},
$$
where the pivot columns and zeros are omitted and the stars are replaced by solid black circles. A Ferrers diagram represents partitions as patterns of dots, with the $i$-th column
having the same number of dots as the $i$-th term $\gamma_i$ in the partition $\#$dots $=\gamma_1+\dots+\gamma_l$, where $\gamma_1\le\dots\le \gamma_l$ and $\gamma_i\in\mathbb{N}_{>0}$.
As noted above, the number of dots in the Ferrers diagram corresponding to the pivot vector $v=(v_1,\ldots,v_n)$ is given by
$\sum\limits_{i=1}^{n}\big(\left(1-v_i\right)\cdot \sum\limits_{j=1}^{i} v_j \big)$. Note that different pivot vectors can produce the same Ferrers diagram, i.e.\
initial zeroes and trailing ones do not change the Ferrers diagram. For example $v=(1,0,1,0,0,0,1,1,1,0,0)\in\mathbb{F}_2^{11}$ yields the same Ferrers diagram as shown above.

The general idea of the multilevel or Echelon-Ferrers construction is to construct constant dimension codes $\mathcal{C}_v\subseteq \EF{v}{q}$
for different pivot vectors $v$ and combine them to $\mathcal{C}= \bigcup_{v\in\mathcal{H}} \mathcal{C}_v$, where $\mathcal{H}\subseteq \mathbb{F}_2^n$.
Now let us dive into the details.

First note that the subspace distance $d_S(U,U')$ between two subspaces $U$ and $U'$ of $\mathbb{F}_q^n$ can be expressed
via the rank of their generator matrices:
\begin{eqnarray}
  \label{eq_d_s_rk}
  d_S(U,U')&=&\dim(U+U')-\dim(U\cap U')=2\dim(U+U')-\dim(U)-dim(U')\nonumber\\
  &=&2\rk\!\left(\begin{smallmatrix}\tau(U)\\ \tau(U')\end{smallmatrix}\right)-\rk(\tau(U))-\rk(\tau(U')).
\end{eqnarray}
For $U,U'\in \G{n}{k}{q}$ this simplifies to $d_S(U,U')=2\rk\!\left(\begin{smallmatrix}\tau(U)\\ \tau(U')\end{smallmatrix}\right)-2k$. If moreover
$U,U'\in\EF{v}{q}$ for some pivot vector $v\in\mathbb{F}_2^n$, then this can be further simplified. To this end let $\widehat{\tau}(U)$ denote the
$k\times (n-k)$ matrix that arises from $\tau(U)$ by removing the pivot columns, where $U\in\G{n}{k}{q}$. Using the rank distance
$d_R(A,A'):=\rk(A-A')$ for two matrices of the same size, we have
\begin{equation}
  d_S(U,U')=2d_R(\widehat{\tau}(U),\widehat{\tau}(U)')
\end{equation}
for all $U,U'\in \EF{v}{q}$ for some pivot vector $v$. So-called \emph{rank metric} of sets of $m\times n$ matrices in $\mathbb{F}_q^{m\times n}$ with respect
to the rank distance have been studied since the seventies \cite{delsarte1978bilinear}. If $d_r\le m\le n$ then the maximum number of elements
in $\mathbb{F}_q^{m\times n}$ with pairwise rank distance at least $d_r$ is $q^{n(m-d_r+1)}$, see e.g.~\cite{delsarte1978bilinear}. This upper
bound can be achieved for all parameters and the corresponding codes are called \emph{maximum rank distance} (MRD) codes. Moreover, there even exists
a \emph{linear} MRD code $\mathcal{M}$ in all cases, where we call $\mathcal{M}\subseteq\mathbb{F}_q^{m\times n}$ linear if $\mathcal{M}$ is a
subspace of $\mathbb{F}_q^{m\times n}$. Our situation is a bit more involved since the Echelon-Ferrers diagram of a given pivot vector forces
some restrictions on the matrices $A$ of a rank distance code $\mathcal{M}$. To this end we define the \emph{support} of an $m\times n$ matrix $A$
as the set of its non-zero entries, i.e., $\operatorname{supp}(A)=\left\{(i,j)\in[m]\times [n]\,:\, a_{i,j}\neq 0\right\}$. Given a pivot vector $v$
the elements $A$ of a rank metric code $\mathcal{M}$ have to satisfy that $(i,j)\in\operatorname{supp}(A)$ implies that the corresponding
Echelon-Ferrers diagram contains a dot at position $(i,j)$.
More formally, for a given $m \times n$ Ferrers diagram $\mathcal F$, an $(\mathcal F, \delta)_q$ {\em Ferrers diagram rank-metric} (FDRM) {\em code} $\mathcal C$
is a set of $m \times n$ matrices in $\mathbb F^{m \times n}_q$ with minimum rank distance $\delta={\min}\{\rk(A-B):A,B \in \mathcal C, A\neq B\}$,
and for each $m \times n$ matrix in $\mathcal C$, all entries not in $\mathcal F$ are zero. If $\mathcal C$ forms a $k$-dimensional $\mathbb F_q$-linear subspace of
$\mathbb F^{m \times n}_q$, then it is called {\em linear} and such a code is denoted by an $[\mathcal F, k, \delta]_q$ code.
If $\mathcal F$ is a {\em full} $m\times n$ diagram with $mn$ dots, then its corresponding FDRM code is just a classical rank metric codes.
Before we give an example let us state the two crucial theorems for the Echelon-Ferrers construction.

\begin{Theorem} (see \cite{etzion2009error})
  \label{thm_echelon_ferrers}
  For integers $k,n,delta$ with $1\le k\le n$ and $1\le \delta\le \min\{k,n-k\}$, let $\mathcal{H}$ be a
  binary constant weight code of length $n$, weight $k$, and minimum Hamming distance $2\delta$.
  For each $h\in \mathcal{H}$ let $\mathcal{C}_h\subseteq\EF{h}{q}$ be an $(n,\star,2\delta;k)_q$ constant dimension code. Then, $\bigcup_{h\in\mathcal{H}} \,\mathcal{C}_h$ is
  a constant dimension code of dimension $k$ having a subspace distance of at least $2\delta$.
\end{Theorem}

The code $\mathcal{H}$ is also called \emph{skeleton code}. For the building blocks $\mathcal{C}_h$ we have the following upper
bound:

\begin{Theorem} (see \cite{etzion2009error})
  \label{thm_upper_bound_ef}
  For integers $1\le k\le n$ let $v\in\mathbb{F}_2^n$ be a vector of weight $k$. If
  $\mathcal{C}_v\subseteq \EF{v}{q}$ is a subspace code having a minimum subspace
  distance of at least $2\delta$, then
\[
    \#\mathcal{C}_v \le q^{\min\{\nu_i\,:\, 0\le i\le \delta-1\}},
\]
  where $\nu_i$ is the number of dots in the Echelon-Ferrers diagram $\mathcal{F}$, that corresponds to $v$, which are neither contained in the
  first $i$ rows nor contained in the rightmost $\delta-1-i$ columns.
\end{Theorem}

Theorem \ref{thm_upper_bound_ef} shows that for any $[\mathcal F, k, \delta]_q$ code, $k\leq \min \{\nu_i:0\leq i\leq \delta-1\}$. The authors of \cite{etzion2009error} conjecture
that Theorem~\ref{thm_upper_bound_ef} is tight for all parameters $q$, $\mathcal{F}$, and $\delta$, which is still unrebutted. As already mentioned in the introduction, constructions
settling the conjecture in several cases are given e.g.\ in \cite{antrobus2019state,antrobus2019maximal,etzion2016optimal,zhang2019constructions,
liu2019constructions,liu2019several,randrianarisoa2019some}.

In order to illustrate Theorem~\ref{thm_upper_bound_ef} let us consider our example of an Echelon-Ferrers diagram again and choose a minimum subspace distance of $8$:
$$
  \begin{array}{llllll}
    \bb & \bb & \bb & \rb & \rb & \rb \\
        & \bb & \bb & \rb & \rb & \rb \\
        &     &     &     & \rb & \rb \\
        &     &     &     & \rb & \rb \\
        &     &     &     & \rb & \rb
  \end{array}
  \quad
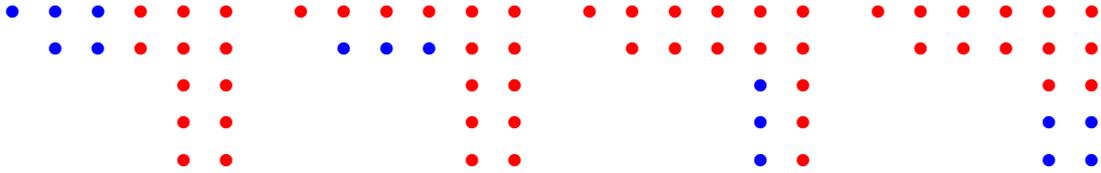
  \begin{array}{llllll}
    \rb & \rb & \rb & \rb & \rb & \rb \\
        & \bb & \bb & \bb & \rb & \rb \\
        &     &     &     & \rb & \rb \\
        &     &     &     & \rb & \rb \\
        &     &     &     & \rb & \rb
  \end{array}
  \quad
  \begin{array}{llllll}
    \rb & \rb & \rb & \rb & \rb & \rb \\
        & \rb & \rb & \rb & \rb & \rb \\
        &     &     &     & \bb & \rb \\
        &     &     &     & \bb & \rb \\
        &     &     &     & \bb & \rb
  \end{array}
  \quad
  \begin{array}{llllll}
    \rb & \rb & \rb & \rb & \rb & \rb \\
        & \rb & \rb & \rb & \rb & \rb \\
        &     &     &     & \rb & \rb \\
        &     &     &     & \bb & \bb \\
        &     &     &     & \bb & \bb
  \end{array}
$$
i.e., we have $\nu_0=5$, $\nu_1=3$, $\nu_2=3$, and $\nu_3=4$, so that $\#\mathcal{C}_v\le q^3$. As an example we give a rank metric code matching the
upper bound for the pivot vector $v=(1,1,0,0,0,1,1,1,0,0)\in\mathbb{F}_2^{10}$ with Echelon-Ferrers diagram
$$
  \begin{array}{lllll}
     \bb & \bb & \bb & \bullet & \bullet \\
     \bb & \bb & \bb & \bullet & \bullet \\
         &     &     & \bb     & \bb     \\
         &     &     & \bb     & \bb     \\
         &     &     & \bb     & \bb
  \end{array}.
$$
Note that we have removed a dot from our initial example and shortened the pivot vector. However, the upper bound remains the same and the subsequent linear rank metric code
easily transfers to the original example. Consider the lower right blue $3\times 2$ rectangle of dots. This subdiagram corresponds to an MRD code, i.e., for each field size
$q$ there exists a linear rank distance code $\mathcal{M}\subseteq \mathbb{F}_q^{3\times 2}$ of cardinality $q^3$ and minimum rank distance $2$. Since $\mathcal{M}$ is linear,
we can assume the existence of three matrices $M_1,M_2,M_3\in\mathbb{F}_q^{3\times 2}$ with $\langle M_1,M_2,M_3\rangle=\mathcal{M}$ and the three matrices are of the shape
$$
  M_1=\begin{pmatrix} 1 & \star\\
                      0 & \star\\
                      0 & \star \end{pmatrix},
  \quad
  M_2=\begin{pmatrix} 0 & \star\\
                      1 & \star\\
                      0 & \star \end{pmatrix},\text{ and }
  M_3=\begin{pmatrix} 0 & \star\\
                      0 & \star\\
                      1 & \star \end{pmatrix}.
$$
With this, $\langle S_1,S_2,S_3\rangle$ is the desired linear rank metric code of cardinality $q^3$ and minimum rank distance $4$, where
$S_i=\begin{pmatrix} M_i^\top & \star \\ 0&M_i\end{pmatrix}$ for $1\le i\le 3$ (actually, we can replace the latter $\star$-entry by a zero matrix.)

\medskip

As an example for a complete Echelon-Ferrers construction we consider the skeleton code
$$
  \mathcal{H}=\left\{11111000000000,00001111100000,00010100011100,00100000100111\right\}\subseteq\mathbb{F}_2^{14}
$$
with corresponding Echelon-Ferrers diagrams
$$
  \begin{array}{lllllllll}
     \bullet & \bullet & \bullet & \bullet & \bullet & \bullet & \bullet & \bullet & \bullet \\
     \bullet & \bullet & \bullet & \bullet & \bullet & \bullet & \bullet & \bullet & \bullet \\
     \bullet & \bullet & \bullet & \bullet & \bullet & \bullet & \bullet & \bullet & \bullet \\
     \bullet & \bullet & \bullet & \bullet & \bullet & \bullet & \bullet & \bullet & \bullet \\
     \bullet & \bullet & \bullet & \bullet & \bullet & \bullet & \bullet & \bullet & \bullet
  \end{array}
  \,\,
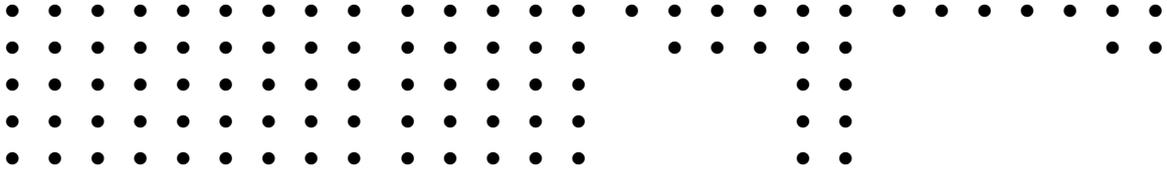
  \begin{array}{lllll}
     \bullet & \bullet & \bullet & \bullet & \bullet \\
     \bullet & \bullet & \bullet & \bullet & \bullet \\
     \bullet & \bullet & \bullet & \bullet & \bullet \\
     \bullet & \bullet & \bullet & \bullet & \bullet \\
     \bullet & \bullet & \bullet & \bullet & \bullet
  \end{array}
  \,\,
  \begin{array}{llllll}
    \bullet & \bullet & \bullet & \bullet & \bullet & \bullet \\
            & \bullet & \bullet & \bullet & \bullet & \bullet \\
            &         &         &         & \bullet & \bullet \\
            &         &         &         & \bullet & \bullet \\
            &         &         &         & \bullet & \bullet
  \end{array}
  \,\,
  \begin{array}{lllllll}
    \bullet & \bullet & \bullet & \bullet & \bullet & \bullet & \bullet \\
            &         &         &         &         & \bullet & \bullet \\
            &         &         &         &         &         &         \\
            &         &         &         &         &         &         \\
            &         &         &         &         &         &
  \end{array}
$$
Theorem~\ref{thm_upper_bound_ef} gives the upper bounds $q^{18}$, $q^{10}$, $q^3$, and $q^0$ for the four subcodes $\mathcal{C}_h$, respectively.
Since the first two Echelon-Ferrers diagrams are rectangular they can be realized by MRD codes. The third Echelon-Ferrers diagram is exactly the one
that we have treated before. Since the fourth diagram can be realized in a trivial way by a zero matrix we obtain
\begin{equation}
  A_q(14,8;5)\ge M_q(14,8;5)\ge q^{18}+q^{10}+q^3+q^0,
\end{equation}
cf.~\cite[Table 1]{gorla2017subspace}.

\section{An algorithm for the maximum clique problem with polynomial weights}
\label{sec_weighted_max_clique}

Let $G=(V,E)$ be an undirected  graph with vertex set $V$ and edge set $E$. A \emph{clique} of $G$ is a subset $U$ of $V$ such that $\big\{\{a,b\}\,:\,
a,b\in U\text{ with }a\neq b\big\}\subseteq E$. The problem of finding a clique of maximum possible cardinality is NP-complete for arbitrary graphs
and polynomial time solvable for perfect graphs, see e.g.~\cite{pardalos1994maximum}. Since this problem occurs in many applications a lot of
different algorithms have been proposed, see e.g.~\cite{wood1997algorithm,ostergaard2002fast}. A variant of the maximum clique problem is the
\emph{weighted maximum clique problem}, where we have a weight $w(v)$ for each vertex $v\in V$ and aim to maximize $w(U)=\sum_{u\in U} w(u)$ over
the set of cliques $U$ of $G$. Here we want to study the situation where the weights $w(v)$ are polynomials and we want to find for each integer
$i\ge 2$ a clique $U$ that maximizes $w(U)$ evaluated at $i$ simultaneously.\footnote{The assumption $i\ge 2$ comes from our subsequent application,
where only sizes of finite fields can be attained. This assumption is not crucial for the following considerations. In order to ease the notation
we take this assumption nevertheless and leave the necessary small modifications for the more general case to the reader.}

Assuming that the upper bound of Theorem~\ref{thm_upper_bound_ef} is tight, the determination of $M_q(n,d;k)$ parametric in $q$ is of that type. The
first problem we have to face is that there is no total ordering of polynomials evaluated at positive integers. If the evaluation point is fixed
then we are in the situation of real numbers, where we have a total ordering. For two polynomials $f$, $g$ and a positive integer $i$ we write
$f \succeq_i g$ if $f(i)\ge g(i)$ and $f \succ_i g$ if $f(i)> g(i)$. If $f(x)=\sum_{l=0}^n f_l\cdot x^l$ and $g(x)=\sum_{l=0}^n g_l\cdot x^l$ for
some large enough integer $n$, then we write $f\succ_\infty g$ if we have $f_j>g_j$ for the largest index $j$ where $f_j$ and $g_j$ differ. We write
$f\succeq_\infty g$ if either $f\succ_\infty g$ or $f=g$, noting that also $\succeq_\infty$ is a total ordering. As an abbreviation we
write $f\succeq g$ if $f \succeq_q g$ for all $q\in \mathbb{N}$ with $q\ge 2$ and $f\succ g$ if $f\succ_q g$ for all $q\in \mathbb{N}$ with $q\ge 2$. However,
for the polynomials \begin{eqnarray*}
  f(q)&=&q^{28}+q^{24}+q^{22}+8q^{20}+q^{19}+2q^{18}+3q^{17}+5q^{16}+3q^{15}+3q^{14}+4q^{13}\\&&+2q^{12}+3q^{11}+5q^{10}+6q^9+5q^8+4q^7+3q^6+5q^5+q^3+q^2+q^0
\end{eqnarray*}
and
\begin{eqnarray*}
  g(q)&=&q^{28}+q^{24}+q^{22}+8q^{20}+q^{19}+3q^{18}+q^{17}+4q^{16}+4q^{15}+5q^{14}+q^{13}\\&&+4q^{12}+5q^{11}+6q^{10}+3q^9+5q^8+3q^7+4q^6+q^5+2q^4+q^2+q^1+q^0,
\end{eqnarray*}
which occur at the determination of $M_q(12,4;5)$, we have $f\succ_2 g$ and $g\succ_q f$ for all $q\ge 3$. I.e., we have neither $f\succeq g$ nor $f\preceq g$.

Since we need to compare polynomials in the subsequent algorithm, we define the function \texttt{IsStrictlyBetter}$(f,g)$ such that it is true
iff there exists at least one integer $i\ge 2$ such that $f\succ_i g$. Since \texttt{IsStrictlyBetter}$(f,g)=$\texttt{IsStrictlyBetter}$(f+h,g+h)$
for every polynomial $h$, we can assume that $f(x)=\sum_{l=0}^n f_lx^l$ and $g(x)=\sum_{l=0}^n g_lx^l$ are given such that $f_l,g_l\ge 0$ and $f_lg_l=0$
for all $0\le l\le n$. If $f\neq g$, let $j$ be the largest index with $f_j+g_j>0$. If $f_j>0$, then \texttt{IsStrictlyBetter}$(f,g)$ is true since
$f\succ_{\infty} g$, i.e., $f\succ_{q} g$ for all sufficiently large $q$. If $g_j>0$ and $j=0$, then \texttt{IsStrictlyBetter}$(f,g)$ is false.
If $g_j>0$ and $j\ge 1$, then let $\lambda$ be the largest real number such that $\sum_{l=0}^{j-1} f_l\cdot\lambda^{j-1}=g_j\lambda^j$, i.e.,
$\lambda=\sum_{l=0}^{j-1} f_l/g_j$. For all $q\ge \lambda$ we have $f\preceq_q g$, so that $f\succ_q g$ needs to be checked for $2\le q<\lambda$ only.

Let $ub$ be an upper bound on the maximum (unweighted) clique of $G$ and $\overline{w}(v)$ such that $\overline{w}(v)\succeq w(v)$ for all $v\in V$ and
$v_1,\dots,v_{\# V}$ be an ordering of the vertices in $V$ such that $\overline{w}(v_i)\succeq \overline{w}(v_{i+1})$ for all indices $1\le i<\# V$. We
remark that if the $w(v)$ all are monomials, then we can take $\overline{w}(v)=w(v)$ for all $v\in V$ and the desired ordering exists. Otherwise, if
$w(v)=\sum_{i=0}^n f_iq^i$, then we can e.g.\ set $\overline{w}(v)=q^n\cdot \sum_{i=0}^n\left|f_i\right|$ for arbitrary $q$. Similar as $w(U)=\sum_{u\in U} w(u)$ we
also write $\overline{w}(U)=\sum_{u\in U}\overline{w}(u)$ for each $U\subseteq V$. Since $\succeq$ is a total ordering for all $\overline{w}(v)$,
where $v\in V$, we write $\min\{\overline{w}(u)\,:\, u\in U\}$ for $\overline{w}(u')$ where $u'\in U$ with $\overline{w}(u)\succeq\overline{w}(u')$ for
all $u\in U$. Given the described ordering we can compute an upper bound for each clique $C'$ containing a sub clique $C$.

\SetKwFunction{dive}{Dive}
\SetKwFunction{ub}{UB}
\SetKwFunction{isb}{IsStrictlyBetter}
\SetKwFunction{newrecord}{NewRecord}
\begin{algorithm}[htp]
  \KwIn{graph $G=(V,E)$ with an ordering of the vertices in $V$ as described above, two weight functions $w\colon V\to\mathbb{R}_{\ge 0}[x]$ and
  $\overline{w}\colon V\to\mathbb{R}_{\ge 0}[x]$, a clique $U$ in $G$, an upper bound $ub$ on the maximum clique size in $G$}
  \KwOut{An upper bound $f$ for the weight of every clique extension of $U$}
  $f\longleftarrow w(U)$\;
  $\hat{U}\longleftarrow U$\;
  \For{$i$ from $1$ to $\# V$}
  {
    \If{$v_i\notin U$ \textbf{and} $\Big\{\left\{x,v_i\right\}\,:\,x\in U\Big\}\subseteq E$ \textbf{and} $\#\hat{U}<ub$}
    {
      $f\longleftarrow f+\overline{w}(v_i)$\;
      $\hat{U}\longleftarrow \hat{U}\cup\left\{v_i\right\}$\;
    }
  }
  \Return{$f$}\;
  \caption{\texttt{UB}: upper bound for the weight of an extended clique}\label{alg:clique_ub}
\end{algorithm}

\begin{Lemma}
  \label{lemma_clique_extension}
  Let $G=(V,E)$ be an undirected graph, $w\colon V\to\mathbb{R}_{\ge 0}[x]$ and  $\overline{w}\colon V\to\mathbb{R}_{\ge 0}[x]$ be weight functions satisfying
  $\overline{w}(v)\succeq w(v)$ for all $v\in V$ and $v_1,\dots,v_{\# V}$ be an ordering of the vertices in $V$ such that
  $\overline{w}(v_i)\succeq \overline{w}(v_{i+1})$ for all indices $1\le i<\# V$. If $U$ and $U'$ are cliques in $G$ with $U\subseteq U'$, then we
  have $w(U')\preceq f$, where $f$ is the polynomial returned by Algorithm \ref{alg:clique_ub}.
\end{Lemma}
\begin{Proof}
  Let $cand$ be the set of vertices $v$ in $V\backslash U$ with $\Big\{\left\{x,v\right\}\,:\,x\in U\Big\}\subseteq E$: Since $U'$ is a clique containing
  $U$ we have $U'\backslash U\subseteq cand$. If $\# cand\le ub-\# U$, then from $\#U'\le ub$ we conclude $U'\subseteq U\cup cand=\hat{U}$, so that
  $$
    f=w(U)+\overline{w}(cand)\succeq w(U)+w(cand)\succeq w(U').
  $$
  Otherwise we have $\# \hat{U}\backslash U\ge \# U'\backslash U$. So, due to the assumed ordering of the vertices in $V$ we have $\overline{w}(\hat{U}\backslash U)
  \succeq \overline{w}(U'\backslash U)$, so that
  $$
    f=w(U)+\overline{w}(\hat{U}\backslash U)\succeq w(U)+\overline{w}(U'\backslash U)\succeq w(U').
  $$
\end{Proof}

As an abbreviation we write $\widehat{w}(U)$ for the polynomial $f$ returned by Algorithm~\ref{alg:clique_ub} applied with $U$ whenever the other parameters
are clear from the context.

Given an additional parameter $1\le max\_dive\le ub$ our strategy is to indirectly consider all cliques of size at most $max\_dive$ in $G$.

\begin{algorithm}[htp]
  \KwIn{graph $G=(V,E)$ with an ordering of the vertices in $V$ as described above, two weight functions $w\colon V\to\mathbb{R}_{\ge 0}[x]$ and
  $\overline{w}\colon V\to\mathbb{R}_{\ge 0}[x]$, an upper bound $ub$ on the maximum clique size in $G$, and a parameter $1\le max\_dive\le ub$}
  \KwOut{A list $\mathcal{U}$ of cliques of $G$ that contains a weight maximum clique $U$ with respect to $w(U)[q]$ and $\#U\le max\_dive$ for each
  integer $q\ge 2$ and a list $\widehat{\mathcal{U}}$ of cliques of $G$ that yields a general upper bound on the maximum weight of a clique in $G$}
  \tcp{global data structures:}
  $\mathcal{U} \longleftarrow \emptyset$\;
  $\widehat{\mathcal{U}} \longleftarrow \emptyset$\;
  \tcp{local data structures:}
  $sol \longleftarrow \emptyset$\;
  \dive($G$, $sol$, $w$, $\overline{w}$, $ub$, $max\_dive$)\;
  \Return{$\mathcal{U}$}\;
  \caption{Framework for the maximum weight clique algorithm}\label{alg:maximum_weight_clique}
\end{algorithm}

\begin{algorithm}[htp]
  \KwIn{clique $sol\subseteq V$ and the input data from Algorithm~\ref{alg:maximum_weight_clique}}
  \KwOut{-}
  \newrecord($sol$, $w$, $\overline{w}$, $ub$, $max\_dive$)\;
  Let $1\le l\le \# V$ be the smallest index such that the elements in $sol$ have strictly smaller indices; return if no such index exists\;
  \If{$\#sol\ge max\_dive$}
  {
    \Return{}\;
  }
  \For{$i$ from $l$ to $\# V$}
  {
    \If{$\Big\{\left\{x,v_i\right\}\,:\,x\in sol\Big\}\subseteq E$}
    {
      $f \longleftarrow w(sol)+\left(max\_dive-\#sol\right)\cdot \overline{w}(v_i)$\;
      $\widehat{f} \longleftarrow w(sol)+\left(ub-\#sol\right)\cdot \overline{w}(v_i)$\;
      \If{\isb($f$,$w(U)$)$=$false for at least one $U\in\mathcal{U}$ \textbf{and} \isb($\widehat{f}$,$\widehat{w}(\widehat{U})$)$=$false for at least one $\widehat{U}\in\widehat{\mathcal{U}}$}
      {
        \Return{}\;
      }
      $cand\longleftarrow sol\cup\left\{v_i\right\}$\;
      $f'\longleftarrow$ \ub($G$, $cand$, $w$, $\overline{w}$, $max\_dive$)\;
      $\widehat{f}'\longleftarrow$ \ub($G$, $cand$, $w$, $\overline{w}$, $ub$)\;
      \If{\isb($f'$,$w(U)$)$=$true for all $U\in\mathcal{U}$ \textbf{or} \isb($\widehat{f}'$,$\widehat{w}(\widehat{U})$)$=$true for all $\widehat{U}\in\widehat{\mathcal{U}}$}
      {
        \dive($G$, $sol\cup\left\{v_i\right\}$, $w$, $\overline{w}$, $ub$,$max\_dive$)\;
      }
    }
  }
  \Return{}\;
  \caption{Subroutine \texttt{Dive}}\label{alg:dive}
\end{algorithm}

\begin{algorithm}[htp]
  \KwIn{clique $sol\subseteq V$ and the input data from Algorithm~\ref{alg:maximum_weight_clique}}
  \KwOut{-}
  \tcp{The function just updates $\mathcal{U}$ and $\widehat{\mathcal{U}}$}
  \If{\isb($w(sol)$,$w(U)$)$=$true for all $U\in\mathcal{U}$}
  {
    $\mathcal{U}\longleftarrow\mathcal{U}\cup\{sol\}$\;
    \For{$U\in\mathcal{U}$}
    {
      \If{\isb($w(U)$,$w(U')$)$=$false for at least one $U'\in\mathcal{U}\backslash \{U\}$}
      {
        remove $U$ from $\mathcal{U}$\;
      }
    }
  }
  \If{\isb($\widehat{w}(sol)$,$\widehat{w}(U)$)$=$true for all $\widehat{U}\in\widehat{\mathcal{U}}$}
  {
    $\widehat{\mathcal{U}}\longleftarrow\widehat{\mathcal{U}}\cup\{sol\}$\;
    \For{$U\in\widehat{\mathcal{U}}$}
    {
      \If{\isb($\widehat{w}(U)$,$\widehat{w}(U')$)$=$false for at least one $U'\in\widehat{\mathcal{U}}\backslash \{U\}$}
      {
        remove $U$ from $\widehat{\mathcal{U}}$\;
      }
    }
  }
  \Return{}\;
  \caption{Subroutine \texttt{NewRecord}}\label{alg:new_record}
\end{algorithm}

\begin{Proposition}
  \label{prop_max_weighted_clique}
  Let $G=(V,E)$ be an undirected graph, $w\colon V\to\mathbb{R}_{\ge 0}[x]$ and  $\overline{w}\colon V\to\mathbb{R}_{\ge 0}[x]$ be weight functions satisfying
  $\overline{w}(v)\succeq w(v)$ for all $v\in V$ and $v_1,\dots,v_{\# V}$ be an ordering of the vertices in $V$ such that
  $\overline{w}(v_i)\succeq \overline{w}(v_{i+1})$ for all indices $1\le i<\# V$.
  If $1\le max\_dive\le ub$ are integers such that the maximum clique size in $G$ is at most $ub$, then Algorithm~\ref{alg:maximum_weight_clique} computes a set
  $\mathcal{U}$ of cliques of $G$ such that for each clique $C$ of $G$ with $\# C\le max\_dive$ and each integer $q\ge 2$ there
  exists an element $U\in\mathcal{U}$ with $w(U)\succeq_q w(C)$. Moreover, Algorithm~\ref{alg:maximum_weight_clique} computes a set $\widehat{\mathcal{U}}$ of cliques
  of $G$ such that for each clique $C$ of $G$ 
  and each integer $q\ge 2$ there exists an element $\widehat{U}\in\widehat{\mathcal{U}}$ with $\widehat{w}(\widehat{U})\succeq_q w(C)$.
\end{Proposition}
\begin{Proof}
  Due to the condition $\Big\{\left\{x,v_i\right\}\,:\,x\in sol\Big\}\subseteq E$ and the recursive calls of the subroutine \texttt{Dive} the set $sol$ always
  is a clique in $G$. Moreover, we have $\#sol\le max\_dive$. After the initialization of $\mathcal{U}$ and $\widehat{\mathcal{U}}$ in Algorithm~\ref{alg:maximum_weight_clique},
  those sets are  only changed by the subroutine \texttt{NewRecord}, which is called only at the start of the subroutine \texttt{Dive}. The subroutine \texttt{Dive}
  calls itself where the cardinality of $sol$ is increased by exactly $1$ in each recursion and in the initial call in Algorithm~\ref{alg:maximum_weight_clique}. Thus,
  all elements of $\mathcal{U}$ and $\widehat{\mathcal{U}}$ are cliques of maximum size at most $max\_dive$ in $G$ at any time of the algorithm.

  Let $q\ge 2$ be an arbitrary but fixed integer and $C$ an arbitrary clique of $G$. We assume $C=\left\{v_{i_1},\dots,v_{i_{\# C}}\right\}$,
  where $1\le v_{i_1}<\dots < v_{i_{\# C}}\le\# V$, and set $C_j=\left\{v_{i_1},\dots,v_{i_{j}}\right\}$ for all $1\le j\le\# C$. We have to show the existence
  of an element $\widehat{U}\in\widehat{\mathcal{U}}$ with $\widehat{w}(\widehat{U})\succeq_q w(C)$\footnote{We remind the reader that we write $\widehat{w}(U)$ 
  for the polynomial $f$ returned by Algorithm~\ref{alg:clique_ub} applied to $U$.} and, if $\#C\le max\_dive$, the existence of an element
  $U\in\mathcal{U}$ with $w(U)\succeq_q w(C)$. Note that $\widehat{w}(U)\succeq_q w(U)$ and for $\#C\ge max\_dive$ we have $\widehat{w}(C_{max\_dive}) \succeq_q w(C)$
  due to Lemma~\ref{lemma_clique_extension}.

  Now let $j$ be the maximum index such that \texttt{Dive} is called with $sol=C_j$. Note that $j\le \min\left\{max\_dive,\#C\right\}$. If $j=\#C$, then
  the subroutine \texttt{NewRecord} is called with $sol=C_j$. Then, either there exists $U\in\mathcal{U}$ with $w(U)\succeq_q w(C)$ or $C$ is added to $\mathcal{U}$ and
  either there exists $\widehat{U}\in\widehat{\mathcal{U}}$ with $\widehat{w}(\widehat{U})\succeq_q \widehat{w}(C)\succeq_q w(C)$ or $C$ is added to $\widehat{\mathcal{U}}$.
  If $j=max\_dive<\#C$, then either $C_j$ is added to $\widehat{\mathcal{U}}$ and $\widehat{w}(C_j)\succeq_q w(C)$ or there exists an element
  $\widehat{U}\in\widehat{\mathcal{U}}$ with $\widehat{w}(\widehat{U})\succeq_q\widehat{w}(C_j)\succeq _q w(C)$. In the remaining cases we have $j<max\_dive$ and $j<\# C$.
  Note that $C_{j+1}=C_j\cup \left\{v_{i_{j+1}}\right\}$. Now, assume that $i_j<i\le i_{j+1}$ is an index with $\Big\{\left\{x,v_i\right\}\,:\,x\in sol\Big\}\subseteq E$ and there exist
  $U\in\mathcal{U}$ with \isb($f$,$w(U)$)$=$false and $\widehat{U}\in\widehat{\mathcal{U}}$ with \isb($\widehat{f}$,$\widehat{w}(\widehat{U})$)$=$false. Thus, we have
  $w(U)\succeq_q f$ and $\widehat{w}(\widehat{U})\succeq_q \widehat{f}$. If $\#C\le max\_dive$, then we have
  \begin{eqnarray*}
    f&\succeq_q& w(C_j)+\left(max\_dive-\#sol\right)\cdot \overline{w}(v_i)\succeq_q w(C_j)+\left(max\_dive-\#sol\right)\cdot \overline{w}(v_{i_{j+1}}) \\
    & \succeq_q& w(C_j)+\sum_{u\in C\backslash C_j} \overline{w}(u)\succeq_q w(C),
  \end{eqnarray*}
  so that $w(U)\succeq_q f\succeq_q w(C)$. Without any assumption on the clique size we have
  \begin{eqnarray*}
    \widehat{f}&\succeq_q& w(C_j)+\left(ub-\#sol\right)\cdot \overline{w}(v_i)
    \succeq_q w(C_j)+\left(\#C-\#sol\right)\cdot \overline{w}(v_{i_{j+1}}) \\
    & \succeq_q& w(C_j)+\sum_{u\in C\backslash C_j} \overline{w}(u)\succeq_q w(C),
  \end{eqnarray*}
  so that $\widehat{w}(\widehat{U})\succeq_q \widehat{f}\succeq_q w(C)$. If no such index $i$ exists, then the loop reaches $i=i_{j+1}$ and we note that
  $\Big\{\left\{x,v_{i_{j+1}}\right\}\,:\,x\in sol\Big\}\subseteq E$. Thus, we have $cand=C_{j+1}$.
  From Lemma~\ref{lemma_clique_extension}
  we can conclude $f'\succeq_q w(C)$ if $\#C\le max\_dive$ and $\widehat{f}'\succeq_q w(C)$ in general. The assumption that \texttt{Dive} is not called with
  $sol=C_{j+1}$ yields that \isb($f'$,$w(U)$)$=$false for an element $U\in\mathcal{U}$ and \isb($\widehat{f}'$,$\widehat{w}(\widehat{U})$)$=$false for an element 
  $\widehat{U}\in\widehat{\mathcal{U}}$. Thus, we have $\widetilde{w}(\widetilde{U})\succeq\widetilde{f}'\succeq w(C)$ and $w(U)\succeq_q f'\succeq_q w(C)$
  if $\#C\le max\_dive$.

  Now, let us sum up the conclusion of the previous case analysis. In any case we have the following. If $\#C\le max\_dive$ then there exists a clique $U$ of $G$ with
  $\#U\le max\_dive$ and $w(U)\succeq_q w(C)$ such that $U$ was added to $\mathcal{U}$ at some point during the execution of Algorithm~\ref{alg:maximum_weight_clique}.
  Similarly, without any assumption on the cardinality of $C$, there exists a clique $\widehat{U}$ of $G$ with $\#\widehat{U}\le max\_dive$ and $\widehat{w}(\widehat{U})
  \succeq_q w(C)$ such that $\widehat{U}$ was added to $\widehat{\mathcal{U}}$ at some point during the execution of Algorithm~\ref{alg:maximum_weight_clique}.

  Finally, we observe that removals from $\mathcal{U}$ or $\widehat{\mathcal{U}}$ are only performed in the subroutine \texttt{NewRecord}. However, $U$ is only removed
  from $\mathcal{U}$ if there exists an element $U'\in \mathcal{U}$ with $w(U')\succeq_q w(U)$, so that we can iteratively replace $U$ by $U'$. Similarly, $\widehat{U}$
  is only removed from $\widehat{\mathcal{U}}$ if there exists an element $U'\in\widehat{\mathcal{U}}$ with $\widehat{w}(U')\succeq_q\widehat{w}(\widehat{U})$, so that
  we can iteratively replace $\widehat{U}$ by $U'$.
\end{Proof}

We can apply Proposition~\ref{prop_max_weighted_clique} and Algorithm~\ref{alg:maximum_weight_clique} in order to compute the exact value of $\overline{M}_q(n,d;k)\ge M_q(n,d;k)$
for all integers $q\ge 2$ for moderate parameters $n$, $d$ and $k$. To this end let $V$ be the set of all binary vectors in $\mathbb{F}_2^n$ of Hamming weight $k$.
For each pair of different $u,v\in V$ we have $\{u,v\}\in E$ iff $d_H(u,v)\ge d$. As weight function we use the upper bound of Theorem~\ref{thm_upper_bound_ef}
for both $w(v)$ and $\overline{w}(v)$ for all $v\in V$. The used value of $max\_dive=ub$ is taken from \cite{brouwer1990new} and 
https://www.win.tue.nl/$\sim$aeb/codes/Andw.html. An example is given by
\begin{eqnarray}
  \overline{M}_q(14,6;4) &=& q^{20}+q^{14}+q^{10}+q^9+q^8+2q^6+2q^5+2q^4+q^3+q^2,
\end{eqnarray}
where we have used $max\_dive=ub=14=A_1(14,6;4)$. It took 45558~iterations, i.e.\ calls of the subroutine \texttt{Dive}, to compute this upper bound in less than a second. An attaining set of
$13<14$ pivot vectors is given by $11110000000000$, $00011110000000$, $00100011100000$, $01000101010000$, $10001001001000$, $00010000111000$, $01001000100100$,
$10000010010100$, $10000100100010$, $00100100001100$, $00101000010010$, $01000010001010$, $00010001000110$. Since these 0/1 vectors are quite long we will represent them as integers
having the corresponding base $2$ representation, i.e., $b_1b_2\dots b_n$ is mapped to the integer $\sum_{i=1}^n b_i\cdot 2^{n-i}$. In our example we obtain
$$
  \{15360, 1920, 2272, 4432, 8776, 1080, 4644, 8340, 8482, 2316, 2578, 4234, 1094\}.
$$
As shown in e.g.\ \cite{gorla2017subspace} 
the upper bound $\overline{M}_q(14,6;4)\ge M_q(14,6;4)$ can indeed be attained, i.e., the upper bound of Theorem~\ref{thm_upper_bound_ef} can be reached for the $13$ used
pivot vectors.

Another example is given by
\begin{eqnarray}
  \overline{M}_q(15,10;6) &=& q^{18}+q^{5}+q^{0},
\end{eqnarray}
where we have used $max\_dive=ub=3=A_1(15,10;6)$. In its first steps Algorithm~\ref{alg:maximum_weight_clique} greedily selects the
pivot vectors $111111000000000\overset{\wedge}{=}32256$, $100000111110000\overset{\wedge}{=}16880$, and $000001000011111\overset{\wedge}{=}543$,
that already give the tightest upper bound. At the point where $sol=\{32256,16880\}$ should be further extended by a node $v_i$, we obtain
$f(q)=q^{18}+q^{5}+q^0$ and $f'(q)=q^{18}+q^{5}+q^{0}$, so that this branch is cut off due to the check applied to $f'$. In the next step the clique
$sol=\{32256\}$ gets cut off. Then algorithm tries the unique one-element clique with polynomial $q^{17}$. After cutting off we have $sol=\emptyset$, where
we can bound with $f(q)=3\cdot q^{16}\leq q^{18}+q^{5}+q^{0}$. Also in this example the upper bound can be attained, i.e., we have $\overline{M}_q(15,10;6)=M_q(15,10;6)$.

We remark that Algorithm~\ref{alg:maximum_weight_clique} does not need too much computation time for all cases where $2k\ge d\ge 10$ and $2k\le n\le 19$. We list the
upper bounds based on Theorem~\ref{thm_upper_bound_ef} in Appendix~\ref{sec_upper_polynomial}. We remark that for all these cases, except for $M_q(19,10;9)$,
the upper bound $M_q(n,d;k)$ is indeed a polynomial that is valid for all field sizes $q\ge 2$, i.e., one skeleton code can be used for all field sizes. In the exceptional
case there is an upper bound for $M_q(19,10;9)$ for all field
sizes $q\ge 3$ corresponding to a skeleton code with $13$ elements, while there is a different upper bound for the binary case $M_2(19,10;9)$ corresponding to a
skeleton code with $A_1(19,10;9)=19$ elements, see Appendix~\ref{sec_upper_polynomial}. In general, it seems that if there are different skeleton codes that
yield different upper bounds for different field sizes, then the skeleton codes for smaller field sizes have larger cardinality. For larger field sizes the cardinality
of the skeleton code yielding the tightest upper bound can have a cardinality which is significantly smaller than $A_1(n,d;k)$.

If $d\le 8$, then Algorithm~\ref{alg:maximum_weight_clique} partially needs quite some computation time. This is due to several facts: For fixed parameters $n$ and $k$ the
sizes of the skeleton codes increase with decreasing distance $d$. Even more importantly, the number of vertices of our graphs can explode. More precisely, the graph $G=(V,E)$
for the determination of an upper bound for $M_q(n,d;k)$ has
\begin{equation}
  \# V \,=\, {n\choose k}
\end{equation}
vertices. If $\#\mathcal{U}>1$ in intermediate steps of Algorithm~\ref{alg:maximum_weight_clique}, i.e., there is no unique current best solution valid for all field sizes,
the derived cuts can be too weak resulting in many traversed partial cliques. An example is given by e.g.\ $M_q(17,8;7)$. However, this can be easily prevented. We can
run Algorithm~\ref{alg:maximum_weight_clique} for all {\lq\lq}small{\rq\rq} field sizes $q\le \Lambda$ separately. Here the somewhat complex function \texttt{IsStrictlyBetter}$(f,g)$
can be replaced by the direct check $f\succ_q g$. For the remaining cases $q>\Lambda$ we can adjust \texttt{IsStrictlyBetter}$(f,g)$ such that it assumes $q\ge \Lambda+1$ when checking
the {\lq\lq}small{\rq\rq} cases directly. Of course the cutting works also better if the algorithm already has found a relatively good solution. Otherwise it may happen that
the algorithm wastes its time in a region of the combinatorial search space with a lot of similar solutions which later are superseded by a better solution in some
different region of the search space. In order to find {\lq\lq}good{\rq\rq} initial partial cliques and to get an idea how $\mathcal{U}$ will be splitted among the field sizes,
one can perform a partial incomplete search by suitably setting the parameter $ub$ to a value strictly smaller than $A_1(n,d;k)$. (Of course, one can also increase the value of $ub$
in several iterations, each time taking the best found clique from the previous run as a starting solution.)

However, some cases remain quite hard. For e.g.\ $M_2(18,8;7)$, i.e., where we already fixed the field size to $q=2$, it took Algorithm~\ref{alg:maximum_weight_clique}
$2\,073\,919\,117$ calls of \texttt{Dive} and 3~hours of computation time on an ordinary laptop to determine the, with respect to Theorem~\ref{thm_upper_bound_ef}, tightest upper
bound for $M_2(18,8;7)$, with a corresponding clique of size $26<33=A_1(18,8;7)$. Sometimes there are many cliques that are equally good. An example is given by
$M_q(12,4;3)$, where Algorithm~\ref{alg:maximum_weight_clique} enumerated several million 
cliques of exactly the same (optimal) weight. It is no surprise that the instances get quite hard for $d\le 8$, since e.g.\
$46\le A_1(18,8;8)\le 49$, $48\le A_1(18,9;8)\le 58$, and $88\le A_1(19,8;9)\le 103$ are the tightest known bounds on $A_1(n,d;k)$, which we can use for $ub$. We remark that
the, with respect to Theorem~\ref{thm_upper_bound_ef}, tightest upper bound for $M_2(18,8;8)$ is indeed given by a skeleton code of cardinality $46$. However, this does not
answer the question whether $A_1(18,8;8)=46$ or $A_1(18,8;8)>46$.

Next, we want to discuss alternative approaches that we can apply in a subproblem, i.e., in the case where a partial clique $u$ is given and we only have to decide whether there
is an extension of $u$ that needs to be added to $\mathcal{U}$. For the unweighted case a well-known standard formulation as an integer linear programming (ILP)
is:
\begin{align*}
  \max \quad& \sum_{v\in V} x_v\\
  \text{s.t.}\quad& x_u+x_v \le 1 &\forall \{u,v\}\in E\\
  & x_v\in\{0,1\} &\forall x\in V
\end{align*}
The maximum clique corresponding to an optimal solution $x^\star$ is given by $\left\{v\in V\,:\, x_v^\star=1\right\}$. However, this ILP model usually has a large integrality
gap, i.e., the target value of the optimal solution of its continuous relaxation is much larger than the one of the original binary problem. If we have an independent set
$I\subseteq V$, i.e., a set of vertices such that no two are joined by an edge in $E$, then we can add the extra constraint $\sum_{i\in I} x_i\le 1$. If we have a set $\mathcal{I}$
of independent sets of $G=(V,E)$ such that for each edge $e\in E$ there exists an independent set $I_e\in\mathcal{E}$ with $e\subseteq I_e$, then the maximum clique size
is also attained by:
\begin{align*}
  \max \quad& \sum_{v\in V} x_v\\
  \text{s.t.}\quad& \sum_{i\in I} x_i\le 1 &\forall I\in\mathcal{I}\\
  & x_v\in\{0,1\} &\forall x\in V
\end{align*}
Of course the formulation gets better if the independent sets get large. In our situation we can choose for every subset $S$ of $\{1,\dots,n\}$ of cardinality
$t:=k-\tfrac{d}{2}+1$ the independent set $I_S$ as the set of vertices which have a $1$ in their binary representation as a pivot vector at all positions contained
in $S$. Note that two elements from $u,v\in I_S$ coincide in at least $t$ positions so that $d_H(u,v)<d$. Even terminating the solution process of the above ILP
after an arbitrary amount of time gives an upper bound on the maximum (unweighted) clique size of a given graph $G=(V,E)$.


The ILP approach is not limited to the unweighted maximum clique problem. Given weights $w_v$ for each vertex, we just have to maximize $\sum_{v\in V} w_vx_v$ instead
of $\sum_{v\in V}x_v$. However, in our situation the weights, if not polynomials anyway, can be quite large, which causes numerical problems. This is also true for
most available implementations of weighted maximum clique algorithms, which use integers of a restricted size to store the weights. In the situation of a subgraph $G_u$
for some partial clique $u$, the weights might be small enough so that we can apply the ILP mode for the weighted maximum clique problem directly, assuming that we have
fixed the field size to some small number. An example is given by the determination of an upper bound for $M_2(13,4;4)$. Since the leading coefficient is $q^{27}$, the weights
can get as large as $134\,217\,728$ even in the binary case. However, if we fix the first $17$ elements of a finally optimal skeleton code, which will have size $55$ in the end,
the maximum possible weight is $4096$, which is small enough for a reliable numerical evaluation. We remark that this approach might be essentially useful for the situations where we
have a lot of optimal cliques of equal weight.

For a small example, i.e., $M_q(14,8;5)$, we want to demonstrate how the ILP formulation for the weighted maximum clique problem can be utilized to solve the parametric
case. Of course, we cannot use the polynomial weights directly. Instead of this, we introduce integer counting variables $a_i$ that count the number of chosen
vertices whose weight polynomial is $q^i$:
\begin{align*}
  \max \quad& \sum_i c_ia_i \\
  \text{s.t.}\quad& \sum_{i\in I} x_i\le 1 &\forall I\in\mathcal{I}\\
  & -a_i+\sum_{v\in V\,:\, w(v)=q^i} x_v =0&\forall i\\
  & \sum_i a_i \le A_1(n,d;k)\\
  & x_v\in\{0,1\} &\forall x\in V\\
  & a_i\in \mathbb{N} &\forall i,
\end{align*}
where the set $\mathcal{I}$ of independent sets is constructed as described above. For the target function we will choose different coefficients $c_i$ in
different evaluations. If the maximum target value is given my $\pi$, then we have concluded the valid inequality $\sum_{i}c_ia_i\le \pi$. In some cases
we will impose further assumptions.

In our example we have $A_1(14,8;5)=4$ and the maximum exponent of $w(v)=q^i$ is given by $18$, i.e., the sums over $i$ can be restricted to run from $0$ to $18$.
By $m_i$ we denote the number vertices $v\in V$ with $w(v)=q^i$:
\begin{center}
  \setlength{\tabcolsep}{0.97mm}
  \begin{tabular}{llllllllllllllllllll}
  \hline
  $\mathbf{i}$   & 18 & 17 & 16 & 15 & 14 & 13 & 12 & 11 & 10 & 9 &  8   &  7  &   6 &  5  &  4  &  3  &   2 & 1   &  0\\
  $\mathbf{m_i}$ &  1 &  1 &  5 &  5 & 15 & 15 & 35 & 35 & 70 & 70 & 121 & 119 & 177 & 167 & 215 & 192 & 232 & 163 & 364\\
  \hline
  \end{tabular}
\end{center}
By an explicit construction it is known that $M_q(14,8;5)\ge q^{18}+q^{10}+q^3+q^0$ for all field sizes $q\ge 2$, see e.g.~\cite[Table 1]{gorla2017subspace} or 
Appendix~\ref{subsec_mindist_8}. For $a_{18}=0$ we have $a_{17}+a_{16}+a_{15}+a_{14}+a_{13}\le 1$ and $a_{17}+a_{16}+a_{15}+a_{14}+a_{13}+a_{12}+a_{11}+a_{10}+a_9\le 2$.
Since
$$
  q^{18}+q^{10}+q^3+q^0 \succ q^{17}+q^{12}+2q^8
$$
we have $a_{18}=1$ and conclude $\sum_{i=11}^{17}a_i=0$, $a_{10}+a_9+a_8+a_7\le 1$, $a_{10}+a_9+a_8+a_7+a_6+a_5\le 2$. Since
$$
  q^{18}+q^{10}+q^3+q^0 \succ q^{18}+q^9+q^6+q^4
$$
we can assume $a_{10}=1$ and conclude $\sum_{i=4}^9 a_i=0$ and $a_3+a_2\le 1$. Since
$$
  q^{18}+q^{10}+q^3+q^0 \succ q^{18}+q^{10}+q^2+q^1
$$
we can assume $a_3=1$ and conclude $a_2+a_1=0$, so that $M_q(14,8;5)\le q^{18}+q^{10}+q^3+q^0$ for all field sizes $q\ge 2$.

We remark, that all presented results in this paper are verified by exact integer computations, i.e., without using linear programming
formulations.

\section{Constructions for Ferrers diagram rank-metric codes}
\label{sec_constructions}

In Section~\ref{sec_weighted_max_clique} we have used Theorem~\ref{thm_upper_bound_ef} to upper bound $\#\mathcal{C}_v$ in the Echelon-Ferrers construction, so that Algorithm~\ref{alg:maximum_weight_clique} computes upper bounds. The aim of this section is to summarize some constructions for Ferrers diagram rank-metric codes from the literature that give lower bounds on $\#\mathcal{C}_v$ for all possible pivot vectors $v$ (given some parameters $n$, $d$, and $k$).
Choosing the resulting polynomial as weight function $w(v)$, Algorithm~\ref{alg:maximum_weight_clique} computes lower bounds for $M_q(n,d;k)$.

For convenience, a Ferrers diagram $\mathcal F$ is identified with the cardinalities of its columns. Given positive integers $m$, $n$ and $1\leq \gamma_0 \leq \gamma_1 \leq \cdots \leq \gamma_{n-1}\leq m$, there exists a unique Ferrers diagram $\mathcal F$ of size $m\times n$ such that the $(j+1)$-th column of $\mathcal F$ has cardinality $\gamma_j$ for any $0\leq j\leq n-1$. In this case we write $\mathcal F=[\gamma_0,\gamma_1,\ldots,\gamma_{n-1}]$. An FDRM code attaining the upper bound in Theorem~\ref{thm_upper_bound_ef} is called {\em optimal}. So far all known FDRM codes over $\mathbb F_q$ with the largest possible dimension are optimal.

\subsection{Constructions based on subcodes of MRD codes}

MRD codes play an important role in the constructions for FDRM codes. Examining subcodes of MRD codes, one can construct optimal FDRM codes with minimum rank distance $\delta$ whose optimality can be obtained by deleting its rightmost $\delta-1$ columns. This approach produces most of known optimal FDRM codes. The interested reader is referred to Lemma 2.1 in \cite{liu2019several} for the basic idea of this method.

Gabidulin codes are a classical class of MRD codes. By exploring subcodes of restricted Gabidulin codes, Liu, Chang and Feng \cite{liu2019several} presented the following construction that unifies many known constructions for optimal FDRM codes.

\begin{Theorem}\label{the:from sys MRD-new2} \cite[Theorem 2.8]{liu2019several}
Let $l$ be a positive integer and $1=t_0<t_1<t_2<\cdots<t_l$ be integers such that $t_1\mid t_2\mid \cdots\mid t_l$. When $l>1$, let $t_{2}=s_{2}t_{1}$. Let $r$ be a nonnegative integer and $\delta$, $n$, $k$ be positive integers satisfying $r+1\leq \delta\leq n-r$, $t_{l-1}<n-r\leq t_l$, $k=n-\delta+1$ and $k\leq t_1$. Let $\mathcal F=[\gamma_0,\gamma_1,\ldots,\gamma_{n-1}]$ be an $m\times n$ Ferrers diagram $(m=\gamma_{n-1})$ satisfying
\begin{itemize}
\item[$(1)$] $\gamma_{k-1}\leq wt_1$,
\item[$(2)$] $\gamma_{k}\geq wt_1$ for $k<t_1$ and $\delta\geq 2$,
\item[$(3)$] $\gamma_{t_{\theta}}\geq t_{\theta+1}$ for $1\leq \theta \leq l-1$,
\item[$(4)$] $\gamma_{n-r+h}\geq t_l+\sum_{j=0}^{h} \gamma_j$ for $0\leq h\leq r-1$,
\end{itemize}
for some $w\in\{1,2,\ldots,s_2\}$ and for $w=1$ if $l=1$. Then there is an optimal $[\mathcal F, \sum_{i=0}^{k-1} \gamma_i, \delta]_q$ code for any prime power $q$.
\end{Theorem}

In Appendices \ref{sec_upper_polynomial} and \ref{the_best_known_dimension_of_given_FDRMCs}, we examine the lower bounds for $M_q(n,d,k)$ for all $4\le d\le 2k$, and $ 2k\le n\le 19$. Their corresponding optimal FDRM codes that are used to produce subspace codes $\mathcal{C}_v$ in the Echelon-Ferrers construction all have small numbers of rows and columns. For this reason, to apply Theorem \ref{the:from sys MRD-new2}, it is often required that $l=1$ or $w=1$.

Taking $l=1$ and $t_1=n-r$ in Theorem \ref{the:from sys MRD-new2}, we have the following theorem.

\begin{Theorem} \cite[Theorem 3.13]{liu2019constructions} \label{thm:subcodes from Gab}
Let $\delta$, $n$ and $r$ be positive integers satisfying $r+1\leq \delta \leq n-r$. Let $\mathcal{F}$ be an $m\times n$ Ferrers diagram satisfying that
\begin{itemize}
\item[$(1)$] $\gamma_{n-\delta}\leq n-r$,
\item[$(2)$] $\gamma_{n-\delta+1}\geq n-r$,
\item[$(3)$] $\gamma_{n-r+i}\geq n-r+\sum_{j=0}^{i} \gamma_j$ for $0\leq i\leq r-1$.
\end{itemize}
Then there exists an optimal $[\mathcal F, \sum_{i=0}^{n-\delta} \gamma_i, \delta]_q$ code for any prime power $q$.
\end{Theorem}

Theorem \ref{thm:subcodes from Gab} with $r=0$ (resp. $r=1$) can be seen as a generalization of the following Theorem \ref{thm:shortening} (resp. Theorem \ref{thm:from subcodes}). We remark that Theorem \ref{thm:shortening} was first presented by Etzion and Silberstein \cite{etzion2009error}, and its proof was simplified in \cite{etzion2016optimal} by means of shortening systematic MRD codes.

\begin{Theorem} \cite[Theorem 3]{etzion2016optimal} \label{thm:shortening}
Let $m\geq n$ and $\mathcal F=[\gamma_0,\gamma_1,\ldots,\gamma_{n-1}]$ be an $m \times n$ Ferrers diagram satisfying $\gamma_{n-\delta+1}\geq n$. Then there exists an optimal $[\mathcal F, \sum_{i=0}^{n-\delta} \gamma_i, \delta]_q$ code for any prime power $q$.
\end{Theorem}

\begin{Theorem} \cite[Theorem 8]{etzion2016optimal} \label{thm:from subcodes}
Let $2\leq \delta \leq n-1$ and $\mathcal{F}=[\gamma_0,\gamma_1,\ldots,\gamma_{n-1}]$ be an $m\times n$ Ferrers diagram satisfying that $\gamma_{n-\delta+1}\geq n-1$. Then there exists an $[\mathcal F, k, \delta]_q$ code for any prime power $q$, where $k=\min\{m-n+1, \gamma_0\} + \sum_{i=1}^{n-\delta} \gamma_i$. Furthermore, when $\gamma_{n-1}\geq n-1+\gamma_0$, the resulting FDRM code is optimal.
\end{Theorem}

Taking $w=1$ and $r=0$ in Theorem \ref{the:from sys MRD-new2}, we obtain the following theorem.

\begin{Theorem} \cite[Theorem 3.2]{zhang2019constructions} \label{the:from sys MRD-new1}
Let $l$ be a positive integer. Let $1=t_0<t_1<t_2<\cdots<t_l$ be integers such that $t_1\mid t_2\mid \cdots\mid t_l$.  Let $n$ and $\delta$ be positive integers satisfying $t_{l-1}<n\leq t_l$ and $n-t_1+1< \delta\leq n$. Let $\mathcal F$ be an $m\times n$ Ferrers diagram satisfying
\begin{itemize}
\item[$(1)$] $\gamma_{n-\delta}\leq t_1$,
\item[$(2)$] $\gamma_{n-\delta+1}\geq t_1$,
\item[$(3)$] $\gamma_{t_\theta}\geq t_{\theta+1}$ for $1\leq \theta \leq l-1$,
\end{itemize}
 Then there exists an optimal $[\mathcal F, \sum_{i=0}^{n-\delta} \gamma_i, \delta]_q$ code for any prime power $q$.
\end{Theorem}

On the other hand, by examining subcodes of different MRD codes other than Gabidulin codes, it is possible to obtain new optimal FDRM codes. Using a description on generator matrices of a class of systematic MRD codes presented in \cite{antrobus2019maximal}, Liu, Chang and Feng \cite{liu2019several} gave the following class of optimal FDRM codes.

\begin{Theorem} \cite[Theorem 2.3]{liu2019several}  \label{cfdrm}
Let $m\geq n\geq \delta\geq 2$ and $k=n-\delta+1$. If an $m\times n$ Ferrers diagram $\mathcal F=[\gamma_0,\gamma_1,\ldots,\gamma_{n-1}]$ satisfies
\begin{itemize}
\item[$(1)$] $\gamma_{k}\geq n$ or $\gamma_{k}-k\geq \gamma_i-i$ for each $i=0,1,\ldots,k-1$,
\item[$(2)$] $\gamma_{k+1}\geq n$,
\end{itemize}
then there exists an optimal $[\mathcal F,\sum_{i=0}^{k-1} \gamma_i,\delta]_q$ code for any prime power $q$.
\end{Theorem}

\subsection{Constructions from MDS codes}

FDRM codes can be constructed via maximum distance separable (MDS) codes. It is well known that a $[v,v-d+1,d]_{q}$ MDS code exists for any $q\geq v-1$ or $d\in \{1,2,v\}$.

A {\em diagonal} of an $m\times n$ Ferrers diagram $\mathcal F$ with $m\geq n$ is a consecutive sequence of entries, going upwards diagonally from the rightmost column to either the leftmost column or the first row. Let $D_i$, $1\leq i\leq m$, denote the $i$-th diagonal in $\mathcal F$, where $i$ counts the diagonals from the top to the bottom and let $\theta_i$ denote the number of dots on $D_i$ in $\mathcal F$.

\begin{Theorem} \cite[Construction 1]{etzion2016optimal} \label{thm:from MDS}
Let $m\geq n$ and $\mathcal F$ be an $m \times n$ Ferrers diagram. Let $\delta$ be an integer such that $0<\delta \leq n$, and $\theta_{max}=\max_{1\leq i\leq m}{\theta_i}$. Then there exists an $[\mathcal F, k, \delta]_q$ code for any prime power $q\geq \theta_{max}-1$, where $k=\sum_{i=1}^{m} {\max\{0, \theta_i-\delta +1\}}$.
\end{Theorem}

Applying Theorems \ref{thm:shortening} and \ref{thm:from MDS}, we get the following result.

\begin{Theorem}  \label{cor:delta=3}
\begin{itemize}
\item[$(1)$] \cite{etzion2009error} Let $\delta\in\{1,2\}$. There exists an optimal $[\mathcal F, \sum_{i=0}^{n-\delta} \gamma_i, \delta]_q$ code for any Ferrers diagram $\mathcal F$ and any prime power $q$.
\item[$(2)$] \cite[Theorem 11]{etzion2016optimal} Let $n\geq 3$. There exists an optimal $[\mathcal F, k, 3]_q$ code for any $n\times n$ Ferrers diagram $\mathcal F$ and any prime power $q$.
\end{itemize}
\end{Theorem}

\subsection{New FDRM codes by combining old ones} Another flexible way to obtain FDRM codes is to assemble small FDRM codes. This approach sometimes gives rise to optimal FDRM codes with minimum rank distance $\delta$ whose optimality cannot be achieved by deleting its rightmost $\delta-1$ columns.

\begin{Theorem} \cite[Theorem 9]{etzion2016optimal} \label{thm:combine with same dim}
Let $\mathcal F_i$ for $i=1,2$ be an $m_i \times n_i$ Ferrers diagram, and $\mathcal C_i$ be an $[\mathcal F_i, k_i, \delta_i]_q$ code. Let $\mathcal D$ be an $m_3 \times n_3$ full Ferrers diagram with $m_3n_3$ dots, where $m_3 \geq m_1$ and $n_3 \geq n_2$. Let
\begin{center}
$\mathcal F=\left(
  \begin{array}{cc}
    \mathcal F_1 & \mathcal D \\
      & \mathcal F_2 \\
  \end{array}
\right)$
\end{center}
be an $m \times n$ Ferrers diagram, where $m=m_2+m_3$ and $n=n_1+n_3$. Then there exists an $[\mathcal F, \min\{k_1,k_2\}, \delta_1+\delta_2]_q$ code.
\end{Theorem}

As an application of Theorem \ref{thm:combine with same dim}, we obtained a $[[2,2,2,5,5], 3, 4]_q$ code for any prime power $q$ at the end of Section \ref{sec_multilevel}.

\begin{Theorem} \cite[Theorem 4.14]{liu2019constructions} \label{thm:com-3}
Let $m=m_1+m_3$, $n=n_1+n_3$ and $\delta \leq m_1+1$. Let
\[\mathcal F=
\begin{array}{c@{\hspace{-5pt}}c@{\hspace{-5pt}}c}
&\begin{array}{cc}
\overbrace{\rule{15mm}{0mm}}^{n_1}&
\overbrace{\rule{15mm}{0mm}}^{n_3}
\end{array}
\\
&
\begin{array}{cccccc}
     \bullet & \cdots & \bullet & \bullet & \cdots & \bullet \\
     \vdots & \mathcal F_1 & \vdots & \vdots & \mathcal F_4 & \vdots \\
     \circ & \cdots & \bullet & \bullet & \cdots & \bullet \\
      &  & \bullet & \bullet & \cdots & \bullet \\
      &  &  & \vdots & \mathcal F_3 & \vdots \\
      &  &  & \circ & \cdots & \bullet
   \end{array}
& \begin{array}{l}
\left.\rule{0mm}{8mm}\right\}m_1\vspace{0.2cm}\\
\left.\rule{0mm}{8mm}\right\}m_3
\end{array}
\end{array}=[\gamma_0,\gamma_1,\ldots,\gamma_{n-1}]
\]
be an $m \times n$ Ferrers diagram, where $\mathcal F_1$ is an $m_1 \times n_1$ Ferrers diagram, $\mathcal F_3$ is an $m_3 \times n_3$ Ferrers diagram, and $\mathcal F_4$ is an $m_1 \times n_3$ full Ferrers diagram. Suppose that $\mathcal{F}$ satisfies:
\begin{itemize}
\item[$(1)$] if $\delta<m_1+1$, then $n_3 \geq m_1;$
\item[$(2)$] $1+m_1+n_3\leq \max\{n_1,m_3\};$
\item[$(3)$] $\alpha_{m_1+n_3-\delta+2}\geq m_1+n_3;$
\item[$(4)$] $\rho_{\delta-2}-n_3\geq m_3$,
\end{itemize}
where $\rho_i$ denotes the number of dots in the $i$-th row of ${\mathcal F}$, $0\leq i\leq m_1+m_3-1$, and $\alpha_{m_1+n_3-\delta+2}$ denotes the $(m_1+n_3-\delta+2)$-th smallest number in the set $\{\rho_i-n_3:0\leq i\leq m_1-1\}\cup\{\gamma_j-m_1:n_1\leq j\leq n-1\}$. Then there exists an optimal $[\mathcal F, \sum_{i=\delta-1}^{m_1+m_3-1}\rho_i, \delta]_q$ code $\mathcal C$ for any prime power $q$.
\end{Theorem}

Finally we quote the following three sporadic optimal FDRM codes for later use.

\begin{Theorem} \label{thm:small}
\begin{itemize}
\item[$(1)$] \cite[Example III.16]{antrobus2019maximal} There exists an optimal $[[2,2,4,4,6,6],8,4]_q$ code.
\item[$(2)$] \cite[Section VIII]{etzion2016optimal} There exists an optimal $[[3,3,3,5],6,3]_q$ code.
\item[$(3)$] \cite[Example 4.15]{liu2019constructions} There exists an optimal $[[2,2,2,3,6],5,3]_q$ code.
\end{itemize}
\end{Theorem}

\section{Partial spreads}
\label{sec_partial_spread}

A partial spread is an $(n,\star,2k;k)$ constant dimension code, i.e., a constant dimension code with the maximum possible minimum subspace distance,
given the dimension $k$ of the codewords. The aim of this subsection is to analytically determine $M_q(n,d;k)$ for the case of partial spreads.

\begin{Lemma}
  \label{lemma_upper_ef_bound_partial_spread}
  For integers $1\le k\le n$ let $v\in\mathbb{F}_2^n$ be a vector of weight $k$.
  Let $\mathcal{C}_v\subseteq \EF{v}{q}$ be a subspace code having a minimum subspace
  distance of $2k$ and
  $$
    j:=\max\{1\le i\le n\,:\, v_i=1\}.
  $$
  If $j\le n-k$, then $w(v)=q^{n-j}$ and $w(v)\le q^{n-j}$ otherwise.
\end{Lemma}
\begin{proof}
  Let $\mathcal{F}$ be the Echelon-Ferrers diagram corresponding to the pivot vector $v$. Since removing the first $k-1$ rows yields
  exactly $n-j$ dots in the last row, Theorem~\ref{thm_upper_bound_ef} yields $w(v)\le q^{n-j}$. Moreover, $\mathcal{F}$ contains all
  dots of a rectangular $k\times (n-j)$ Echelon-Ferrers diagram. For this rectangular Echelon-Ferrers diagram we are in
  the MRD situation so that $w(v)\ge q^{n-j}$ if $n-j\ge k$, which is satisfied for $j\le n-k$.
\end{proof}

\begin{Theorem}
  If $n\equiv 0\pmod k$ we have
  $$
    M_q(n,2k;k)=\sum_{i=0}^{\tfrac{n}{k}-1}q^{ik}=\frac{q^n-1}{q^k-1}
  $$
  and
  $$
    M_q(n,2k;k)=1+\sum_{i=1}^{\left\lfloor \tfrac{n}{k}\right\rfloor-1} q^{n-ik}
    =\frac{q^n-q^{k+(n\,\operatorname{mod}\,k)}+q^k-1}{q^k-1}
  $$
  otherwise.
\end{Theorem}
\begin{proof}
  Since every $k$-subspace consists of $\gaussm{k}{1}{q}$ points and in a partial $k$-spread two different elements have no point in common, we have the upper bound
  $$
    M_q(n,2k;k)\le A_q(n,2k;k)\le \gaussm{n}{1}{q}/ \gaussm{k}{1}{q}=\frac{q^n-1}{q^k-1}.
  $$
  If $k$ divides $n$, then the latter term equals $\sum_{i=0}^{n/k-1}q^{ik}$.

  For each $0\le i\le \left\lfloor n/k\right\rfloor-1$ we define the pivot vector
  $$
    v^i:=0^{ik}1^k0^{n-(i+1)k}\in\mathbb{F}_2^n,
  $$
  where $a^b$ denotes the concatenation of $b$ times the symbol $a$ for $a\in\{0,1\}$. Note that the Echelon-Ferrers diagram corresponding to $v_i$ is rectangular,
  so that we can choose MRD codes $\mathcal{C}_{v^i}\subseteq \EF{v^i}{q}$ in the Echelon-Ferrers construction. For $i=\left\lfloor n/k\right\rfloor-1$ we have
  $\#\mathcal{C}_{v^i}=1$ and
  $$
    \#\mathcal{C}_{v^i}=q^{n-(i+1)k}
  $$
  otherwise. This gives matching lower bounds for $M_q(n,d;k)$ in all cases, since the $v^i$ have Hamming weight $k$ and pairwise Hamming distance $2k$.

  In remains to prove the upper bound for $M_q(n,d;k)$ for the cases where $n\not\equiv 0\pmod k$. To this end we assume that $x^0,\dots,x^{r-1}\in \mathbb{F}_2^n$
  are binary vectors of Hamming weight $k$ and pairwise Hamming weight $2k$ such that
  $$
    \sum_{i=0}^{r-1} w(x^i)\,>\, \sum_{i=0}^{\left\lfloor n/k\right\rfloor-1} w(v^i).
  $$
  Finally, we will end up with a contradiction, which proves the upper bound. Note that the upper bound $\#\mathcal{C}_{v^i}\le w(v^i)$ is indeed attained.

  First we observe $r\le \left\lfloor n/k\right\rfloor$ since no two vectors $x^i$ share a common $1$, each vector consists of exactly $k$ ones, and there are only
  $n$ positions for the ones. Since the weight function $w$ is non-negative we can assume $r=\left\lfloor n/k\right\rfloor$.

  If $x,x'\in\mathbb{F}_2^n$ both have Hamming weight $k$ and $x$ arises from $x'$ by shifting one $1$ to the left, then we have $w(x)\ge w(x')$. Thus we can
  assume that the vectors $x^0,\dots,x^{r-1}$ contain their $r\cdot k$ ones in the first $r\cdot k$ positions, as it is the case for the vectors $v^0,\dots,v^{r-1}$.

  By $j(x^i)$ we denote the coordinate of the last one, i.e., $j(x^i)=\max\{1\le h\le n\,:\,x_h^i=1\}$. If $w(x^i)\ge n-k$, then we can assume that the $k$ ones of
  $x^i$ are at the positions $\{j,j-1,\dots,j-k+1\}$ since otherwise we could swap the missing positions from other vector $x^{i'}$, which keeps $w(x^i)$ fix, see
  Lemma~\ref{lemma_upper_ef_bound_partial_spread}, and does not increase $w(x^{i'})$ since we apply a sequence of left shifts to ones.

  Now let $1\le i\le n$ such that $x^i$ contains a one in position $r\cdot k$, which is the last position where a one can occur. Let $S\subseteq \{1,\dots rk\}$ be
  the positions where $x^i$ contains ones. If $S=\{rk,rk-1,\dots,rk-k+1\}$ then we are done, since then there exists a one-to-one correspondence between the
  $x^i$ and the $v^i$. Otherwise, there exists an index $1\le h\le n$ such that $x^h$ contains $t\ge 1$ ones in positions that are contained in $\{rk,rk-1,\dots,rk-k+1\}$.
  From Lemma~\ref{lemma_upper_ef_bound_partial_spread} we conclude $w(x^i),w(x^h)\le q^{k-1}$. Now let $x$ arise from $x^i$ by removing $t$ ones in coordinates at most $n-k$
  and adding $t$ ones in the positions where $x^h$ has its $t$ ones with coordinates strictly larger than $n-k$. The let $x'$ arise from $x^h$ the other way round, i.e.,
  the $t$ ones in coordinates strictly larger than $n-k$ a removed and the ones that are removed from $x^i$ are added. Using Lemma~\ref{lemma_upper_ef_bound_partial_spread}
  we conclude $w(x')\ge q^k$ since $j(x')\le n-k$. Since $w(x)\ge 1$ we have
  $$
    w(x^i)+w(x^h)\le 2q^{k-1}\overset{q\ge 2}{\le} q^k<q^k+1\le w(x)+w(x').
  $$
  Thus, we can replace $x^i$ and $x^h$ by $x$ and $x'$. After at most $k-1$ such replacements we are in the situation where $x^i$, i.e., the vector with $j(x^i)=rk$,
  contains its $k$ ones exactly in the positions $\{rk,rk-1,\dots,rk-k+1\}$. Again we can conclude that there exists a one-to-one correspondence between the
  $x^i$ and the $v^i$, which yields our final contradiction.
\end{proof}

Note that the special choice of the pivot vectors $v_i$ was also mentioned in \cite[Observation 3.4]{kurz2017improved}, see also \cite{skachek2010recursive} for a
more general construction. Despite the simplicity of the skeleton code achieving $M_q(n,2k;k)$, so far we only know $A_q(n,2k;k)>M_q(n,2k;k)$ for the cases
where $q=2$, $k=3$, and $n\equiv 2\pmod 3$, see \cite{el2010maximum}. In the other direction, it has been shown that for $k>\gaussm{r}{1}{q}$ we have
$A_q(lk+r,2k;k)=M_q(lk+r,2k;k)$ for all $l\ge 2$ \cite{nuastase2017maximum}. For the tightest known upper bounds on $A_q(n,2k;k)$ in general, we refer to
\cite{kurz2017improved,honold2018partial}.



\appendix

\section{Upper and lower bounds for $M_q(n,d;k)$}
\label{sec_upper_polynomial}

In this section we determine upper and lower bounds for $M_q(n,d;k)$ for all parameters satisfying $4\le d\le 2k$, and $ 2k\le n\le 19$.

To present upper bounds we apply Algorithm~\ref{alg:maximum_weight_clique} or one of the refinements described in Section~\ref{sec_weighted_max_clique}.
We organize the obtained results in different subsections according to the corresponding minimum subspace distance $d$. For $d=4$ we limit our considerations
to $n\le 14$ and refer to \cite{lifted_codes} for $n>14$. For $d>4$ it is not known whether \cite[Conjecture 1]{etzion2009error} is true. To this end we denote
by $\overline{M}_q(n,d;k)$ maximum possible cardinality of an $(n, \star, d; k)_q$ code within the class of the multilevel construction assuming that
\cite[Conjecture 1]{etzion2009error} is true and only consider bounds for $\overline{M}_q(n,d;k)$. For $d\ge 10$ the plain application of Algorithm~\ref{alg:maximum_weight_clique}
was sufficient to determine the, with respect to Theorem~\ref{thm_upper_bound_ef}, tightest upper bound for $M_q(n,d;k)$. In some cases, where $d\le 8$ and $q\le 4$, we were not able to
determine this {\lq\lq}tight{\rq\rq} upper bound. Then we give both an upper and a lower bound for $\overline{M}_q(n,d;k)$. We also state the output of
Algorithm~\ref{alg:maximum_weight_clique}, i.e., the list of cliques $\mathcal{U}$ as $\mathcal{U}_{n,d,k}$. Here, the vertices, which are pivot vectors $v\in \mathbb{F}_2^n$
with Hamming weight $k$, are given as integers whose representation in base $2$ equals the 0/1-representation, see Section~\ref{sec_weighted_max_clique}

To show lower bounds for $M_q(n,d;k)$, all pivot vectors are colored. Let X be the integer representation of a pivot vector. Then we use the following notation.
\begin{itemize}
\item The black X means that its corresponding optimal FDRM code comes from Theorem \ref{thm:shortening} or \ref{cor:delta=3}.
\item The cyan {\color{cyan}X} means that its corresponding optimal FDRM code comes from Theorem \ref{thm:from subcodes}.
\item The magenta {\color{magenta}X$_r$} means that its corresponding optimal FDRM code comes from Theorem \ref{thm:subcodes from Gab} with the given parameter $r$.
\item The yellow {\color{yellow}X$_{l,t_1,t_2}$} means that its corresponding optimal FDRM code comes from Theorem \ref{the:from sys MRD-new1} with the given parameters $l$, $t_1$ and $t_2$.
\item The green {\color{green}X} means that its corresponding optimal FDRM code comes from Theorem \ref{thm:combine with same dim}.
\item The blue {\color{blue}X} means that its corresponding optimal FDRM code comes from Theorem \ref{thm:small}.
\item The bold {$\textbf{X}$} means that its corresponding optimal FDRM code comes from Theorem \ref{thm:com-3}.
\item The italic $\emph{X}$ means that its corresponding optimal FDRM code can be obtained by removing some pending dots. For example to determine $M_2(15,6;7)$, we use the pivot vector $5214$ that is written in italic type and colored black. Its corresponding Ferrers diagram is $[6,5,2,1,1,1,1]$ which contains one pending dot. Delete the pending dot to obtain a Ferrers diagram $[6,5,2,1,1,1]$. We use the notation $5214:[6,5,2,1,1,1,1]\rightarrow[6,5,2,1,1,1]$ to represent this operation. The desired optimality is guaranteed by the theorem depending on the color of $\emph{X}$.
\item The red {\color{red}X} means that the optimality of its corresponding FDRM code is still unknown. The best known dimension of its corresponding FDRM code is shown in Appendix~\ref{the_best_known_dimension_of_given_FDRMCs}. Take $(12,6,5)_q$ for example. The pivot vector $1256$ is marked in red and its corresponding Ferrers diagram is $[6,4,4,4,3]$. An optimal FDRM code in $[6,4,4,4,3]$ is of dimension 11 while we can only construct a $([5,4,4,4,3],3;10)$ code by Theorem \ref{cor:delta=3} which leads to a $([6,4,4,4,3],3;10)$ code by adding a pending dot.
\end{itemize}

\subsection{Minimum subspace distance $4$}
\label{subsec_mindist_4}

\begin{eqnarray*}
  M_q(4,4;2) &=& q^2+q^0
\end{eqnarray*}
\begin{eqnarray*}
  \mathcal{U}_{4,4,2} &=& \Big\{12,3\Big\}
\end{eqnarray*}

\bigskip

\begin{eqnarray*}
  M_q(5,4;2) &=& q^3+q^0
\end{eqnarray*}
\begin{eqnarray*}
  \mathcal{U}_{5,4,2} &=& \Big\{24,3\Big\}
\end{eqnarray*}

\bigskip

\begin{eqnarray*}
  M_q(6,4;2) &=& q^4+q^2+q^0
\end{eqnarray*}
\begin{eqnarray*}
  \mathcal{U}_{6,4,2} &=& \Big\{48,12,3\Big\}
\end{eqnarray*}

\bigskip

\begin{eqnarray*}
  M_q(6,4;3) &=& q^6+q^2+q^1+q^0
\end{eqnarray*}
\begin{eqnarray*}
  \mathcal{U}_{6,4,3} &=& \Big\{56,38,21,11\Big\}
\end{eqnarray*}

\bigskip

\begin{eqnarray*}
  M_q(7,4;2) &=& q^5+q^3+q^0
\end{eqnarray*}
\begin{eqnarray*}
  \mathcal{U}_{7,4,2} &=& \Big\{96,24,3\Big\}
\end{eqnarray*}

\bigskip

\begin{eqnarray*}
  M_q(7,4;3) &=& q^8+q^4+q^3+q^2+2q^1+q^0
\end{eqnarray*}
\begin{eqnarray*}
  \mathcal{U}_{7,4,3} &=& \Big\{112,44,74,25,22,69,35\Big\}
\end{eqnarray*}

\bigskip

\begin{eqnarray*}
  M_q(8,4;2) &=& q^6+q^4+q^2+q^0
\end{eqnarray*}
\begin{eqnarray*}
  \mathcal{U}_{8,4,2} &=& \Big\{192,48,12,3\Big\}
\end{eqnarray*}

\bigskip

\begin{eqnarray*}
  M_q(8,4;3) &=& q^{10}+q^6+q^5+2q^4+q^3+q^2
\end{eqnarray*}
\begin{eqnarray*}
  \mathcal{U}_{8,4,3} &=& \Big\{224,56,84,140,146,74,38\Big\}
\end{eqnarray*}

\bigskip

\begin{eqnarray*}
  M_q(8,4;4) &=& q^{12}+q^8+q^6+q^5+7q^4+q^3+q^2+q^0
\end{eqnarray*}
\begin{eqnarray*}
  \mathcal{U}_{8,4,4} &=& \Big\{240,204,170,105,60,90,102,150,153,165,195,85,51,15\Big\}
\end{eqnarray*}

\bigskip

\begin{eqnarray*}
  M_q(9,4;2) &=& q^7+q^5+q^3+q^0
\end{eqnarray*}
\begin{eqnarray*}
  \mathcal{U}_{9,4,2} &=& \Big\{384,96,24,3\Big\}
\end{eqnarray*}

\bigskip

\begin{eqnarray*}
  M_q(9,4;3) &=& q^{12}+q^8+q^7+2q^6+q^5+q^4+q^0
\end{eqnarray*}
\begin{eqnarray*}
  \mathcal{U}_{9,4,3} &=& \Big\{448,112,168,280,292,148,76,7\Big\}
\end{eqnarray*}

\bigskip

\begin{eqnarray*}
  M_q(9,4;4) &=& q^{15}+q^{11}+q^9+4q^8+4q^7+q^6+q^4+q^0
\end{eqnarray*}
\begin{eqnarray*}
  \mathcal{U}_{9,4,4} &=& \Big\{480,408,212,120,308,332,338,172,178,202,390,298,101,15\Big\}
\end{eqnarray*}

\bigskip

\begin{eqnarray*}
  M_q(10,4;2) &=& q^8+q^6+q^4+q^2+q^0
\end{eqnarray*}
\begin{eqnarray*}
  \mathcal{U}_{10,4,2} &=& \Big\{768,192,48,12,3\Big\}
\end{eqnarray*}

\bigskip

\begin{eqnarray*}
  M_q(10,4;3) &=& q^{14}+q^{10}+q^9+2q^8+q^7+q^6+q^2+q^1+q^0
\end{eqnarray*}
\begin{eqnarray*}
  \mathcal{U}_{10,4,3} &=& \Big\{896,224,336,560,584,296,152,38,21,11\Big\}
\end{eqnarray*}

\bigskip

\begin{eqnarray*}
  M_q(10,4;4) &=& q^{18}+q^{14}+2q^{12}+3q^{11}+4q^{10}+q^9+q^8+q^6+2q^4+q^2+q^0
\end{eqnarray*}
\begin{eqnarray*}
  \mathcal{U}_{10,4,4} &=& \Big\{960,816,240,424,616,664,676,344,356,404,780,596,204,771,60,\\
  && 195,51,15\Big\}
\end{eqnarray*}

\bigskip

For $q\ge 3$ we have
\begin{eqnarray*}
  M_q(10,4;5) &=& q^{20}+q^{16}+q^{14}+8q^{12}+q^{11}+q^{10}+q^9+q^8+q^7+q^6+q^4+q^3\\
  && +q^2+q^1+q^0
\end{eqnarray*}
with
\begin{eqnarray*}
  \mathcal{U}_{10,4,5} &=& \Big\{992,920,852,632,692,716,722,812,818,842,902,465,682,425,\\
  && 614,357,227,542,285,155,87,47\Big\}
\end{eqnarray*}
and for $q=2$ we have
\begin{eqnarray*}
  M_2(10,4;5) &=& 2^{20}+2^{16}+2^{14}+8\dot 2^{12}+2^{11}+2^{10}+2^9+2^8+2\cdot 2^6+2\cdot 2^5\\
  && +3\cdot 2^4+2\cdot 2^2+2^1+2^0=1\,167\,355
\end{eqnarray*}
with
\begin{eqnarray*}
  \mathcal{U}_{10,4,5} &=& \Big\{992,920,852,632,692,716,722,812,818,842,902,465,682,425,\\
  && 614,229,355,309,333,179,203,542,157,283,87,47\Big\}.
\end{eqnarray*}

\bigskip

\begin{eqnarray*}
  M_q(11,4;2) &=& q^9+q^7+q^5+q^3+q^0
\end{eqnarray*}
\begin{eqnarray*}
  \mathcal{U}_{11,4,2} &=& \Big\{1536,384,96,24,3\Big\}
\end{eqnarray*}

\bigskip

\begin{eqnarray*}
  M_q(11,4;3) &=& q^{16}+q^{12}+q^{11}+2q^{10}+q^9+q^8+q^4+q^3+2q^2+q^1+q^0
\end{eqnarray*}
\begin{eqnarray*}
  \mathcal{U}_{11,4,3} &=& \Big\{1792,448,672,1120,1168,592,304,44,74,25,134,69,35\Big\}
\end{eqnarray*}

\bigskip

\begin{eqnarray*}
  M_q(11,4;4) &=& q^{21}+q^{17}+2q^{15}+3q^{14}+4q^{13}+q^{12}+q^{11}+q^9+q^8+2q^7\\
  && +2q^6+q^5+q^4+q^3+q^2+2q^1+q^0
\end{eqnarray*}
\begin{eqnarray*}
  \mathcal{U}_{11,4,4} &=& \Big\{1920,1632,480,848,1232,1328,1352,688,712,808,1560,1192,\\
  && 408,1542,120,390,773,1157,1283,643,101,86,51,46,75,29\Big\}
\end{eqnarray*}

\bigskip

For $q\ge 3$ we have
\begin{eqnarray*}
  M_q(11,4;5) &=& q^{24}+q^{20}+q^{18}+8q^{16}+q^{15}+3q^{14}+q^{13}+4q^{12}+3q^{11}\\
  && +2q^9+2q^8+3q^7+3q^6+3q^5+q^4+2q^3+2q^2+q^0
\end{eqnarray*}
with
\begin{eqnarray*}
  \mathcal{U}_{11,4,5} &=& \Big\{1984,1840,1704,752,932,1384,1432,1624,1636,1684,1804,\\
  && 1442,852,866,914,1362,905,1228,1249,1795,465,714,1413,\\
  && 454,709,617,1084,357,570,675,313,1109,1171,310,565,\\
  && 595,331,173,1067,110,158,527\Big\}
\end{eqnarray*}
and for $q=2$ we have
\begin{eqnarray*}
  M_2(11,4;5) &=& 2^{24}+2^{20}+2^{18}+8\cdot 2^{16}+2^{15}+3\cdot 2^{14}+6\cdot 2^{12}+3\cdot 2^{11}+2^{10}\\
  && +2^9+3\cdot 2^8+4\cdot 2^7+3\cdot 2^6+4\cdot 2^5+2^4+3\cdot 2^3+2^1+2^0 \\
  &=& 18\,728\,043
\end{eqnarray*}
with
\begin{eqnarray*}
  \mathcal{U}_{11,4,5} &=& \Big\{1984,1840,1704,752,932,1384,1432,1624,1636,1684,1804,\\
  && 1442,852,866,914,905,1228,1234,1249,1361,1795,714,1350,\\
  && 1413,433,709,451,617,1084,346,357,570,675,217,230,\\
  && 398,310,565,595,1163,1075,173,285,299,151,79\Big\}.
\end{eqnarray*}

\bigskip

\begin{eqnarray*}
  M_q(12,4;2) &=& q^{10}+q^8+q^6+q^4+q^2+q^0
\end{eqnarray*}
\begin{eqnarray*}
  \mathcal{U}_{12,4,2} &=& \Big\{3072,768,192,48,12,3\Big\}
\end{eqnarray*}

\bigskip

\begin{eqnarray*}
  M_q(12,4;3) &=& q^{18}+q^{14}+q^{13}+2q^{12}+q^{11}+q^{10}+q^6+q^5+2q^4+2q^3\\
  && +2q^2+q^1+q^0
\end{eqnarray*}
\begin{eqnarray*}
  \mathcal{U}_{12,4,3} &=& \Big\{3584,896,1344,2240,2336,1184,608,56,84,140,146,74,\\
  && 273,38,521,1029,2051\Big\}
\end{eqnarray*}

\bigskip

\begin{eqnarray*}
  M_q(12,4;4) &=& q^{24}+q^{20}+2q^{18}+3q^{17}+4q^{16}+q^{15}+q^{14}+2q^{12}+2q^{10}\\
  && +3q^9+5q^8+q^7+2q^6+q^5+7q^4+q^3+q^2+q^0
\end{eqnarray*}
\begin{eqnarray*}
  \mathcal{U}_{12,4,4} &=& \Big\{3840,3264,960,1696,2464,2656,2704,1376,1424,1616,3120,2384,\\
  && 816,240,3084,780,1546,2314,2566,2569,204,1286,1289,1541,3075,\\
  && 2309,170,771,105,60,90,102,150,153,165,195,85,51,15\Big\}
\end{eqnarray*}

\bigskip

For $q\ge 3$ we have
\begin{eqnarray*}
  M_q(12,4;5) &=& q^{28}+q^{24}+q^{22}+8q^{20}+q^{19}+3q^{18}+q^{17}+4q^{16}+4q^{15}+5q^{14}+q^{13}\\
  && +4q^{12}+5q^{11}+6q^{10}+3q^9+5q^8+3q^7+4q^6+q^5+2q^4+q^2+q^1+q^0
\end{eqnarray*}
with
\begin{eqnarray*}
  \mathcal{U}_{12,4,5} &=& \Big\{3968,3680,3408,1504,1744,1840,2888,3248,3272,3368,3608,1860,\\
  && 2728,2756,2852,1700,920,962,2834,3590,1802,2452,2466,2497,929,\\
  && 1420,1426,2705,3333,1673,632,849,2258,3203,690,1353,2164,2353,\\
  && 2374,1132,1138,1194,1222,1318,1603,1557,2154,2595,241,348,614,\\
  && 2213,2573,357,654,2137,188,314,1081,2315,775,203,1054,539,87,47\Big\}
\end{eqnarray*}
and for $q=2$ we have
\begin{eqnarray*}
  M_2(12,4;5) &=& 2^{28}+2^{24}+2^{22}+8\cdot 2^{20}+2^{19}+2\cdot 2^{18}+3\cdot 2^{17}+5\cdot 2^{16}+3\cdot 2^{15} \\
  && +3\cdot 2^{14}+4\cdot 2^{13}+2\cdot 2^{12}+3\cdot 2^{11}+5\cdot 2^{10}+6\cdot 2^9+5\cdot 2^8+4\cdot 2^7\\
  && +3\cdot 2^6+5\cdot 2^5+2^3+2^2+2^0 = 299\,769\,965
\end{eqnarray*}
with
\begin{eqnarray*}
  \mathcal{U}_{12,4,5} &=& \Big\{3968,3680,3408,992,1744,1840,2888,3248,3272,3368,3608,1860,\\
  && 2728,2852,1700,2500,2754,1474,2456,2708,2834,3590,1428,1802,\\
  && 2466,908,1441,3333,913,1673,2401,2641,632,3203,690,1353,2164,\\
  && 809,1132,1194,1318,1603,370,1137,1221,1557,2154,2595,348,\\
  && 1114,2213,2318,2573,233,598,613,1299,188,230,2105,211,309,\\
  && 395,558,775,2119,539,1039\Big\}.
\end{eqnarray*}

\bigskip

For $q\ge 4$ we have
\begin{eqnarray*}
  M_q(12,4;6) &=& q^{30}+q^{26}+q^{24}+8q^{22}+2q^{20}+q^{19}+8q^{18}+q^{17}+q^{16}+q^{15}+8q^{14}\\
  && +q^{13}+2q^{12}+q^{11}+7q^{10}+q^9+2q^8+q^7+8q^6+2q^5+7q^4+q^3\\
  && +q^2+q^0
\end{eqnarray*}
with
\begin{eqnarray*}
  \mathcal{U}_{12,4,6} &=& \Big\{4032,3888,3752,3312,3432,3480,3492,3672,3684,3732,3852,2978,\\
  && 3412,1953,1008,1890,1938,2898,2913,2961,3276,3843,1873,2762,\\
  && 1737,972,1482,1734,2502,2505,2757,3132,3267,1477,963,2618,\\
  && 1593,828,1338,1590,2358,2361,2613,3123,1333,819,2222,1197,\\
  && 252,683,1134,1182,2142,2157,2205,3087,423,1117,243,363,411,\\
  && 603,615,663,783,343,207,63\Big\},
\end{eqnarray*}
for $q=3$ we have
\begin{eqnarray*}
  M_3(12,4;6) &=& 3^{30}+3^{26}+3^{24}+8\cdot 3^{22}+2\cdot 3^{20}+3^{19}+8\cdot 3^{18}+3^{17}+4\cdot 3^{15}\\
  && +7\cdot 3^{14}+8\cdot 3^{13}+3\cdot 3^{12}+10\cdot 3^{11}+4\cdot 3^{10}+2\cdot 3^9+2\cdot 3^8+3^7\\
  && +8\cdot 3^6+2\cdot 3^5+7\cdot 3^4+3^3+3^2+3^0=208\,977\,947\,565\,256
\end{eqnarray*}
with
\begin{eqnarray*}
  \mathcal{U}_{12,4,6} &=& \Big\{4032,3888,3752,3312,3432,3480,3492,3672,3684,3732,3852,\\
  && 2978,3412,1953,1008,1890,1938,2898,2913,2961,3276,3843,1873,\\
  && 1738,2506,2758,2761,972,1481,1733,2666,2714,3132,3267,1450,\\
  && 1478,1641,1689,1702,2473,2501,2725,963,2406,2454,1370,1381,\\
  && 1429,1594,1622,2362,2393,2614,2617,2645,828,1337,1589,3123,\\
  && 1334,2357,819,2222,1197,252,683,1134,1182,2142,2157,2205,\\
  && 3087,423,1117,243,363,411,603,615,663,783,343,207,63\Big\},
\end{eqnarray*}
and for $q=2$ we have
\begin{eqnarray*}
  M_2(12,4;6) &=& 2^{30}+2^{26}+2^{24}+8\cdot 2^{22}+2^{20}+3\cdot 2^{19}+7\cdot 2^{18}+3\cdot 2^{17}+2\cdot 2^{16}\\
  && +2\cdot 2^{15}+6\cdot 2^{14}+6\cdot 2^{13}+3\cdot 2^{12}+6\cdot 2^{11}+4\cdot 2^{10}+4\cdot 2^9+5\cdot 2^8\\
  && +2^7+6\cdot 2^6+3\cdot 2^5+7\cdot 2^4+2^3+2^2+2^0=1\,196\,408\,797
\end{eqnarray*}
with
\begin{eqnarray*}
  \mathcal{U}_{12,4,6} &=& \Big\{4032,3888,3752,3312,3432,3480,3492,3672,3684,3732,3852,\\
  && 2978,1953,2900,3410,1008,1890,1938,2913,2961,3276,3843,1873,\\
  && 2770,2890,1492,3397,1738,2761,972,1481,1733,2502,3132,3267,\\
  && 1450,1641,1689,1702,2473,2725,963,2282,2702,1594,1622,2362,\\
  && 2393,2614,2617,828,1337,1589,3123,1334,1358,2261,2357,474,\\
  && 485,819,1415,2631,1197,252,683,1182,2142,2157,3087,622,\\
  && 1117,2203,243,363,413,603,663,783,1127,343,207,63\Big\}.
\end{eqnarray*}

\bigskip

\begin{eqnarray*}
  M_q(13,4;2) &=& q^{11}+q^9+q^7+q^5+q^3+q^0
\end{eqnarray*}
\begin{eqnarray*}
  \mathcal{U}_{13,4,2} &=& \Big\{6144,1536,384,96,24,3\Big\}
\end{eqnarray*}

\bigskip

\begin{eqnarray*}
  M_q(13,4;3) &=& q^{20}+q^{16}+q^{15}+2q^{14}+q^{13}+q^{12}+q^8+q^7+2q^6+2q^5\\
  && +3q^4+2q^3+2q^2+q^1+q^0
\end{eqnarray*}
\begin{eqnarray*}
  \mathcal{U}_{13,4,3} &=& \Big\{7168,1792,2688,4480,4672,2368,1216,112,168,280,292,148,\\
  && 546,76,1042,1057,529,2058,4102,4105,2053,67\Big\}
\end{eqnarray*}

\bigskip

\begin{eqnarray*}
  M_q(13,4;4) &=& q^{27}+q^{23}+2q^{21}+3q^{20}+4q^{19}+q^{18}+q^{17}+2q^{15}+2q^{13}\\
  && +3q^{12}+5q^{11}+q^{10}+3q^9+6q^8+7q^7+5q^6+3q^5+3q^4+q^3+q^0
\end{eqnarray*}
\begin{eqnarray*}
  \mathcal{U}_{13,4,4} &=& \Big\{7680,6528,1920,3392,4928,5312,5408,2752,2848,3232,6240,\\
  && 4768,1632,480,6168,1560,3092,4628,5132,5138,408,2572,2578,\\
  && 3082,6150,4618,212,785,1542,120,308,332,338,1169,1289,\\
  && 172,178,202,390,649,2129,2309,298,4145,4169,4229,4355,\\
  && 581,2089,2179,102,1061,1091,547,15\Big\}
\end{eqnarray*}

\bigskip

For $q\ge 3$ we have
\begin{eqnarray*}
  M_q(13,4;5) &=& q^{32}+q^{28}+q^{26}+8q^{24}+q^{23}+3q^{22}+q^{21}+4q^{20}+4q^{19}\\
  && +5q^{18}+q^{17}+9q^{16}+8q^{15}+9q^{14}+6q^{13}+7q^{12}+5q^{11}+q^{10}+5q^9\\
  && +3q^8+q^7+3q^6+4q^5+3q^4+q^3+3q^2
\end{eqnarray*}
with
\begin{eqnarray*}
  \mathcal{U}_{13,4,5} &=& \Big\{7936,7360,6816,3008,3488,3680,5776,6496,6544,6736,7216,\\
  && 3720,5456,5512,5704,3400,1840,1924,5668,7180,3604,4904,4932,\\
  && 4994,1858,2840,2852,5410,6666,3346,1264,1698,4516,4801,5281,\\
  && 6406,6409,6661,7171,1380,1473,2706,3217,4328,4706,4748,4881,\\
  && 740,929,1801,2264,2388,2444,2636,3333,5254,696,1617,2274,\\
  && 3114,3142,4308,1228,2228,2609,2819,4426,5146,5189,376,466,\\
  && 714,1308,4274,4630,426,1550,2217,2245,4209,409,617,4156,\\
  && 661,355,1078,1081,309,333,2131,4235,391,1099,2078,583,\\
  && 110,539,2087\Big\}
\end{eqnarray*}
and for $q=2$ we have
\begin{eqnarray*}
  M_2(13,4;5) &=& 2^{32}+2^{28}+2^{26}+8\cdot 2^{24}+2^{23}+2\cdot 2^{22}+3\cdot 2^{21}+5\cdot 2^{20}+3\cdot 2^{19}\\
  && +3\cdot 2^{18}+3\cdot 2^{17}+8\cdot 2^{16}+8\cdot 2^{15}+9\cdot 2^{14}+5\cdot 2^{13}+8\cdot 2^{12}+9\cdot 2^{11}\\
  && +2^{10}+7\cdot 2^9+2\cdot 2^8+2\cdot 2^7+2\cdot 2^6+2^5+3\cdot 2^4+2\cdot 2^3+3\cdot 2^2+2^0\\
  &=& 4\,796\,825\,069
\end{eqnarray*}
with
\begin{eqnarray*}
  \mathcal{U}_{13,4,5} &=& \Big\{7936,7360,6816,1984,3488,3680,5776,6496,6544,6736,7216,\\
  && 3720,5456,5704,3400,5000,5508,2948,4912,5416,5668,7180,2856,\\
  && 3604,4932,1816,2882,6666,1826,3346,4802,2753,4336,5218,5281,\\
  && 6406,6409,6661,7171,1256,1380,2706,2833,3217,4514,4545,5258,\\
  && 740,929,1618,2264,2388,2636,3206,3333,4705,696,1236,2274,\\
  && 2442,3114,1417,1669,2228,4300,4426,4636,4867,5189,376,466,\\
  && 841,1202,1329,1577,2598,5142,5145,428,618,790,1347,1550,\\
  && 2161,2217,4389,1084,2339,405,4154,597,4243,651,563,2078,\\
  && 2123,118,199,109,283,1063,4111\Big\}.
\end{eqnarray*}

\bigskip

For $q\ge 3$ we have
\begin{eqnarray*}
  M_q(13,4;6) &=& q^{35}+q^{31}+q^{29}+8q^{27}+2q^{25}+4q^{24}+5q^{23}+q^{22}+4q^{21}+8q^{20}\\
  && +14q^{19}+5q^{18}+5q^{17}+9q^{16}+4q^{15}+q^{14}+6q^{13}+9q^{12}+13q^{11}\\
  && +3q^{10}+5q^9+8q^8+3q^7+4q^6+6q^5+3q^4+3q^3+q^0
\end{eqnarray*}
with
\begin{eqnarray*}
  \mathcal{U}_{13,4,6} &=& \Big\{8064,7776,7504,6624,6864,6960,6984,7344,7368,7464,7704,\\
  && 3908,6824,2016,5828,5924,5954,3748,3778,3874,6552,7686,5794,\\
  && 3521,3857,5524,7429,1944,2964,3468,3474,5057,5537,5777,5897,\\
  && 2977,3284,3380,3721,4948,5004,5010,5452,5514,6264,6534,6789,\\
  && 6915,7299,2898,2954,3402,5330,5426,2764,2860,4788,5236,5292,\\
  && 1656,1925,2676,2738,3180,3186,3242,4810,4906,4716,4722,5226,\\
  && 6246,2666,1478,1814,2396,2417,3161,6229,504,1265,1385,1638,\\
  && 4316,4412,4442,4697,5177,934,1621,1678,2236,2266,2362,2617,\\
  && 4329,4453,6174,6189,6195,6219,867,2277,4282,1363,1581,1587,\\
  && 1611,3143,723,821,845,1206,1229,1326,4679,5159,469,1309,\\
  && 2599,435,459,683,795,374,429,669,1118,1179,2327,414,\\
  && 4247,4367,238,574,2191,123\Big\}.
\end{eqnarray*}
For $q=2$ we have
\begin{eqnarray*} 
  M_2(13,4;6) &\ge & 2^{35}+2^{31}+2^{29}+8\cdot 2^{27}+2^{25}+6\cdot 2^{24}+4\cdot 2^{23}+3\cdot 2^{22}+7\cdot 2^{21}\\
  && +6\cdot 2^{20}+11\cdot 2^{19}+8\cdot 2^{18}+6\cdot 2^{17}+7\cdot 2^{16}+6\cdot 2^{15}+7\cdot 2^{14}\\
  && +7\cdot 2^{13}+9\cdot 2^{12}+8\cdot 2^{11}+7\cdot 2^{10}+5\cdot 2^9+6\cdot 2^8\\
  &=& 38\,328\,704\,000
\end{eqnarray*}
with
\begin{eqnarray*}
  \mathcal{U}_{13,4,6} &=& \Big\{8064,7776,7504,6624,6864,6960,6984,7344,7368,7464,7704,\\
  && 3908,2016,5800,5828,5924,5954,6820,3778,3874,6552,7686,2984,\\
  && 3492,3732,3521,3745,5524,5778,5905,6794,7429,1944,3474,3849,\\
  && 5026,5057,5537,3628,3665,4948,5004,5346,5452,5514,6264,6534,\\
  && 6915,7299,2898,2913,2961,3402,5426,5705,6697,6725,2516,2764,\\
  && 3356,3377,5236,5329,1656,1925,2676,2738,3186,3242,4562,4716,\\
  && 4722,4785,4890,5667,6246,948,2412,2666,3177,3225,5210,6293,\\
  && 970,1478,1702,1814,2830,4458,4465,504,1260,1637,2289,4316,\\
  && 4412,5177,5261,6227,2236,2266,2362,2393,2467,4329,5166,6174,\\
  && 870,1365,1379,1614,1675,2445,4282,729,739,825,2695,3143,\\
  && 485,726,845,1205,1325,1565\Big\}.
\end{eqnarray*}

\bigskip

\begin{eqnarray*}
  M_q(14,4;2) &=& q^{12}+q^{10}+q^8+q^6+q^4+q^2+q^0
\end{eqnarray*}
\begin{eqnarray*}
  \mathcal{U}_{14,4,2} &=& \Big\{12288,3072,768,192,48,12,3\Big\}
\end{eqnarray*}

\bigskip

\begin{eqnarray*}
  M_q(14,4;3) &=& q^{22}+q^{18}+q^{17}+2q^{16}+q^{15}+q^{14}+q^{10}+q^9+2q^8+2q^7\\
  && +3q^6+3q^5+3q^4+2q^3+2q^2+q^1+q^0
\end{eqnarray*}
\begin{eqnarray*}
  \mathcal{U}_{14,4,3} &=& \Big\{14336,3584,5376,8960,9344,4736,2432,224,336,560,584,296,\\
  && 1092,152,2084,2114,1058,4116,4161,8204,8210,8225,1041,4106,\\
  && 134,2057,261,515\Big\}
\end{eqnarray*}

\bigskip

For $q\ge 3$ we have
\begin{eqnarray*}
  M_q(14,4;4) &=& q^{30}+q^{26}+2q^{24}+3q^{23}+4q^{22}+q^{21}+q^{20}+2q^{18}+2q^{16}\\
  && +3q^{15}+5q^{14}+q^{13}+4q^{12}+6q^{11}+10q^{10}+8q^9+9q^8+5q^7+4q^6\\
  && +q^5+2q^4+q^2+q^0
\end{eqnarray*}
with
\begin{eqnarray*}
  \mathcal{U}_{14,4,4} &=& \Big\{15360,13056,3840,6784,9856,10624,10816,5504,5696,6464,12480,\\
  && 9536,3264,960,12336,3120,6184,9256,10264,10276,816,5144,5156,\\
  && 6164,12300,9236,240,424,1570,3084,616,664,676,2338,2578,\\
  && 2593,344,356,404,780,1298,1313,1553,4258,4618,12291,596,\\
  && 2321,8290,8338,8353,8458,8710,8713,204,1162,3075,4178,4193,\\
  && 4241,4358,4361,4613,2122,2182,2185,8273,8453,771,1094,1097,\\
  && 1157,2117,60,195,51,15\Big\}.
\end{eqnarray*}
For $q=2$ we have
\begin{eqnarray*} 
  M_2(14,4;4) &\ge& 2^{30}+2^{26}+2\cdot 2^{24}+3\cdot 2^{23}+4\cdot 2^{22}+2^{21}+2^{20}+2\cdot 2^{18}+2\cdot 2^{16}\\
  && +3\cdot 2^{15}+5\cdot 2^{14}+6\cdot 2^{12}+7\cdot 2^{11}+9\cdot 2^{10}+7\cdot 2^9+8\cdot 2^8+5\cdot 2^7\\
  && +3\cdot 2^6 \,=\, 1\,220\,384\,064
\end{eqnarray*}
\begin{eqnarray*}
  \mathcal{U}_{14,4,4} &=& \Big\{15360,13056,3840,6784,9856,10624,10816,5504,5696,6464,12480,\\
  && 9536,3264,960,12336,3120,6184,9256,10264,10276,816,5144,5156,\\
  && 6164,12300,240,424,1556,1570,3084,9234,616,664,676,2338,\\
  && 2578,2593,8468,344,356,780,4258,4370,4618,8481,8721,12291,\\
  && 1185,1297,1545,8290,8458,8710,9221,204,1162,1286,3075,4193,\\
  && 4241,4361,4613,2122,2129,2182,2185,2309,771,4166,8265\Big\}.
\end{eqnarray*}

\bigskip

For $q\ge 3$ we have 
\begin{eqnarray*}
  M_q(14,4;5) &=& q^{36}+q^{32}+q^{30}+8q^{28}+q^{27}+3q^{26}+q^{25}+4q^{24}+4q^{23}+5q^{22}\\
  && +q^{21}+9q^{20}+9q^{19}+12q^{18}+9q^{17}+14q^{16}+6q^{15}+6q^{14}+7q^{13}\\
  && +5q^{12}+4q^{11}+5q^{10}+4q^9+6q^8+8q^7+3q^6+3q^5+4q^4+2q^3+q^2
\end{eqnarray*}
with
\begin{eqnarray*}
  \mathcal{U}_{14,4,5} &=& \Big\{15872,14720,13632,6016,6976,7360,11552,12992,13088,13472,14432,\\
  && 7440,10912,11024,11408,6800,3680,3848,11336,14360,7208,9808,9864,\\
  && 9988,3716,5680,5704,10820,13332,6692,992,3396,9032,9602,10562,\\
  && 12812,12818,13322,14342,2760,2946,3457,5412,6434,8656,9412,9496,\\
  && 9762,1480,1858,3602,4528,4776,4888,5272,6412,9089,10433,10762,\\
  && 12561,1392,3234,4548,6228,6305,6665,8616,10380,11269,1729,2288,\\
  && 2408,2456,2468,2849,5218,5638,6282,7171,8852,10292,10505,12425,\\
  && 1428,1809,2616,5385,8548,9313,852,908,4705,5201,9260,12357,2641,\\
  && 3100,3121,4434,4869,5253,8418,1234,8312,8498,8753,12323,690,\\
  && 1322,1577,4739,710,806,4204,8771,10259,618,2326,4262,8346,451,\\
  && 665,677,2138,2150,8462,220,233,316,345,1166,4154,8278,779,\\
  && 1078,2567,589,1287,4149,542,2093,4171,2197,8327,115,8221,1051\Big\}.
\end{eqnarray*}
For $q=2$ we have
\begin{eqnarray*}
  M_2(14,4;5) &\ge & 2^{36}+2^{32}+2^{30}+8\cdot 2^{28}+2^{27}+3\cdot 2^{26}+2^{25}+4\cdot 2^{24}+4\cdot 2^{23}\\
  && +5\cdot 2^{22}+2^{21}+9\cdot 2^{20}+9\cdot 2^{19}+12\cdot 2^{18}+9\cdot 2^{17}+14\cdot 2^{16}+6\cdot 2^{15}\\
  && +6\cdot 2^{14}+7\cdot 2^{13}+5\cdot 2^{12}+4\cdot 2^{11}+5\cdot 2^{10}+4\cdot 2^9+6\cdot 2^8+8\cdot 2^7\\
  && +3\cdot 2^6+3\cdot 2^5+4\cdot 2^4+2\cdot 2^3+2^2 \\
  &=& 76\,748\,289\,908
\end{eqnarray*}
with
\begin{eqnarray*}
  \mathcal{U}_{14,4,5} &=& \Big\{15872,14720,13632,6016,6976,7360,11552,12992,13088,13472,14432,\\
  && 7440,10912,11024,11408,6800,3680,3848,11336,14360,7208,9808,9864,\\
  && 9988,3716,5680,5704,10820,13332,6692,992,3396,9032,9602,10562,\\
  && 12812,12818,13322,14342,2760,2946,3457,5412,6434,8656,9412,9496,\\
  && 9762,1480,1858,3602,4528,4776,4888,5272,6412,9089,10433,10762,\\
  && 12561,1392,3234,4548,6228,6305,6665,8616,10380,11269,1729,2288,\\
  && 2408,2456,2468,2849,5218,5638,6282,7171,8852,10292,10505,12425,\\
  && 1428,1809,2616,5385,8548,9313,852,908,4705,5201,9260,12357,2641,\\
  && 3100,3121,4434,4869,5253,8418,1234,8312,8498,8753,12323,690,\\
  && 1322,1577,4739,710,806,4204,8771,10259,618,2326,4262,8346,451,\\
  && 665,677,2138,2150,8462,220,233,316,345,1166,4154,8278,779,\\
  && 1078,2567,589,1287,4149,542,2093,4171,2197,8327,115,8221,1051\Big\}.
\end{eqnarray*}

\bigskip

For $q\ge 7$ we have 
\begin{eqnarray*}
  M_q(14,4;6) &=& q^{40}+q^{36}+q^{34}+8q^{32}+3q^{30}+3q^{29}+5q^{28}+q^{27}+5q^{26}+9q^{25}+22q^{24}\\
  && +7q^{23}+8q^{22}+7q^{21}+5q^{20}+q^{19}+7q^{18}+9q^{17}+20q^{16}+5q^{15}+20q^{14}\\
  && +4q^{13}+10q^{12}+8q^{11}+16q^{10}+12q^9+11q^8+3q^7+2q^6+2q^4+q^2+q^0
\end{eqnarray*}
with
\begin{eqnarray*}
  \mathcal{U}_{14,4,6} &=& \Big\{16128,15552,15008,13248,13728,13920,13968,14688,14736,14928,15408,\\
  && 4032,7816,13648,11656,11848,11908,7496,7556,7748,13104,15372,11588,\\
  && 3888,7042,7714,11048,14858,5928,6936,6948,10114,11074,11137,11554,\\
  && 11794,11809,5954,6017,6568,6760,6977,7442,7457,7697,9896,10008,\\
  && 10020,10904,11028,12528,13068,13578,13830,13833,14598,14601,14853,\\
  && 15363,5796,5908,6804,10049,10660,10852,11537,3312,3852,5528,5720,\\
  && 9576,10472,10584,13573,5352,5476,6360,6372,6484,9620,9812,9432,\\
  && 9444,10452,12492,13059,5332,1008,2786,3276,3843,4792,6322,12458,\\
  && 1761,4578,4818,8632,8824,8884,9394,10354,10417,2529,2732,3242,4472,\\
  && 4532,4724,5234,5297,6257,8658,8906,8913,12348,12378,12390,12393,\\
  && 12438,12441,12453,12483,2506,2761,4561,8564,9329,972,1452,1481,\\
  && 1644,1692,1734,2412,2460,2652,3132,3162,3174,3177,3222,3225,3237,\\
  && 3267,4805,6286,12373,8645,9358,10318,10381,937,963,1372,2618,\\
  && 3157,4771,5198,5261,6221,12339,874,922,934,1593,2707,8611,8803,\\
  && 9293,828,857,869,917,1338,1379,1427,1590,1619,1675,2358,2361,\\
  && 2613,3123,4654,6187,854,1333,2387,4491,4683,8494,8734,8749,\\
  && 8839,9259,10267,10279,819,2439,2631,4382,4397,4637,5147,5159,\\
  && 6167,8523,12303,4423,8477,9239,252,3087,243,783,207,63\Big\}.
\end{eqnarray*}
For $q\in\{2,3,4,5\}$ we have
\begin{eqnarray*}
  M_q(14,4;6) &\ge& q^{40}+q^{36}+q^{34}+8q^{32}+3q^{30}+3q^{29}+5q^{28}+q^{27}+5q^{26}+9q^{25}+22q^{24}\\
  && +7q^{23}+8q^{22}+7q^{21}+5q^{20}+q^{19}+7q^{18}+9q^{17}+20q^{16}+5q^{15}+20q^{14}\\
  && +4q^{13}+10q^{12}+8q^{11}+16q^{10}+12q^9+11q^8+3q^7+2q^6+2q^4+q^2+q^0
\end{eqnarray*}
with
\begin{eqnarray*}
  \mathcal{U}_{14,4,6} &=& \Big\{16128,15552,15008,13248,13728,13920,13968,14688,14736,14928,15408,\\
  && 4032,7816,13648,11656,11848,11908,7496,7556,7748,13104,15372,11588,\\
  && 3888,7042,7714,11048,14858,5928,6936,6948,10114,11074,11137,11554,\\
  && 11794,11809,5954,6017,6568,6760,6977,7442,7457,7697,9896,10008,\\
  && 10020,10904,11028,12528,13068,13578,13830,13833,14598,14601,14853,\\
  && 15363,5796,5908,6804,10049,10660,10852,11537,3312,3852,5528,5720,\\
  && 9576,10472,10584,13573,5352,5476,6360,6372,6484,9620,9812,9432,\\
  && 9444,10452,12492,13059,5332,1008,2786,3276,3843,4792,6322,12458,\\
  && 1761,4578,4818,8632,8824,8884,9394,10354,10417,2529,2732,3242,4472,\\
  && 4532,4724,5234,5297,6257,8658,8906,8913,12348,12378,12390,12393,\\
  && 12438,12441,12453,12483,2506,2761,4561,8564,9329,972,1452,1481,\\
  && 1644,1692,1734,2412,2460,2652,3132,3162,3174,3177,3222,3225,3237,\\
  && 3267,4805,6286,12373,8645,9358,10318,10381,937,963,1372,2618,\\
  && 3157,4771,5198,5261,6221,12339,874,922,934,1593,2707,8611,8803,\\
  && 9293,828,857,869,917,1338,1379,1427,1590,1619,1675,2358,2361,\\
  && 2613,3123,4654,6187,854,1333,2387,4491,4683,8494,8734,8749,\\
  && 8839,9259,10267,10279,819,2439,2631,4382,4397,4637,5147,5159,\\
  && 6167,8523,12303,4423,8477,9239,252,3087,243,783,207,63\Big\}.
\end{eqnarray*}

\bigskip

For $q=7$ we have 
\begin{eqnarray*}
  M_q(14,4;7) &=& q^{42}+q^{38}+q^{36}+8q^{34}+2q^{32}+9q^{30}+q^{29}+8q^{28}+q^{27}+27q^{26}+q^{25}\\
  && +2q^{24}+q^{23}+9q^{22}+2q^{21}+9q^{20}+q^{19}+27q^{18}+2q^{17}+9q^{16}+2q^{15}\\
  && +27q^{14}+20q^{13}+19q^{12}+10q^{11}+16q^{10}+2q^9+q^8+q^7+2q^6+q^5\\
  && +q^4+q^3+q^2+q^1+q^0
\end{eqnarray*}
with
\begin{eqnarray*}
  \mathcal{U}_{14,4,7} &=& \Big\{16256,15968,15696,14816,15056,15152,15176,15536,15560,15656,\\
  && 15896,14148,15016,10208,11972,12068,12098,13988,14018,14114,14744,\\
  && 15878,8001,6096,7620,7956,11938,13716,13761,14097,15621,7841,4016,\\
  && 4040,6056,7076,7106,7586,7820,7826,7946,10136,11156,11201,11660,\\
  && 11666,11681,11921,12041,13196,13202,13217,13706,13961,14456,14726,\\
  && 14981,15107,15491,7569,11146,13428,7049,9848,10118,10868,11372,\\
  && 11378,12908,12914,13418,14438,6021,7281,3971,5496,6516,7260,10858,\\
  && 12636,12657,13401,14421,6761,2936,3320,4856,6380,6386,6506,6716,\\
  && 6746,7226,8696,9830,10460,10481,10556,10586,10601,10841,11321,\\
  && 12476,12506,12521,12602,12857,14366,14381,14387,14411,5733,6489,\\
  && 1908,3683,5478,5718,9558,9573,9813,10426,13383,5461,6329,1772,1777,\\
  && 1898,2917,3301,3411,3637,3661,4838,4963,5347,5678,5683,5707,6727,\\
  && 8678,8918,8933,9014,9038,9398,9422,9518,9758,9773,11303,12839,1522,\\
  && 1881,2902,3286,3382,3406,4581,4821,4917,4941,5301,5325,5421,5661,\\
  && 7191,9043,9427,9523,9547,10567,1500,1513,1754,2531,2771,2867,2891,\\
  && 3251,3275,3371,3611,4566,5406,6439,8661,8878,9501,10775,12567,1863,\\
  && 2742,2766,2862,3246,4781,8883,8907,9003,9387,956,2485,2509,2731,\\
  && 2845,3229,4526,4531,4555,4766,4891,5275,8606,8621,8861,12431,\\
  && 1703,4509,2459,1431,911,8318,4221,2171,1143,623,351,191\Big\}.
\end{eqnarray*}
For $q\in\{2,3,4,5\}$ we have
\begin{eqnarray*}
  M_q(14,4;7) &\ge & q^{42}+q^{38}+q^{36}+8q^{34}+2q^{32}+9q^{30}+q^{29}+8q^{28}+q^{27}+27q^{26}+q^{25}\\
  && +2q^{24}+q^{23}+9q^{22}+2q^{21}+9q^{20}+q^{19}+27q^{18}+2q^{17}+9q^{16}+2q^{15}\\
  && +27q^{14}+20q^{13}+19q^{12}+10q^{11}+16q^{10}+2q^9+q^8+q^7+2q^6+q^5\\
  && +q^4+q^3+q^2+q^1+q^0
\end{eqnarray*}
with
\begin{eqnarray*}
  \mathcal{U}_{14,4,7} &=& \Big\{16256,15968,15696,14816,15056,15152,15176,15536,15560,15656,\\
  && 15896,14148,15016,10208,11972,12068,12098,13988,14018,14114,14744,\\
  && 15878,8001,6096,7620,7956,11938,13716,13761,14097,15621,7841,4016,\\
  && 4040,6056,7076,7106,7586,7820,7826,7946,10136,11156,11201,11660,\\
  && 11666,11681,11921,12041,13196,13202,13217,13706,13961,14456,14726,\\
  && 14981,15107,15491,7569,11146,13428,7049,9848,10118,10868,11372,\\
  && 11378,12908,12914,13418,14438,6021,7281,3971,5496,6516,7260,10858,\\
  && 12636,12657,13401,14421,6761,2936,3320,4856,6380,6386,6506,6716,\\
  && 6746,7226,8696,9830,10460,10481,10556,10586,10601,10841,11321,\\
  && 12476,12506,12521,12602,12857,14366,14381,14387,14411,5733,6489,\\
  && 1908,3683,5478,5718,9558,9573,9813,10426,13383,5461,6329,1772,1777,\\
  && 1898,2917,3301,3411,3637,3661,4838,4963,5347,5678,5683,5707,6727,\\
  && 8678,8918,8933,9014,9038,9398,9422,9518,9758,9773,11303,12839,1522,\\
  && 1881,2902,3286,3382,3406,4581,4821,4917,4941,5301,5325,5421,5661,\\
  && 7191,9043,9427,9523,9547,10567,1500,1513,1754,2531,2771,2867,2891,\\
  && 3251,3275,3371,3611,4566,5406,6439,8661,8878,9501,10775,12567,1863,\\
  && 2742,2766,2862,3246,4781,8883,8907,9003,9387,956,2485,2509,2731,\\
  && 2845,3229,4526,4531,4555,4766,4891,5275,8606,8621,8861,12431,\\
  && 1703,4509,2459,1431,911,8318,4221,2171,1143,623,351,191\Big\}.
\end{eqnarray*}

\subsection{Minimum subspace distance $6$}
\label{subsec_mindist_6}

\begin{eqnarray*}
  M_q(6,6;3) &=& q^3+q^0
\end{eqnarray*}
\begin{eqnarray*}
  \mathcal{U}_{6,6,3} &=& \Big\{56,7\Big\}
\end{eqnarray*}

\bigskip

\begin{eqnarray*}
  M_q(7,6;3) &=& q^4+q^0
\end{eqnarray*}
\begin{eqnarray*}
  \mathcal{U}_{7,6,3} &=& \Big\{112,7\Big\}
\end{eqnarray*}

\bigskip

\begin{eqnarray*}
  M_q(8,6;3) &=& q^5+q^0
\end{eqnarray*}
\begin{eqnarray*}
  \mathcal{U}_{8,6,3} &=& \Big\{224,7\Big\}
\end{eqnarray*}

\bigskip

\begin{eqnarray*}
  M_q(8,6;4) &=& q^8+q^0
\end{eqnarray*}
\begin{eqnarray*}
  \mathcal{U}_{8,6,4} &=& \Big\{240,15\Big\}
\end{eqnarray*}

\bigskip

\begin{eqnarray*}
  M_q(9,6;3) &=& q^6+q^3+q^0
\end{eqnarray*}
\begin{eqnarray*}
  \mathcal{U}_{9,6,3} &=& \Big\{448,56,7\Big\}
\end{eqnarray*}

\bigskip

\begin{eqnarray*}
  M_q(9,6;4) &=& q^{10}+q^3+q^0
\end{eqnarray*}
\begin{eqnarray*}
  \mathcal{U}_{9,6,4} &=& \Big\{480,{\color{green}284},39\Big\}
\end{eqnarray*}

\bigskip

\begin{eqnarray*}
  M_q(10,6;3) &=& q^7+q^4+q^0
\end{eqnarray*}
\begin{eqnarray*}
  \mathcal{U}_{10,6,3} &=& \Big\{896,112,7\Big\}
\end{eqnarray*}

\bigskip

\begin{eqnarray*}
  M_q(10,6;4) &=& q^{12}+q^6+q^2+q^1+q^0
\end{eqnarray*}
\begin{eqnarray*}
  \mathcal{U}_{10,6,4} &=& \Big\{960,{\color{blue}312},{\color{green}102,149},523\Big\}
\end{eqnarray*}

\bigskip

\begin{eqnarray*}
  M_q(10,6;5) &=& q^{15}+q^6+2q^2+2q^1
\end{eqnarray*}
\begin{eqnarray*}
  \mathcal{U}_{10,6,5} &=& \Big\{992,796,{\color{green}205},339,182,{\color{green}555}\Big\}
\end{eqnarray*}

\bigskip

\begin{eqnarray*}
  M_q(11,6;3) &=& q^8+q^5+q^0
\end{eqnarray*}
\begin{eqnarray*}
  \mathcal{U}_{11,6,3} &=& \Big\{1792,224,7\Big\}
\end{eqnarray*}

\bigskip

\begin{eqnarray*}
  M_q(11,6;4) &=& q^{14}+q^8+q^4+q^3+q^2+q^0
\end{eqnarray*}
\begin{eqnarray*}
  \mathcal{U}_{11,6,4} &=& \Big\{1920,240,{\color{cyan}300},586,{\color{green}1049},135\Big\}
\end{eqnarray*}

\bigskip

\begin{eqnarray*}
  M_q(11,6;5) &=& q^{18}+q^9+q^8+q^5+q^4+3q^3+3q^2
\end{eqnarray*}
\begin{eqnarray*}
  \mathcal{U}_{11,6,5} &=& \Big\{1984,824,{\color{cyan}1204,1129,358,597},675,1550,218,397,1299\Big\}
\end{eqnarray*}

\bigskip

\begin{eqnarray*}
  M_q(12,6;3) &=& q^9+q^6+q^3+q^0
\end{eqnarray*}
\begin{eqnarray*}
  \mathcal{U}_{12,6,3} &=& \Big\{3584,448,56,7\Big\}
\end{eqnarray*}

\bigskip

\begin{eqnarray*}
  M_q(12,6;4) &=& q^{16}+q^{10}+q^6+q^5+q^4+q^2+q^1+2q^0
\end{eqnarray*}
\begin{eqnarray*}
  \mathcal{U}_{12,6,4} &=& \Big\{3840,480,{\color{cyan}568},1108,2194,1161,2085,270,579\Big\}
\end{eqnarray*}

\bigskip

\begin{eqnarray*}
  \overline{M}_q(12,6;5) &=& q^{21}+q^{12}+q^{11}+q^9+3q^7+2q^6+2q^5
\end{eqnarray*}
\begin{eqnarray*}
  \mathcal{U}_{12,6,5} &=& \Big\{3968,880,{\color{red}1256},{\color{cyan}2260},2378,2604,3122,1308,1606,{\color{cyan}422},666\Big\}
\end{eqnarray*}

\bigskip

\begin{eqnarray*}
  \overline{M}_q(12,6;6) &=& q^{24}+q^{15}+2q^{10}+4q^8+4q^7+6q^6+2q^5+q^3+q^0
\end{eqnarray*}
\begin{eqnarray*}
  \mathcal{U}_{12,6,6} &=& \Big\{4032,3640,{\color{red}1830},2482,1393,2412,2837,3237,937,1436,2702,\\
  && 3158,756,1258,1683,{\color{red}2265},2659,3339,858,1613,455,63\Big\}
\end{eqnarray*}

\bigskip

\begin{eqnarray*}
  M_q(13,6;3) &=& q^{10}+q^7+q^4+q^0
\end{eqnarray*}
\begin{eqnarray*}
  \mathcal{U}_{13,6,3} &=& \Big\{7168,896,112,7\Big\}
\end{eqnarray*}

\bigskip

\begin{eqnarray*}
  M_q(13,6;4) &=& q^{18}+q^{12}+q^8+q^7+q^6+q^4+3q^3+2q^2+q^1+q^0
\end{eqnarray*}
\begin{eqnarray*}
  \mathcal{U}_{13,6,4} &=& \Big\{7680,960,1136,2216,4388,2322,{\color{green}540},4170,4241,1158,1289,\\
  && 2117,547\Big\}
\end{eqnarray*}

\bigskip

\begin{eqnarray*}
  M_q(13,6;5) &=& q^{24}+q^{15}+q^{14}+q^{12}+3q^{10}+2q^9+2q^8+q^5+q^4+q^2
\end{eqnarray*}
\begin{eqnarray*}
  \mathcal{U}_{13,6,5} &=& \Big\{7936,992,3280,5288,4440,4756,6244,2444,2616,1332,1612,\\
  && 1411,2627,{\color{green}4147}\Big\}
\end{eqnarray*}

\bigskip

\begin{eqnarray*}
  \overline{M}_q(13,6;6) &=& q^{28}+q^{19}+q^{16}+q^{14}+2q^{13}+3q^{12}+9q^{11}+q^{10}+q^8\\
  && +q^6+q^0
\end{eqnarray*}
\begin{eqnarray*}
  \mathcal{U}_{13,6,6} &=& \Big\{8064,7280,{\color{red}4968},3660,2786,{\color{cyan}4820},1873,2520,2868,1508,1720,\\
  && 3370,5322,5404,5670,6316,6470,6682,{\color{cyan}4529},3222,{\color{blue}909},63\Big\}
\end{eqnarray*}

\bigskip

\begin{eqnarray*}
  M_q(14,6;3) &=& q^{11}+q^8+q^5+q^0
\end{eqnarray*}
\begin{eqnarray*}
  \mathcal{U}_{14,6,3} &=& \Big\{14336,1792,224,7\Big\}
\end{eqnarray*}

\bigskip

\begin{eqnarray*}
  M_q(14,6;4) &=& q^{20}+q^{14}+q^{10}+q^9+q^8+2q^6+2q^5+2q^4+q^3+q^2
\end{eqnarray*}
\begin{eqnarray*}
  \mathcal{U}_{14,6,4} &=& \Big\{15360,1920,2272,4432,8776,{\color{cyan}1080},4644,8340,8482,2316,2578,\\
  && 4234,1094\Big\}
\end{eqnarray*}

\bigskip

\begin{eqnarray*}
  M_q(14,6;5) &=& q^{27}+q^{18}+q^{17}+q^{15}+3q^{13}+2q^{12}+2q^{11}+q^8+2q^7\\
  && +2q^5+2q^4+q^3+q^2
\end{eqnarray*}
\begin{eqnarray*}
  \mathcal{U}_{14,6,5} &=& \Big\{15872,1984,6560,10576,8880,9512,12488,4888,5232,2664,3224,\\
  && 5382,2694,{\color{green}8965},3141,9347,4675,8294,4245,{\color{green}307}\Big\}
\end{eqnarray*}

\bigskip

\begin{eqnarray*}
  \overline{M}_q(14,6;6) &=& q^{32}+q^{23}+q^{20}+3q^{18}+5q^{17}+7q^{15}+q^{14}+2q^{12}+q^{10}\\
  && +q^9+q^8+q^7+q^3+q^2+2q^1+2q^0
\end{eqnarray*}
\begin{eqnarray*}
  \mathcal{U}_{14,6,6} &=& \Big\{16128,14560,3792,{\color{red}5040},{\color{cyan}9160},13464,3496,5572,5736,9584,9892,\\
  && 6488,6796,7220,10644,10808,11340,12884,2916,7299,12588,10819,\\
  && 1820,1827,12563,8363,4199,663,{\color{green}1115},252,2319\Big\}
\end{eqnarray*}

\bigskip

\begin{eqnarray*}
  \overline{M}_q(14,6;7) &=& q^{35}+q^{26}+q^{21}+2q^{19}+3q^{18}+9q^{17}+4q^{16}+3q^{14}+q^{12}\\
  && +q^{11}+2q^{10}+2q^9+q^8+q^7+q^5+3q^4+2q^2+q^1+3q^0
\end{eqnarray*}
\begin{eqnarray*}
  \mathcal{U}_{14,6,7} &=& \Big\{16256,15472,13160,11850,14540,11092,13852,14886,6881,7497,9956,\\
  && 10722,10936,11564,13010,13638,14618,9688,10034,12724,13482,5925,7699,\\
  && 11414,3505,{\color{red}3725},1987,5017,5333,9102,6535,2859,{\color{blue}4467},1657,3175,\\
  && 6205,2267,4687,695,493,1311,8318\Big\}
\end{eqnarray*}

\bigskip

\begin{eqnarray*}
  M_q(15,6;3) &=& q^{12}+q^9+q^6+q^3+q^0
\end{eqnarray*}
\begin{eqnarray*}
  \mathcal{U}_{15,6,3} &=& \Big\{28672,3584,448,56,7\Big\}
\end{eqnarray*}

\bigskip

\begin{eqnarray*}
  M_q(15,6;4) &=& q^{22}+q^{16}+q^{12}+q^{11}+q^{10}+2q^8+2q^7+2q^6+q^5+q^4
\end{eqnarray*}
\begin{eqnarray*}
  \mathcal{U}_{15,6,4} &=& \Big\{30720,3840,4544,8864,17552,2160,9288,16680,16964,4632,5156,\\
  && 8468,2188\Big\}
\end{eqnarray*}

\bigskip

\begin{eqnarray*}
  M_q(15,6;5) &=& q^{30}+q^{21}+q^{20}+q^{18}+3q^{16}+2q^{15}+2q^{14}+q^{11}+2q^{10}\\
  && +5q^8+q^6+q^5+4q^4+q^3
\end{eqnarray*}
\begin{eqnarray*}
  \mathcal{U}_{15,6,5} &=& \Big\{31744,3968,13120,21152,17760,19024,24976,9776,10464,5328,6448,\\
  && 6668,9484,17930,12426,16588,18697,20742,25093,3142,1193,451,662,\\
  && 4197,8281,2595\Big\}
\end{eqnarray*}

\bigskip

\begin{eqnarray*}
  M_q(15,6;6) &=& q^{36}+q^{27}+q^{24}+3q^{22}+5q^{21}+8q^{19}+2q^{16}+q^{15}+2q^{14}\\
  && +2q^{13}+3q^{12}+2q^{11}+q^{10}+3q^6+q^5+6q^4+q^3
\end{eqnarray*}
\begin{eqnarray*}
  \mathcal{U}_{15,6,6} &=& \Big\{32256,29120,7584,10080,18320,26928,6992,11144,11472,19168,19784,\\
  && 5832,12976,13592,14440,21288,21616,22680,25768,22790,25176,3640,\\
  && 11525,13446,26755,28709,6789,10822,13059,7235,17958,17221,504,\\
  && 1253,1366,1419,723,2230,4638,{\color{green}8494},9267,{\color{green}16590},17437\Big\}
\end{eqnarray*}

\bigskip

For $q\ge 4$ we have
\begin{eqnarray*}
  \overline{M}_q(15,6;7) &=& q^{40}+q^{31}+2q^{26}+q^{24}+6q^{23}+6q^{22}+2q^{21}+4q^{20}+3q^{19}+2q^{18}\\
  && +q^{16}+3q^{15}+2q^{14}+3q^{13}+2q^{12}+2q^{11}+q^{10}+q^9+4q^8+2q^7\\
  && +2q^6+q^4+2q^3+2q^2
\end{eqnarray*}
with
\begin{eqnarray*}
  \mathcal{U}_{15,6,7} &=& \Big\{32512,30944,{\color{red}20176},29848,14020,14936,{\color{red}19880},21424,22120,25544,\\
  && 25968,11714,23180,23604,26274,27028,27724,7841,21953,15402,19300,\\
  && 22866,28972,13154,14729,27185,10129,29206,6026,11557,{\color{red}18204},19331,\\
  && 3890,3913,{\color{cyan}6566},6933,11022,19555,25867,5362,{\color{cyan}9449},14407,5433,2545,\\
  && 2794,12469,24787,4572,8634,{\color{red}3229},20651,1653,941,1390,16695,16975\Big\},
\end{eqnarray*}
for $q=3$ we have
\begin{eqnarray*}
  \overline{M}_3(15,6;7) &=& 3^{40}+3^{31}+2\cdot 3^{26}+3^{24}+5\cdot 3^{23}+9\cdot 3^{22}+3\cdot 3^{21}+3\cdot 3^{20}+2\cdot 3^{19}\\
  && +3^{18}+3\cdot 3^{17}+3^{16}+2\cdot 3^{15}+3\cdot 3^{14}+3\cdot 3^{12}+3^{11}+3\cdot 3^{10}+3^9\\
  && +2\cdot 3^8+2\cdot 3^7+2\cdot 3^6+3^5+2\cdot 3^4+4\cdot 3^3\\
  &=& 12\,158\,289\,296\,788\,694\,436
\end{eqnarray*}
with
\begin{eqnarray*}
  \mathcal{U}_{15,6,7} &=& \Big\{32512,30944,20176,29848,14020,19880,21448,21872,25520,26216,11938,\\
  && 15442,22932,23096,23628,26050,26968,27276,27700,7617,14993,22177,\\
  && 14730,19300,28972,13154,15401,29206,6034,10121,23107,{\color{red}7462,18204},\\
  && 26915,3889,3914,13589,6925,{\color{red}9457},21771,5721,5354,9530,19591,8922,\\
  && 2546,2793,6261,10350,4537,{\color{green}16636},12455,981,1654,942,1389,{\color{green}4446,8765}\Big\},
\end{eqnarray*}
and for $q=2$ we have
\begin{eqnarray*}
  M_2(15,6;7) &=& 2^{40}+2^{31}+2\cdot 2{26}+9\cdot 2^{23}+6\cdot 2^{22}+2\cdot 2^{21}+4\cdot 2^{20}+3\cdot 2^{19}+3\cdot 2^{18}\\
  && +2^{17}+2^{16}+3\cdot 2^{15}+4\cdot 2^{14}+2^{13}+2\cdot 2^{12}+2\cdot 2^{11}+2\cdot 2^{10}+2^9\\
  && +3\cdot 2^8+4\cdot 2^6+2\cdot 2^5+2\cdot 2^4+2^3+2^2\\
  &=& 1\,101\,905\,124\,972
\end{eqnarray*}
with
\begin{eqnarray*}
  \mathcal{U}_{15,6,7} &=& \Big\{32512,30944,20176,29848,14018,19880,21424,21956,22120,25544,25968,\\
  && 26276,29268,22872,23180,23604,27028,27192,27724,7841,15441,11170,\\
  && 15402,19300,28972,7570,14729,27779,10129,13153,14662,23107,11798,\\
  && 11557,13082,18204,3913,{\color{cyan}5926},10949,13837,6933,21771,28723,9449,\\
  && 10458,9614,25351,3302,2545,4586,4825,3229,6253,16636,{\color{cyan}17067},\emph{5214},\\
  && 8814,982,1651,{\color{green}8541},{\color{green}2366}\Big\},
\end{eqnarray*}
where $5214:[6,5,2,1,1,1,1]\rightarrow[6,5,2,1,1,1]$.

\bigskip

\begin{eqnarray*}
  M_q(16,6;3) &=& q^{13}+q^{10}+q^7+q^4+q^0
\end{eqnarray*}
\begin{eqnarray*}
  \mathcal{U}_{16,6,3} &=& \Big\{57344,7168,896,112,7\Big\}
\end{eqnarray*}

\bigskip

\begin{eqnarray*}
  M_q(16,6;4) &=& q^{24}+q^{18}+q^{14}+q^{13}+q^{12}+2q^{10}+2q^9+2q^8+q^7+q^6+q^0
\end{eqnarray*}
\begin{eqnarray*}
  \mathcal{U}_{16,6,4} &=& \Big\{61440,7680,9088,17728,35104,4320,18576,33360,33928,9264,10312,\\
  && 16936,4376,15\Big\}
\end{eqnarray*}

\bigskip

For $q\ge 3$ we have
\begin{eqnarray*}
  M_q(16,6;5) &=& q^{33}+q^{24}+q^{23}+q^{21}+3q^{19}+2q^{18}+2q^{17}+q^{14}+2q^{13}\\
  && +5q^{11}+2q^9+3q^8+5q^7+3q^6+q^5+q^4+q^0
\end{eqnarray*}
with
\begin{eqnarray*}
  \mathcal{U}_{16,6,5} &=& \Big\{63488,7936,26240,42304,35520,38048,49952,19552,20928,10656,12896,\\
  && 13336,18968,35860,24852,33176,37388,41490,50186,6290,21509,2380,\\
  && 8969,11267,902,1330,{\color{cyan}8394},16556,49233,689,37123,41093,6185,{\color{green}32870},31\Big\}
\end{eqnarray*}
and for $q=2$ we have
\begin{eqnarray*}
  M_2(16,6;5) &=& 2^{33}+2^{24}+2^{23}+2^{21}+3\cdot 2^{19}+2\cdot 2^{18}+2\cdot 2^{17}+2^{14}+2\cdot 2^{13}\\
  && +5\cdot 2^{11}+2^9+6\cdot 2^8+4\cdot 2^7+5\cdot 2^6+2^5+2^0=8\,619\,602\,785
\end{eqnarray*}
with
\begin{eqnarray*}
  \mathcal{U}_{16,6,5} &=& \Big\{63488,7936,26240,42304,35520,38048,49952,19552,20928,10656,12896,\\
  && 13336,18968,35860,24852,33176,37388,41490,50186,21509,2386,4753,\\
  && 6282,8969,10316,11267,1414,16556,20530,49233,1225,1329,16966,\\
  && 37123,41093,34857,31\Big\}.
\end{eqnarray*}

\bigskip

For $q\ge 3$ we have
\begin{eqnarray*}
  \overline{M}_q(16,6;6) &=& q^{40}+q^{31}+q^{28}+3q^{26}+5q^{25}+8q^{23}+2q^{20}+q^{19}+2q^{18}+3q^{17}+3q^{16}\\
  && +5q^{15}+3q^{14}+q^{12}+2q^{11}+5q^{10}+3q^9+5q^8+2q^7+2q^6+q^5+q^0
\end{eqnarray*}
with
\begin{eqnarray*}
  \mathcal{U}_{16,6,6} &=& \Big\{64512,58240,15168,20160,36640,53856,13984,22288,22944,38336,\\
  && 39568,11664,25952,27184,28880,42576,43232,45360,51536,45580,50352,\\
  && 7280,26892,53514,23049,26122,51718,14474,35980,57417,7430,11781,\\
  && 13577,21580,53381,28710,38403,43267,1008,17068,33482,2505,2842,\\
  && 9414,17221,17795,{\color{red}4508},18538,{\color{red}33132},1641,8613,8857,9276,34069,2723,\\
  && 36950,4451,34873,4661,16415\Big\}.
\end{eqnarray*}
For $q=2$ we have
\begin{eqnarray*}
  \overline{M}_2(16,6;6) &=& 2^{40}+2^{31}+2^{28}+3\cdot 2^{26}+5\cdot 2^{25}+8\cdot 2^{23}+2\cdot 2^{20}+2^{19}+2\cdot 2^{18}\\
   && +2\cdot 2^{17}+7\cdot 2^{16}+2^{15}+3\cdot 2^{14}+3\cdot 2^{13}+2^{12}+2^{11}+5\cdot 2^{10}\\
   && +5\cdot 2^8+3\cdot 2^7+2\cdot 2^6+2^5+2^4+2^1\\
   &=& 1\,102\,367\,740\,722
\end{eqnarray*}
with
\begin{eqnarray*}
  \mathcal{U}_{16,6,6} &=& \Big\{64512,58240,15168,20160,36640,53856,13984,22288,22944,38336,\\
  && 39568,11664,25952,27184,28880,42576,43232,45360,51536,45580,50352,\\
  && 7280,28938,51722,26764,50444,11785,23045,26118,39177,43270,53382,\\
  && 57417,42122,13573,21641,38403,14467,19715,35973,1008,{\color{cyan}4810},2844,\\
  && {\color{cyan}4902},7182,17193,33605,1644,10330,18534,20540,{\color{cyan}\emph{33130}},8979,{\color{cyan}16789},\\
  && {\color{green}32988},{\color{blue}1251},1369,{\color{red}2233},{\color{green}8374,4183}\Big\},
\end{eqnarray*}
where $33130:[10,4,3,3,2,1]\rightarrow[10,4,3,3,2].$

\bigskip

For $q\ge 4$ we have
\begin{eqnarray*}
  \overline{M}_q(16,6;7) &=& q^{45}+q^{36}+2q^{31}+q^{30}+q^{29}+4q^{28}+6q^{27}+q^{26}+2q^{25}+3q^{24}+4q^{23}\\
  && +q^{22}+3q^{21}+6q^{20}+2q^{18}+3q^{17}+3q^{15}+3q^{14}+3q^{13}+4q^{12}+4q^{11}\\
  && +3q^{10}+q^9+2q^8+2q^6+2q^4+q^2
\end{eqnarray*}
with
\begin{eqnarray*}
  \mathcal{U}_{16,6,7} &=& \Big\{65024,61888,15776,59696,{\color{red}20416},40272,39816,42896,43744,58536,\\
  && 29480,29808,30872,53936,54564,55400,15172,14024,57944,27916,46346,\\
  && 52372,23314,36642,47238,58118,59461,{\color{red}20024},29829,61475,7825,11858,\\
  && 23622,{\color{cyan}36428},38497,50953,44057,51843,13862,23077,45589,10705,13187,\\
  && 25923,{\color{cyan}17841},18858,41324,5916,{\color{red}17268},{\color{cyan}\emph{34246}},6386,21134,35237,37084,\\
  && 10893,14411,20825,24806,3433,{\color{red}17129},{\color{cyan}\emph{17626}},{\color{cyan}33619},8890,35102,5291,\\
  && 17943,{\color{cyan}4455},49213,10295\Big\},
\end{eqnarray*}
where $34246:[9,5,4,4,4,1,1]\rightarrow[9,5,4,4,4,1]$ and $17626:[8,5,3,3,2,2,1]\rightarrow[8,5,3,3$, $2,2]$, and for $q=3$ we have
\begin{eqnarray*}
  \overline{M}_3(16,6;7) &=& 3^{45}+3^{36}+2\cdot 3^{31}+3^{30}+8\cdot 3^{28}+6\cdot 3^{27}+2\cdot 3^{25}+3^{24}+2\cdot 3^{23}\\
  && +3\cdot 3^{21}+5\cdot 3^{20}+3\cdot 3^{18}+2\cdot 3^{17}+5\cdot 3^{16}+3^{15}+3\cdot 3^{14}+5\cdot 3^{13}\\
  && +4\cdot 3^{12}+2\cdot 3^{11}+3\cdot 3^{10}+3^8+2\cdot 3^7+2\cdot 3^6+3^4+2\cdot 3^3+3^0\\
  &=& 2\,954\,464\,473\,407\,763\,119\,113
\end{eqnarray*}
with
\begin{eqnarray*}
  \mathcal{U}_{16,6,7} &=& \Big\{65024,61888,23968,59696,36800,14176,15240,15568,23376,26512,27360,\\
  && 27976,58536,45744,46360,47208,54056,54384,55448,22216,57944,58630,\\
  && 47238,55557,{\color{red}20024},57989,59459,{\color{red}7942},29254,45827,46149,54403,27685,\\
  && {\color{cyan}36020},51750,11907,36138,14869,19654,28732,{\color{cyan}34609},35612,18245,23051,\\
  && {\color{red}34412},35497,5797,10693,19731,\emph{34458},35929,7267,20886,{\color{red}33272},37222,\\
  && 2994,24995,6457,9782,34189,{\color{cyan}17107,\emph{9005}},12491,4730,10334,974,{\color{green}16621},\\
  && 41019,5151\Big\},
\end{eqnarray*}
where $34458:[9,5,5,4,2,2,1]\rightarrow[7,5,5,4,2,2,1]$ and $9005:[7,4,4,2,1,1]\rightarrow[7,4,4,2,1]$.\\
For $q=2$ we have
\begin{eqnarray*}
  \overline{M}_2(16,6;7) &\le& 2^{45}+2^{35}+3\cdot 2^{33}+3\cdot 2^{32}+9\cdot 2^{31}+14\cdot 2^{30}+22\cdot 2^{29}+41\cdot 2^{28}\\
  && +28\cdot 2^{27}=35\,318\,321\,381\,376
\end{eqnarray*}
using $max\_dive=5$, $ub=122$ and
\begin{eqnarray*}
  \overline{M}_2(16,6;7) &\ge& 2^{45}+2^{36}+2^{31}+3\cdot 2^{30}+3\cdot 2^{29}+3\cdot 2^{28}+3\cdot 2^{27}+3\cdot 2^{26}+4\cdot 2^{25}\\
  && +2^{24}+2\cdot 2^{23}+5\cdot 2^{22}+3\cdot 2^{21}+5\cdot 2^{19}+2\cdot 2^{18}+2^{17}+2^{16}+6\cdot 2^{15}\\
  && +3\cdot 2^{14}+2^{13}+2\cdot 2^{12}+2\cdot 2^{11}+3\cdot 2^{10}+4\cdot 2^9+3\cdot 2^7+2\cdot 2^5+2^0 \\
  &=& 35\,261\,678\,822\,849
\end{eqnarray*}
with
\begin{eqnarray*}
  \mathcal{U}_{16,6,7} &=& \Big\{65024,61888,59696,15712,23456,40336,28040,42912,52448,29872,47272,\\
  && 54568,22352,39752,43728,27460,30808,57960,14024,45844,53912,58452,\\
  && 55558,29450,59525,12050,{\color{cyan}20116},29957,38500,46214,36408,44106,61475,\\
  && 14897,23683,27686,36613,46105,20041,25489,39026,{\color{cyan}18898},{\color{red}7694},21190,\\
  && {\color{red}34252},37763,50499,51731,\emph{18090},22629,{\color{cyan}35494},41370,{\color{red}33649},41318,9827,\\
  && 10659,4600,6541,{\color{red}17212},5333,{\color{cyan}5430},12627,18617,{\color{cyan}9389},24779,33971,\\
  && {\color{cyan}34909},41487,17439\Big\},
\end{eqnarray*}
where $18090:[8,5,5,4,3,2,1]\rightarrow[7,5,5,4,3,2,1]$.

\bigskip

For $q\ge 4$ we have
\begin{eqnarray*}
  \overline{M}_q(16,6;8) &=& q^{48}+q^{39}+q^{34}+q^{32}+4q^{31}+7q^{30}+4q^{29}+4q^{28}+3q^{27}+5q^{26}\\
  && +5q^{25}+2q^{23}+2q^{22}+7q^{21}+5q^{20}+q^{19}+q^{18}+3q^{16}+8q^{15}+2q^{14}\\
  && +q^{13}+4q^{12}+q^{11}+5q^{10}+4q^9+2q^8+2q^7+2q^6+2q^3
\end{eqnarray*}
with
\begin{eqnarray*}
  \mathcal{U}_{16,6,8} &=& \Big\{65280,63712,59088,62604,54704,56408,59800,62008,48274,52904,54216,\\
  && 55956,58728,60468,61780,52676,54884,58276,59980,30145,30370,40641,\\
  && 52080,31314,31369,43970,28257,{\color{red}31110},44449,46666,55596,31025,40290,\\
  && 45969,47433,47654,14192,60427,29978,30229,20369,23365,{\color{cyan}26438},27434,\\
  && 38790,45482,50972,23717,\emph{24078},39706,40213,54051,12172,15468,7977,\\
  && 37618,42214,19698,25433,26837,36438,{\color{cyan}39097},41705,43130,57523,39118,\\
  && 50515,37349,13020,18921,28779,49404,13607,18125,\emph{34485},35254,\emph{42077},\\
  && 51303,6611,11031,20893,37950,9883,\emph{17838,18782},19005,2019,33646,\\
  && 4783,{\color{green}8655}\Big\},
\end{eqnarray*}
where $24078:[7,6,6,6,6,1,1,1]\rightarrow[7,6,6,6,6,1,1]$, $34485:[8,4,4,3,2,2,1]\rightarrow[7,4,4,3$, $2,2,1]$, $42077:[8,7,5,2,1,1,1]\rightarrow[7,7,5,2,1,1,1]$, $17838:[7,4,3,3,2,1,1,1]\rightarrow[7,4,3$, $3,2,1,1]$ and $18782:[7,5,3,2,1,1,1,1]\rightarrow[7,5,3,2,1,1,1]$,
and for $q=3$ we have
\begin{eqnarray*}
  M_3(16,6;8) &=& 3^{48}+3^{39}+3^{34}+3^{32}+4\cdot 3^{31}+7\cdot 3^{30}+4\cdot 3^{29}+4\cdot 3^{28}+3\cdot 3^{27}\\
  && +5\cdot 3^{26}+5\cdot 3^{25}+2\cdot 3^{23}+2\cdot 3^{22}+7\cdot 3^{21}+5\cdot 3^{20}+3^{19}+3^{18}\\
  && +3\cdot 3^{16}+7\cdot 3^{15}+5\cdot 3^{14}+2\cdot 3^{13}+4\cdot 3^{12}+3^{11}+3\cdot 3^{10}\\
  && +6\cdot 3^9+3^8+2\cdot 3^7+2\cdot 3^6+2\cdot 3^3+3^2 \\
  &=& 79\,770\,518\,480\,353\,684\,929\,435
\end{eqnarray*}
with
\begin{eqnarray*}
  \mathcal{U}_{16,6,8} &=& \Big\{65280,63712,59088,62604,54704,56408,59800,62008,48274,52904,\\
  && 54216,55956,58728,60468,61780,52676,54884,58276,59980,30145,\\
  && 30370,40641,52080,31314,31369,43970,28257,31110,44449,46666,\\
  && 55596,31025,40290,45969,47433,47654,14192,60427,29978,30229,\\
  && 20369,23365,26438,27434,38790,45482,50972,23717,24078,39706,\\
  && 40213,54051,12172,15468,7977,37618,42214,25074,25433,26837,\\
  && 36438,39097,41705,43130,\emph{21718},25785,39118,42547,50515,\emph{19174},\\
  && 37349,13020,18921,28779,49404,13607,18125,35254,42077,6611,\\
  && 11031,\emph{18042},20893,37950,49819,17838,18782,19005,2019,33646,\\
  && 4783,8655,{\color{green}3231}\Big\},
\end{eqnarray*}
where $21718:[7,6,5,3,3,2,1,1]\rightarrow[7,6,5,3,3,2,1]$, $19174:[7,5,4,3,3,3,1,1]\rightarrow[7$, $5,4,3,3,3,1]$ and $18042:[7,4,4,2,2,2,2,1]\rightarrow[7,4,4,2,2,2,2]$.
For $q=2$ we have
\begin{eqnarray*}
  \overline{M}_2(16,6;8) &\le& 2^{48}+2^{39}+2^{34}+2^{33}+7\cdot 2^{32}+17\cdot 2^{31}+33\cdot 2^{30}+66\cdot 2^{29} \\
  && +11\cdot 2^{28}=282\,190\,894\,071\,808
\end{eqnarray*}
using $max\_dive=5$, $ub=138$ and
\begin{eqnarray*}
  \overline{M}_2(16,6;8) &\ge& 2^{48}+2^{39}+2^{34}+2^{32}+4\cdot 2^{31}+7\cdot 2^{30}+4\cdot 2^{29}+4\cdot 2^{28}+3\cdot 2^{27}\\
  && +5\cdot 2^{26}+5\cdot 2^{25}+6\cdot 2^{22}+7\cdot 2^{21}+6\cdot 2^{20}+2\cdot 2^{18}+2\cdot 2^{16}+7\cdot 2^{15}\\
  && +5\cdot 2^{14}+4\cdot 2^{13}+5\cdot 2^{12}+3\cdot 2^{10}+4\cdot 2^9+2^8+4\cdot 2^7+3\cdot 2^6+2^3\\
  &=& 282\,066\,487\,846\,856
\end{eqnarray*}
with
\begin{eqnarray*}
  \mathcal{U}_{16,6,8} &=& \Big\{65280,63712,59088,62604,54704,56408,59800,62008,48274,52904,54216,\\
  && 55956,58728,60468,61780,52676,54884,58276,59980,30145,30370,40641,\\
  && 52080,31314,31369,43970,28257,31110,44449,46666,55596,31025,40290,\\
  && 45969,47433,47654,27850,29978,30229,42802,44569,46193,20369,23365,\\
  && 26438,27434,38790,45482,50972,23717,24078,39706,40213,44302,54051,\\
  && 14121,15468,37618,42214,25074,25433,26837,36438,39097,41705,43130,\\
  && 11911,21718,25785,39118,50515,7563,19174,\emph{35725},37349,7731,13020,\\
  && 18921,28779,49404,18125,\emph{34485},35254,18042,20893,37950,49819,17838,\\
  && 10599,18782,19005,41277,2019,14367,33646,4783\Big\},
\end{eqnarray*}
where $35725:[7,5,5,5,2,1,1,1]\rightarrow[7,5,5,5,2,1,1]$ and $34485:[7,6,4,3,1,1,1,1]\rightarrow[7,6,4,3,1,1,1]$.

\bigskip

\begin{eqnarray*}
  M_q(17,6;3) &=& q^{14}+q^{11}+q^8+q^5+q^0
\end{eqnarray*}
\begin{eqnarray*}
  \mathcal{U}_{17,6,3} &=& \Big\{114688,14336,1792,224,7\Big\}
\end{eqnarray*}

\bigskip

\begin{eqnarray*}
  M_q(17,6;4) &=& q^{26}+q^{20}+q^{16}+q^{15}+q^{14}+2q^{12}+2q^{11}+2q^{10}+q^9\\
  && +q^8+q^0
\end{eqnarray*}
\begin{eqnarray*}
  \mathcal{U}_{17,6,4} &=& \Big\{122880,15360,18176,35456,70208,8640,37152,66720,67856,18528,\\
  && 20624,33872,8752,15\Big\}
\end{eqnarray*}

\bigskip

For $q\ge 3$ we have
\begin{eqnarray*}
  M_q(17,6;5) &=& q^{36}+q^{27}+q^{26}+q^{24}+3q^{22}+2q^{21}+2q^{20}+q^{17}+2q^{16}\\
  && +5q^{14}+2q^{12}+4q^{11}+4q^{10}+6q^9+4q^8+q^7+2q^6+q^1+q^0
\end{eqnarray*}
with
\begin{eqnarray*}
  \mathcal{U}_{17,6,5} &=& \Big\{126976,15872,52480,84608,71040,76096,99904,39104,41856,21312,\\
  && 25792,26672,37936,71720,49704,66352,74776,82980,100372,12580,\\
  && 43018,1384,4760,17938,22534,2828,16788,21513,98466,1697,2658,\\
  && 6417,33932,74246,82186,12370,37381,41057,68611,82001,10373,\\
  &&33094,203,55\Big\}
\end{eqnarray*}
and for $q=2$ we have
\begin{eqnarray*}
  M_2(17,6;5) &=& 2^{36}+2^{27}+2^{26}+2^{24}+3\cdot 2^{22}+2\cdot 2^{21}+2\cdot 2^{20}+2^{17}+2\cdot 2^{16}+5\cdot 2^{14}\\
  && +2^{12}+6\cdot 2^{11}+6\cdot 2^{10}+4\cdot 2^9+2\cdot 2^8+3\cdot 2^7+2^6+2^1+2^0\\
  &=& 68\,956\,824\,515
\end{eqnarray*}
with
\begin{eqnarray*}
  \mathcal{U}_{17,6,5} &=& \Big\{126976,15872,52480,84608,71040,76096,99904,39104,41856,21312,\\
  && 25792,26672,37936,71720,49704,66352,74776,82980,100372,43018,1384,\\
  && 2468,12578,16792,17938,22534,2648,12436,21513,37132,41060,98466,\\
  && 1676,4769,6417,74246,33106,68611,9477,65745,81994,49285,779,55\Big\}.
\end{eqnarray*}

\bigskip

For $q\ge 3$ we have
\begin{eqnarray*}
  M_q(17,6;6) &=& q^{44}+q^{35}+q^{32}+3q^{30}+5q^{29}+8q^{27}+2q^{24}+q^{23}+2q^{22}+3q^{21}\\
  && +3q^{20}+6q^{19}+7q^{18}+q^{17}+q^{16}+3q^{15}+4q^{14}+4q^{13}+7q^{12}+4q^{11}\\
  && +7q^{10}+q^9+q^7+q^3+q^1+q^0
\end{eqnarray*}
with
\begin{eqnarray*}
  \mathcal{U}_{17,6,6} &=& \Big\{129024,116480,30336,40320,73280,107712,27968,44576,45888,76672,\\
  && 79136,23328,51904,54368,57760,85152,86464,90720,103072,91160,\\
  && 100704,14560,53784,107028,46092,52244,103442,28948,71960,114828,\\
  && 15377,23562,27148,43160,84498,106762,2016,58371,76806,78345,87045,\\
  && 101385,114769,39430,10002,34136,43269,66964,5004,5684,34442,{\color{cyan}66264},\\
  && 2900,33681,37076,71811,3282,3372,3717,9609,17990,18552,18822,\\
  && 17225,17713,20658,24778,10819,33394,37161,49701,66342,69740,73905,\\
  && 66883,9317,81963,4187,2103\Big\}.
\end{eqnarray*}
For $q=2$ we have
\begin{eqnarray*}
  \overline{M}_2(17,6;6) &\le & 2^{44}+2^{35}+2^{34}+2^{33}+2^{32}+8\cdot 2^{31}+16\cdot 2^{30}+21\cdot 2^{29}+30\cdot 2^{28} \\
  && +44\cdot 2^{27} = 17\,716\,203\,225\,088
\end{eqnarray*}
using $max\_dive=5$, $ub=124$ and
\begin{eqnarray*}
  \overline{M}_2(17,6;6) &\ge & 2^{44}+q^{35}+2^{32}+3\cdot 2^{30}+5\dot 2^{29}+8\cdot 2^{27}+2\cdot 2^{24}+2^{23}+2\cdot 2^{22}+3\cdot 2^{21}\\
  && +3\cdot 2^{20}+6\cdot 2^{19}+7\cdot 2^{18}+2^{17}+2^{16}+3\cdot 2^{15}+4\cdot 2^{14}+4\cdot 2^{13}\\
  && +7\cdot 2^{12}+4\cdot 2^{11}+7\cdot 2^{10}+2^9+2^7+2^3+2^1+2^0\\
  &=& 17\,637\,885\,259\,403
\end{eqnarray*}
with
\begin{eqnarray*}
  \mathcal{U}_{17,6,6} &=& \Big\{129024,116480,30336,40320,73280,107712,27968,44576,45888,76672,\\
  && 79136,23328,51904,54368,57760,85152,86464,90720,103072,91160,\\
  && 100704,14560,53784,107028,46092,52244,103442,28948,71960,114828,\\
  && 15377,23562,27148,43160,84498,106762,2016,58371,76806,78345,87045,\\
  && 101385,114769,39430,10002,34136,43269,66964,5004,5684,34442,66264,\\
  && 2900,33681,37076,71811,3282,3372,3717,9609,17990,18552,18822,\\
  && 17225,17713,20658,24778,10819,33394,37161,49701,66342,69740,73905,\\
  && 66883,9317,81963,4187,2103\Big\}.
\end{eqnarray*}

\bigskip

For $q\ge 4$ we have
\begin{eqnarray*}
  \overline{M}_q(17,6;7) &=& q^{50}+q^{41}+2q^{36}+q^{35}+q^{34}+4q^{33}+6q^{32}+q^{31}+2q^{30}+3q^{29}\\
  && +4q^{28}+q^{27}+3q^{26}+8q^{25}+q^{24}+5q^{23}+6q^{22}+4q^{21}+3q^{20}+3q^{19}\\
  && +3q^{18}+5q^{17}+6q^{16}+2q^{15}+4q^{14}+4q^{13}+q^{12}+5q^{11}+3q^{10}+4q^9\\
  && +2q^8+q^7+q^4+q^3+q^2
\end{eqnarray*}
with
\begin{eqnarray*}
  \mathcal{U}_{17,6,7} &=& \Big\{130048,123776,31552,119392,40832,80544,79632,85792,87488,117072,\\
  && 58960,59616,61744,107872,109128,110800,30344,28048,115888,55832,\\
  && 92692,104744,46628,73284,94476,116236,118922,40048,59658,122950,\\
  && 15650,23716,47244,72856,76994,101906,111107,122921,116869,29793,\\
  && 43809,54537,88114,103686,27724,46154,77065,86673,88137,100033,\\
  && 24067,58499,78981,91162,10180,11832,21410,12776,{\color{cyan}19340},37716,51750,\\
  && {\color{red}66544},71210,19154,{\color{red}34264},72067,82762,106773,28822,35252,36010,\\
  && 36165,41650,75889,22805,74534,25369,37545,49516,{\color{red}66988},5553,\emph{18198},\\
  && 25253,53331,82140,3733,5721,6490,37094,69989,10891,14407,18787,\\
  && 18617,66899,68141,98426,33671,82983,9334,33853,12347,{\color{green}1231}\Big\},
\end{eqnarray*}
where $18198:[8,5,5,5,2,1,1]\rightarrow[8,5,5,5,2,1]$,
and for $q=3$ we have
\begin{eqnarray*}
  \overline{M}_3(17,6;7) &=& 3^{50}+3^{41}+2\cdot 3^{36}+3^{35}+8\cdot 3^{33}+6\cdot 3^{32}+2\cdot 3^{30}+3^{29}+2\cdot 3^{28}\\
  && +5\cdot 3^{26}+5\cdot 3^{25}+2\cdot 3^{24}+4\cdot 3^{23}+4\cdot 3^{22}+4\cdot 3^{21}+3\cdot 3^{20}\\
  && +5\cdot 3^{19}+8\cdot 3^{18}+5\cdot 3^{17}+3\cdot 3^{16}+6\cdot 3^{15}+5\cdot 3^{14}+5\cdot 3^{13}\\
  && +2\cdot 3^{12}+3\cdot 3^{10}+2\cdot 3^9+3^8+2\cdot 3^5+3^4+3^2+3^0 \\
  &=& 717\,934\,867\,043\,892\,179\,334\,175
\end{eqnarray*}
with
\begin{eqnarray*}
  \mathcal{U}_{17,6,7} &=& \Big\{130048,123776,31552,119392,40832,30368,77504,79632,80288,85792,\\
  && 87488,88720,117072,58960,59616,61744,107872,109104,110800,28048,\\
  && 115888,117260,61962,94476,15472,111109,115978,118921,122950,15884,\\
  && 91657,92298,107660,108809,55558,87558,{\color{cyan}36456},54348,58629,104522,\\
  && 23817,30853,39508,77062,44547,78890,90232,116771,39096,39265,94227,\\
  && 36148,38450,46105,72835,101445,{\color{red}18316},27174,{\color{red}33776},36050,41804,45443,\\
  && 70538,{\color{cyan}72233},5988,6882,14618,69146,102566,5848,13510,68044,11433,\\
  && 19331,19989,49961,71957,82757,20724,25033,{\color{cyan}67185},70002,99094,5553,\\
  && 18794,{\color{cyan}34469},74531,99011,10841,21603,\emph{9109},25630,73957,9555,{\color{red}66236},\\
  && 66861,8622,41067,49303,{\color{green}1615},2111\Big\},
\end{eqnarray*}
where $9109:[7,4,4,4,2,1]\rightarrow[6,4,4,4,2,1]$. For $q=2$ we have
\begin{eqnarray*}
  \overline{M}_2(17,6;7) &\le& 2^{50}+2^{41}+2^{40}+3\cdot 2^{38}+6\cdot 2^{37}+7\cdot 2^{36}+20\cdot 2^{35}+28\cdot 2^{34}\\
  && +44\cdot 2^{33}+73\cdot 2^{32}+22\cdot 2^{31} \\
  &=& 1\,133\,235\,710\,984\,192
\end{eqnarray*}
using $max\_dive=5$, $ub=206$ and
\begin{eqnarray*}
  \overline{M}_2(17,6;7) &\ge& 2^{50}+2^{41}+2^{36}+3\cdot 2^{35}+3\cdot 2^{34}+3\cdot 2^{33}+3\cdot 2^{32}+3\cdot 2^{31}+2\cdot 2^{30}\\
  && +2\cdot 2^{29}+2^{28}+3\cdot 2^{27}+4\cdot 2^{26}+2\cdot 2^{25}+6\cdot 2^{24}+3\cdot 2^{23}+3\cdot 2^{22}\\
  && +8\cdot 2^{21}+5\cdot 2^{20}+4\cdot 2^{19}+6\cdot 2^{18}+3\cdot 2^{17}+6\cdot 2^{16}+2\cdot 2^{15}\\
  && +5\cdot 2^{14}+4\cdot 2^{13}+2^{12}+2^{11}+3\cdot 2^{10}+2^9+2\cdot 2^8+2^7+2^6+2^4\\
  && +2^3+2^2+2^0 = 1\,128\,371\,758\,491\,869
\end{eqnarray*}
with
\begin{eqnarray*}
  \mathcal{U}_{17,6,7} &=& \Big\{130048,123776,119392,30528,47808,80672,56080,88512,102208,59744,\\
  && 92752,110928,24224,46496,79504,58928,91360,116912,28048,116008,\\
  && 54480,61964,117258,92428,94264,119046,15472,72984,110730,122953,\\
  && 91654,109061,15882,47369,59526,87180,103945,104524,{\color{cyan}36504},58629,\\
  && 100004,103474,107548,115843,30789,46150,61475,{\color{cyan}69026},71012,76937,\\
  && 79107,83729,{\color{cyan}20044},40069,{\color{cyan}68328,68500},79894,27395,35745,{\color{cyan}36148},57626,\\
  && 14508,{\color{red}17392},21386,39462,72323,85029,10124,74570,99026,10930,13861,\\
  && 22634,23571,49612,90261,37701,41208,25257,35945,70065,83034,\\
  && 100627,5577,{\color{cyan}5910},9923,75875,6745,{\color{cyan}18774,9529},17763,51229,20662,\\
  && {\color{green}8677},20781,98421,66343,3343,{\color{cyan}33339},{\color{green}1262},8287\Big\}.
\end{eqnarray*}

\bigskip

For $q\ge 3$ we have
\begin{eqnarray*}
  \overline{M}_q(17,6;8) &=& q^{54}+q^{45}+q^{40}+2q^{38}+4q^{37}+6q^{36}+7q^{35}+3q^{34}+2q^{33}+5q^{32}\\
  && +6q^{31}+3q^{30}+2q^{29}+7q^{28}+5q^{27}+5q^{26}+3q^{25}+2q^{24}+3q^{23}+6q^{22}\\
  && +5q^{21}+4q^{20}+8q^{19}+6q^{18}+5q^{17}+4q^{16}+3q^{15}+5q^{14}+6q^{13}+3q^{12}\\
  && +4q^{10}+q^9+q^8+2q^7+q^6+q^5+q^3+q^2+q^1
\end{eqnarray*}
with
\begin{eqnarray*}
  \mathcal{U}_{17,6,8} &=& \Big\{130560,127424,125232,{\color{red}81312},118152,56672,59296,117472,119984,\\
  && 56208,60624,63656,116560,119592,124008,60232,62744,{\color{red}80720},93508,\\
  && 109444,120920,123544,62064,96404,112740,87940,96522,54984,89282,\\
  && 94882,105748,111756,46916,47906,63749,{\color{red}77512},111698,125059,31425,\\
  && 120067,123462,85794,108298,62502,{\color{red}79416},89132,91538,105098,111177,\\
  && 123925,31793,{\color{cyan}40612},46737,52792,87649,44641,{\color{cyan}73481},85649,92713,\\
  && 105009,{\color{cyan}28294},47644,103124,24148,36802,44332,90540,99812,27106,\\
  && 45492,48139,{\color{red}71128},84764,86388,{\color{red}29462},30221,78629,100817,106844,\\
  && 37857,39372,88583,90353,26435,29033,42441,{\color{cyan}43418},82890,99250,\\
  && 100009,100714,23146,38314,39154,49656,52302,76854,19892,28890,\\
  && 51853,{\color{cyan}71913},100540,12053,35700,41708,51795,37722,\emph{71246},83257,\\
  && 13541,26748,34716,41785,50403,13195,18284,{\color{cyan}36147},38009,68999,83149,\\
  && 9970,21180,102567,{\color{cyan}7325},20915,49959,75867,17365,74391,6503,53279,\\
  && 5687,2990,41079,{\color{green}66173},1375\Big\},
\end{eqnarray*}
where $71246:[9,6,5,5,3,1,1,1]\rightarrow[8,6,5,5,3,1,1,1].$ For $q=2$ we have
\begin{eqnarray*}
  \overline{M}_2(17,6;8) &\le& 2^{54}+2^{44}+2\cdot 2^{42}+4\cdot 2^{41}+8\cdot 2^{40}+13\cdot 2^{39}+27\cdot 2^{38}+43\cdot 2^{37}\\
  &&+64\cdot 2^{36}+95\cdot 2^{35}+2^{34}\\
  &=& 18\,086\,536\,780\,185\,600
\end{eqnarray*}
using $max\_dive=5$, $ub=259$ and
\begin{eqnarray*}
  \overline{M}_2(17,6;8) &\ge& 2^{54}+2^{45}+2^{40}+2\cdot 2^{38}+4\cdot 2^{37}+6\cdot 2^{36}+7\cdot 2^{35}+3\cdot 2^{34}+2\cdot 2^{33}\\
  && +5\cdot 2^{32}+6\cdot 2^{31}+3\cdot 2^{30}+2\cdot 2^{29}+7\cdot 2^{28}+5\cdot 2^{27}+5\cdot 2^{26}+3\cdot 2^{25}\\
  && +2\cdot 2^{24}+3\cdot 2^{23}+6\cdot 2^{22}+5\cdot 2^{21}+4\cdot 2^{20}+8\cdot 2^{19}+6\cdot 2^{18}+5\cdot 2^{17}\\
  && +4\cdot 2^{16}+3\cdot 2^{15}+5\cdot 2^{14}+6\cdot 2^{13}+3\cdot 2^{12}+4\cdot 2^{10}+2^9+2^8\\
  && +2\cdot 2^7+2^6+2^5+2^3+2^2+2^1\\
  &=& 18\,052\,545\,205\,879\,918
\end{eqnarray*}
with
\begin{eqnarray*}
  \mathcal{U}_{17,6,8} &=& \Big\{130560,127424,125232,81312,118152,56672,59296,117472,119984,\\
  && 56208,60624,63656,116560,119592,124008,60232,62744,80720,93508,\\
  && 109444,120920,123544,62064,96404,112740,87940,96522,54984,89282,\\
  && 94882,105748,111756,46916,47906,63749,77512,111698,125059,31425,\\
  && 120067,123462,85794,108298,62502,79416,89132,91538,105098,111177,\\
  && 123925,31793,40612,46737,52792,87649,44641,73481,85649,92713,\\
  && 105009,28294,47644,103124,24148,36802,44332,90540,99812,27106,\\
  && 45492,48139,71128,84764,86388,29462,30221,78629,100817,106844,\\
  && 37857,39372,88583,90353,26435,29033,42441,43418,82890,99250,\\
  && 100009,100714,23146,38314,39154,49656,52302,76854,19892,28890,\\
  && 51853,71913,100540,12053,35700,41708,51795,37722,71246,83257,\\
  && 13541,26748,34716,41785,50403,13195,18284,36147,38009,68999,83149,\\
  && 9970,21180,102567,7325,20915,49959,75867,17365,74391,6503,53279,\\
  && 5687,2990,41079,66173,1375\Big\}.
\end{eqnarray*}

\bigskip

\begin{eqnarray*}
  M_q(18,6;3) &=& q^{15}+q^{12}+q^9+q^6+q^3+q^0
\end{eqnarray*}
\begin{eqnarray*}
  \mathcal{U}_{18,6,3} &=& \Big\{229376,28672,3584,448,56,7\Big\}
\end{eqnarray*}

\bigskip

\begin{eqnarray*}
  M_q(18,6;4) &=& q^{28}+q^{22}+q^{18}+q^{17}+q^{16}+2q^{14}+2q^{13}+2q^{12}+q^{11}\\
  && +q^{10}+q^3+q^0
\end{eqnarray*}
\begin{eqnarray*}
  \mathcal{U}_{18,6,4} &=& \Big\{245760,30720,36352,70912,140416,17280,74304,133440,135712,37056,\\
  && 41248,67744,17504,{\color{green}284},39\Big\}
\end{eqnarray*}

\bigskip

For $q\ge 3$ we have
\begin{eqnarray*}
  M_q(18,6;5) &=& q^{39}+q^{30}+q^{29}+q^{27}+3q^{25}+2q^{24}+2q^{23}+q^{20}+2q^{19}+5q^{17}\\
  && +2q^{15}+4q^{14}+5q^{13}+5q^{12}+4q^{11}+4q^{10}+3q^9+2q^8+q^7+q^6+q^5\\
  && +q^4+2q^0
\end{eqnarray*}
with
\begin{eqnarray*}
  \mathcal{U}_{18,6,5} &=& \Big\{253952,31744,104960,169216,142080,152192,199808,78208,83712,\\
  && 42624,51584,53344,75872,143440,99408,132704,149552,165960,200744,\\
  && 25160,86036,2768,9520,35876,45068,1480,4932,5656,33576,43026,\\
  && 12834,67864,148492,164372,196930,41281,74762,131492,137222,6433,\\
  && 74257,82113,98466,17554,20746,66188,18949,37009,10377,8390,899,\\
  && 3139,109,131099\Big\}.
\end{eqnarray*}
For $q=2$ we have
\begin{eqnarray*}
  \overline{M}_2(18,6;5) &\le& 2^{39}+2^{30}+2^{29}+2^{28}+4\cdot 2^{27}+4\cdot 2^{26}+6\cdot 2^{25}+16\cdot 2^{24}+26\cdot 2^{23}\\
  && +12\cdot 2^{22}=553\,178\,365\,952
\end{eqnarray*}
using $max\_dive=5$, $ub=72$ and
\begin{eqnarray*}
  \overline{M}_2(18,6;5) &\ge& 2^{39}+2^{30}+2^{29}+2^{27}+3\cdot 2^{25}+2\cdot 2^{24}+2\cdot 2^{23}+2^{20}+2\cdot 2^{19}\\
  && +5\cdot 2^{17}+2^{15}+6\cdot 2^{14}+4\cdot 2^{13}+8\cdot 2^{12}+4\cdot 2^{11}\\
  && +2\cdot 2^{10}+2^9+2\cdot 2^8+2^7+3\cdot 2^6+2^1+2^0 \\
  &=& 551\,654\,600\,003
\end{eqnarray*}
with
\begin{eqnarray*}
  \mathcal{U}_{18,6,5} &=& \Big\{253952,31744,104960,169216,142080,152192,199808,78208,83712,42624,\\
  && 51584,53344,75872,143440,99408,132704,149552,165960,200744,86036,\\
  && 976,9520,19012,24904,35876,45068,5320,43026,74264,196932,2728,3394,\\
  && 4900,6424,25122,83978,148492,164372,5650,98594,137222,139428,74758,\\
  && 82113,49673,37009,131466,197123,1413,20739,{\color{cyan}32966},1067,93\Big\}
\end{eqnarray*}

\bigskip

For $q\ge 3$ we have
\begin{eqnarray*}
  M_q(18,6;6) &=& q^{48}+q^{39}+q^{36}+3q^{34}+5q^{33}+8q^{31}+2q^{28}+q^{27}+2q^{26}+3q^{25}\\
  && +4q^{24}+6q^{23}+6q^{22}+q^{21}+2q^{20}+5q^{19}+8q^{18}+8q^{17}+7q^{16}\\
  && +4q^{15}+6q^{14}+5q^{13}+2q^{12}+2q^{11}+q^8+q^7+2q^6+4q^4+q^0
\end{eqnarray*}
with
\begin{eqnarray*}
  \mathcal{U}_{18,6,6} &=& \Big\{258048,232960,60672,80640,146560,215424,55936,89152,91776,153344,\\
  && 158272,46656,103808,108736,115520,170304,172928,181440,206144,\\
  && 182320,201408,29120,107568,214056,92184,104488,206884,4032,57896,\\
  && 143920,229656,30754,47124,54296,86320,168996,213524,116742,153612,\\
  && 156690,174090,202770,229538,78860,20004,155937,68272,86538,102929,\\
  && 133928,205321,5800,10008,11368,68884,77985,114825,132528,180739,\\
  && 7457,29189,37666,45321,57425,143622,151697,229445,6552,19218,\\
  && 34452,35416,37104,49572,76803,{\color{cyan}9060},10892,19594,132706,66788,67938,\\
  && 74322,99402,134225,139480,5458,17740,18644,35589,74122,6726,17121,\\
  && 9605,66001,135273,37006,34963,132363,{\color{green}8374,16506,65837},66183,8271\Big\}.
\end{eqnarray*}
For $q=2$ we have
\begin{eqnarray*}
  \overline{M}_2(18,6;6) &\le & 2^{48}+2^{39}+2^{38}+2^{37}+2^{36}+8\cdot 2^{35}+16\cdot 2^{34}+21\cdot 2^{33}+30\cdot 2^{32} \\
  && +46\cdot 2^{31}+60\cdot 2^{30} \\
  &=& 283\,527\,971\,078\,144
\end{eqnarray*}
using $max\_dive=5$, $ub=186$ and
\begin{eqnarray*}
  \overline{M}_2(18,6;6) &\ge & 2^{48}+2^{39}+2^{36}+3\cdot 2^{34}+5\cdot 2^{33}+8\cdot 2^{31}+2\cdot 2^{28}+2^{27}+2\cdot 2^{26}\\
  && +3\cdot 2^{25}+4\cdot 2^{24}+6\cdot 2^{23}+6\cdot 2^{22}+2^{21}+2\cdot 2^{20}+5\cdot 2^{19}+8\cdot 2^{18}\\
  && +8\cdot 2^{17}+7\cdot 2^{16}+4\cdot 2^{15}+6\cdot 2^{14}+5\cdot 2^{13}+2\cdot 2^{12}+2\cdot 2^{11}+2^8\\
  && +2^7+2\cdot 2^6+4\cdot 2^4+2^0 \\
  &=& 282\,206\,180\,430\,401
\end{eqnarray*}
with
\begin{eqnarray*}
  \mathcal{U}_{18,6,6} &=& \Big\{258048,232960,60672,80640,146560,215424,55936,89152,91776,153344,\\
  && 158272,46656,103808,108736,115520,170304,172928,181440,206144,\\
  && 182320,201408,29120,107568,214056,92184,104488,206884,4032,57896,\\
  && 143920,229656,30754,47124,54296,86320,168996,213524,116742,153612,\\
  && 156690,174090,202770,229538,78860,20004,155937,68272,86538,102929,\\
  && 133928,205321,5800,10008,11368,68884,77985,114825,132528,180739,\\
  && 7457,29189,37666,45321,57425,143622,151697,229445,6552,19218,\\
  && 34452,35416,37104,49572,76803,9060,10892,19594,132706,66788,67938,\\
  && 74322,99402,134225,139480,5458,17740,18644,35589,74122,6726,17121,\\
  && 9605,66001,135273,37006,34963,132363,8374,16506,65837,66183,8271\Big\}.
\end{eqnarray*}

\bigskip

For $q\ge 5$ we have
\begin{eqnarray*}
  \overline{M}_q(18,6;7) &=& q^{55}+q^{46}+2q^{41}+q^{40}+q^{39}+4q^{38}+6q^{37}+q^{36}+2q^{35}+3q^{34}+4q^{33}\\
  && +q^{32}+3q^{31}+8q^{30}+q^{29}+9q^{28}+7q^{27}+5q^{26}+4q^{25}+8q^{24}+4q^{23}\\
  && +6q^{22}+7q^{21}+7q^{20}+3q^{19}+7q^{18}+5q^{17}+7q^{16}+2q^{15}+4q^{14}+5q^{12}\\
  && +q^{11}+6q^{10}+3q^9+3q^8+3q^7+2q^5+q^3+q^2
\end{eqnarray*}
with
\begin{eqnarray*}
  \mathcal{U}_{18,6,7} &=& \Big\{260096,247552,63104,238784,81664,161088,159264,171584,174976,\\
  && 234144,117920,119232,123488,215744,218256,221600,60688,56096,231776,\\
  && 111664,185384,209488,93256,146568,188952,232472,237844,80096,119316,\\
  && 245900,31300,47432,94488,145712,153988,203812,222214,245842,233738,\\
  && 59586,87618,117258,123025,176228,184593,207372,217609,238083,20360,\\
  && 55448,154130,173346,176274,183301,200066,23664,40321,47633,61702,\\
  && 157962,11984,42820,92212,214113,31747,38680,75432,85521,107785,\\
  && {\color{red}133088},157777,209029,{\color{cyan}68528},103500,104582,{\color{cyan}134932},13732,87301,108613,\\
  && 118819,156421,174121,7842,13250,70504,72020,99986,151778,180774,\\
  && 13921,{\color{cyan}35560},41456,143914,149068,166100,202819,75090,140689,165513,\\
  && 28873,35666,36140,50506,135640,136963,140408,25778,82822,84265,\\
  && 134473,139956,13846,{\color{red}17876},21164,42140,74188,75930,99877,11046,\\
  && 82168,7693,37573,106542,164465,35939,49573,136277,147772,164291,\\
  && 83083,{\color{red}17241},18835,19486,24931,25645,\emph{66790},10827,36986,133767,4917,\\
  && 5305,196915,10525,33579,133230,34071,{\color{green}66205,4495,8407}\Big\},
\end{eqnarray*}
where $66790:[10,5,3,3,3,1,1]\rightarrow[7,5,3,3,3,1,1]$. For $q\in\{3,4\}$ we have
\begin{eqnarray*}
  M_q(18,6;7) &\le & q^{55}+q^{46}+3q^{41}+7q^{40}+26q^{39}+44q^{38}+60q^{37}+94q^{36}+76q^{35}
\end{eqnarray*}
using $max\_dive=5$, $ub=312$ and
\begin{eqnarray*}
  \overline{M}_q(18,6;7) &\ge & q^{55}+q^{46}+2q^{41}+q^{40}+q^{39}+4q^{38}+6q^{37}+q^{36}+2q^{35}+3q^{34}+4q^{33}\\
  && +q^{32}+3q^{31}+8q^{30}+q^{29}+9q^{28}+7q^{27}+5q^{26}+4q^{25}+8q^{24}+4q^{23}\\
  && +6q^{22}+7q^{21}+7q^{20}+3q^{19}+7q^{18}+5q^{17}+7q^{16}+2q^{15}+4q^{14}+5q^{12}\\
  && +q^{11}+6q^{10}+3q^9+3q^8+3q^7+2q^5+q^3+q^2
\end{eqnarray*}
with
\begin{eqnarray*}
  \mathcal{U}_{18,6,7} &=& \Big\{260096,247552,63104,238784,81664,161088,159264,171584,174976,\\
  && 234144,117920,119232,123488,215744,218256,221600,60688,56096,231776,\\
  && 111664,185384,209488,93256,146568,188952,232472,237844,80096,119316,\\
  && 245900,31300,47432,94488,145712,153988,203812,222214,245842,233738,\\
  && 59586,87618,117258,123025,176228,184593,207372,217609,238083,20360,\\
  && 55448,154130,173346,176274,183301,200066,23664,40321,47633,61702,\\
  && 157962,11984,42820,92212,214113,31747,38680,75432,85521,107785,\\
  && 133088,157777,209029,68528,103500,104582,134932,13732,87301,108613,\\
  && 118819,156421,174121,7842,13250,70504,72020,99986,151778,180774,\\
  && 13921,35560,41456,143914,149068,166100,202819,75090,140689,165513,\\
  && 28873,35666,36140,50506,135640,136963,140408,25778,82822,84265,\\
  && 134473,139956,13846,17876,21164,42140,74188,75930,99877,11046,\\
  && 82168,7693,37573,106542,164465,35939,49573,136277,147772,164291,\\
  && 83083,17241,18835,19486,24931,25645,66790,10827,36986,133767,4917,\\
  && 5305,196915,10525,33579,133230,34071,66205,4495,8407\Big\}.
\end{eqnarray*}
For $q=2$ we have
\begin{eqnarray*}
  \overline{M}_2(18,6;7) &\le& 2^{55}+2^{46}+2^{45}+3\cdot 2^{43}+6\cdot 2^{42}+7\cdot 2^{41}+20\cdot 2^{40}+28\cdot 2^{39}+45\cdot 2^{38}\\
  && +75\cdot 2^{37}+108\cdot 2^{36}+17\cdot 2^{35} \\
  &=& 36\,270\,586\,497\,859\,584
\end{eqnarray*}
using $max\_dive=5$, $ub=312$ and
\begin{eqnarray*}
  \overline{M}_2(18,6;7) &\ge& 2^{55}+2^{46}+2^{41}+3\cdot 2^{40}+3\cdot 2^{39}+3\cdot 2^{38}+3\cdot 2^{37}+3\cdot 2^{36}+2\cdot 2^{35}\\
  && +2\cdot 2^{34}+2^{33}+3\cdot 2^{32}+4\cdot 2^{31}+2\cdot 2^{30}+6\cdot 2^{29}+5\cdot 2^{28}+6\cdot 2^{27}\\
  && +8\cdot 2^{26}+8\cdot 2^{25}+6\cdot 2^{24}+7\cdot 2^{23}+4\cdot 2^{22}+6\cdot 2^{21}+4\cdot 2^{20}+3\cdot 2^{19}\\
  && +13\cdot 2^{18}+2\cdot 2^{17}+6\cdot 2^{16}+5\cdot 2^{15}+5\cdot 2^{14}+2\cdot 2^{13}+2\cdot 2^{12}\\
  && +3\cdot 2^{11}+3\cdot 2^{10}+3\cdot 2^8+2^7+2^5+3\cdot 2^4+2^3\\
  &=& 36\,107\,897\,362\,073\,560
\end{eqnarray*}
with
\begin{eqnarray*}
  \mathcal{U}_{18,6,7} &=& \Big\{260096,247552,238784,61056,95616,161344,112160,177024,204416,\\
  && 119488,185504,221856,48448,92992,159008,117856,182720,233824,56096,\\
  && 232016,108960,123928,234516,184856,188528,238092,30944,145968,221460,\\
  && 245906,183308,218122,31764,94738,119052,174360,207890,209048,73008,\\
  && 117258,177161,200008,245833,61713,188931,206948,215096,217605,231686,\\
  && 61578,92300,109573,122950,138052,142024,153874,167458,31241,40088,\\
  && 47622,80138,136656,137000,157841,159788,{\color{red}18400,20248},56323,71490,72296,\\
  && 115252,42772,78924,79953,107153,144646,170058,199457,138337,149140,\\
  && 170117,198052,27722,41832,45268,88102,180522,208931,50770,{\color{cyan}68500},\\
  && 75402,151945,{\color{cyan}35556},75105,172357,11657,11820,13957,{\color{cyan}19846},21624,\\
  && 25048,26161,36425,86193,102787,135906,140114,172198,13490,165169,\\
  && 14629,{\color{cyan}19122},21829,50053,51369,71193,{\color{red}34252},43203,74534,98552,132995,\\
  && {\color{cyan}18796},37489,82290,83139,148582,51285,{\color{red}66505},35603,198795,21070,24995,\\
  && 37934,5018,82989,{\color{cyan}133334},2545,68167,\textbf{131769},{\color{cyan}\emph{66846},132189},4459,33943,\\
  &&41021,3131\Big\},
\end{eqnarray*}
where $66846:[10,5,4,1,1,1,1]\rightarrow[10,5,4,1,1,1]$.

\bigskip

For $q\ge 4$ we have
\begin{eqnarray*}
  \overline{M}_q(18,6;8) &=& q^{60}+q^{51}+q^{46}+2q^{44}+4q^{43}+6q^{42}+7q^{41}+3q^{40}+2q^{39}+7q^{38}\\
  && +5q^{37}+2q^{36}+3q^{35}+6q^{34}+9q^{33}+8q^{32}+4q^{31}+7q^{30}+3q^{29}+8q^{28}\\
  && +9q^{27}+6q^{26}+7q^{25}+6q^{24}+9q^{23}+6q^{22}+8q^{21}+6q^{20}+7q^{19}+5q^{18}\\
  && +q^{17}+q^{16}+4q^{15}+8q^{14}+2q^{13}+5q^{12}+4q^{11}+4q^{10}+2q^9+q^8\\
  && +3q^7+q^5+2q^4+q^0
\end{eqnarray*}
with
\begin{eqnarray*}
  \mathcal{U}_{18,6,8} &=& \Big\{261120,254848,250464,64320,236304,211648,216896,234944,239968,\\
  && 210720,219552,225616,233120,239184,248016,63136,120584,187016,\\
  && 191024,218768,241840,247088,127176,192808,222432,175880,242188,\\
  && 96808,125208,127506,161352,178564,189764,208272,56720,95620,\\
  && 125092,159364,248326,246924,250117,{\color{red}81224},171588,239754,117864,\\
  && 161892,178264,189586,222474,254019,60706,{\color{cyan}81428},105764,119892,\\
  && 124421,{\color{cyan}146600},155170,158152,203888,116418,117537,161937,175297,\\
  && 177250,177425,219210,235011,31938,107944,154892,235561,{\color{red}73601},\\
  && 112710,120963,160518,211203,239637,246297,93321,95281,101320,\\
  && 57840,90980,92753,{\color{cyan}144280},169528,172776,203910,209989,{\color{cyan}77410},\\
  && {\color{cyan}79498},87394,103330,149416,152513,181074,183333,184518,62478,\\
  && 84884,105498,{\color{red}138088},{\color{cyan}141250},221244,39664,43922,54068,{\color{cyan}142164},\\
  && 142241,165780,216083,47653,52764,101713,197616,205641,230085,\\
  && 26513,29396,42849,{\color{red}71396},78305,83416,102776,107798,166328,44148,\\
  && {\color{red}69432,134872},151800,182539,205478,58457,\emph{72146},{\color{red}76024,134628},\\
  && {\color{cyan}140664},143985,149108,184371,20193,{\color{cyan}38601},88118,200205,229710,\\
  && 229779,23321,55373,{\color{cyan}76558},114874,\emph{136626},149873,201302,45835,\\
  && 57741,86669,{\color{cyan}142414},156299,166499,26170,20231,22099,45267,99886,\\
  && 14650,21899,36238,78109,106605,149678,172327,197817,83245,168093,\\
  && 28779,{\color{cyan}36019},38215,74547,90263,7381,10957,{\color{cyan}12750},140823,5980,13479,\\
  && 41566,{\color{cyan}68267,6814},49831,6573,\emph{18782},67943,135515,135727,{\color{green}33259},\\
  &&66767,983\Big\},
\end{eqnarray*}
where $72146:[9,6,6,4,4,4,3,1]\rightarrow[8,6,6,4,4,4,3,1]$, $136626:[10,6,5,4,4,3,3,1]\rightarrow[8,6,5,4,4,3,3,1]$ and $18782:[7,5,3,2,1,1,1,1]\rightarrow[7,5,3,2,1,1,1]$. For $q\in\{2,3\}$ we have
\begin{eqnarray*}
  \overline{M}_q(18,6;8) &\le& q^{60}+q^{51}+q^{46}+q^{44}+20q^{43}+51q^{42}+93q^{41}+138q^{40}+121q^{39}
\end{eqnarray*}
using $max\_dive=5$, $ub=427$ and
\begin{eqnarray*}
  \overline{M}_q(18,6;8) &\ge& q^{60}+q^{51}+q^{46}+2q^{44}+4q^{43}+6q^{42}+7q^{41}+3q^{40}+2q^{39}+7q^{38}\\
  && +5q^{37}+2q^{36}+3q^{35}+6q^{34}+9q^{33}+8q^{32}+4q^{31}+7q^{30}+3q^{29}+8q^{28}\\
  && +9q^{27}+6q^{26}+7q^{25}+6q^{24}+9q^{23}+6q^{22}+8q^{21}+6q^{20}+7q^{19}+5q^{18}\\
  && +q^{17}+q^{16}+4q^{15}+8q^{14}+2q^{13}+5q^{12}+4q^{11}+4q^{10}+2q^9+q^8\\
  && +3q^7+q^5+2q^4+q^0
\end{eqnarray*}
with
\begin{eqnarray*}
  \mathcal{U}_{18,6,8} &=& \Big\{261120,254848,250464,64320,236304,211648,216896,234944,239968,\\
  && 210720,219552,225616,233120,239184,248016,63136,120584,187016,\\
  && 191024,218768,241840,247088,127176,192808,222432,175880,242188,\\
  && 96808,125208,127506,161352,178564,189764,208272,56720,95620,\\
  && 125092,159364,248326,246924,250117,81224,171588,239754,117864,\\
  && 161892,178264,189586,222474,254019,60706,81428,105764,119892,\\
  && 124421,146600,155170,158152,203888,116418,117537,161937,175297,\\
  && 177250,177425,219210,235011,31938,107944,154892,235561,73601,\\
  && 112710,120963,160518,211203,239637,246297,93321,95281,101320,\\
  && 57840,90980,92753,144280,169528,172776,203910,209989,77410,\\
  && 79498,87394,103330,149416,152513,181074,183333,184518,62478,\\
  && 84884,105498,138088,141250,221244,39664,43922,54068,142164,\\
  && 142241,165780,216083,47653,52764,101713,197616,205641,230085,\\
  && 26513,29396,42849,71396,78305,83416,102776,107798,166328,44148,\\
  && 69432,134872,151800,182539,205478,58457,72146,76024,134628,\\
  && 140664,143985,149108,184371,20193,38601,88118,200205,229710,\\
  && 229779,23321,55373,76558,114874,136626,149873,201302,45835,\\
  && 57741,86669,142414,156299,166499,26170,20231,22099,45267,\\
  && 99886,14650,21899,36238,78109,106605,149678,172327,197817,\\
  && 83245,168093,28779,36019,38215,74547,90263,7381,10957,12750,\\
  && 140823,5980,13479,41566,68267,6814,49831,6573,18782,67943,\\
  && 135515,135727,33259,66767,983\Big\}.
\end{eqnarray*}

\bigskip

For $q\ge 4$ we have
\begin{eqnarray*}
  \overline{M}_q(18,6;9) &=& q^{63}+q^{54}+q^{49}+q^{47}+4q^{46}+6q^{45}+5q^{44}+q^{43}+3q^{42}+5q^{41}+11q^{40}\\
  && +q^{39}+7q^{38}+3q^{37}+7q^{36}+5q^{35}+6q^{34}+7q^{33}+4q^{32}+5q^{31}+7q^{30}\\
  && +2q^{29}+7q^{28}+12q^{27}+5q^{26}+8q^{25}+6q^{24}+12q^{23}+6q^{22}+5q^{21}\\
  && +6q^{20}+5q^{19}+5q^{18}+10q^{17}+6q^{16}+4q^{15}+5q^{14}+3q^{13}+4q^{12}\\
  && +2q^{11}+q^{10}+3q^9+q^7+q^6+2q^5+q^3+q^1+q^0
\end{eqnarray*}
with
\begin{eqnarray*}
  \mathcal{U}_{18,6,9} &=& \Big\{261632,258496,249248,256280,240480,242992,252016,254640,236432,\\
  && 240848,243880,247632,250664,255080,227620,236872,239496,248520,\\
  && 251032,235232,163216,224578,256134,193700,219012,224129,242264,\\
  && 255237,64416,97504,114050,129169,161632,187204,187586,193802,\\
  && 226066,226388,252163,211652,97096,122132,124706,127585,190145,\\
  && 194633,219810,112452,127628,191538,125524,208660,216676,233016,\\
  && 236677,242214,242755,159400,171810,177810,183956,250389,110244,\\
  && 121898,178514,184073,218698,254475,60872,63249,89889,186906,191013,\\
  && 211505,211994,93842,95800,120451,175690,95625,106066,{\color{red}155217},177705,\\
  && 215508,102337,175404,177029,184754,206308,221610,246118,154088,\\
  && 221553,32518,81233,104932,112909,146314,209308,232206,60227,63590,\\
  && 87024,93709,117849,{\color{cyan}160409},169428,173684,174513,188636,209138,\\
  && 229880,107474,110954,{\color{red}123934},208035,245971,48248,54753,115956,\\
  && 202521,214332,218163,233678,245933,115609,{\color{cyan}146533},156614,217421,\\
  && 230307,230989,40844,58232,77240,88474,{\color{cyan}108854},117299,135152,167322,\\
  && 168817,173401,181806,206972,31173,54218,79509,161815,202969,205545,\\
  && 40625,54876,84908,89159,167145,55609,61611,108747,{\color{cyan}152293},166630,\\
  && 168636,16073,28020,{\color{red}137155},213662,215147,71594,94299,150326,150919,\\
  && 199507,26533,27377,29902,30908,{\color{red}72583},79143,90734,100149,141114,\\
  && {\color{cyan}\emph{142157},27931},44567,\emph{45854},54055,{\color{cyan}76587},138422,43706,138767,\emph{142493},\\
  && 150622,20330,23254,69340, {\color{cyan}75379},90519,39531,42439,45301,51855,72765,\\
  && 102743,\emph{136621},13292,180511,36211,13235,38107,172143,18263,\emph{35293},\\
  && \emph{18107},135550,{\color{green}67831,8927},16879\Big\},
\end{eqnarray*}
where $146533:[9,6,6,6,6,3,3,1]\rightarrow[9,6,6,6,6,3,3]$, $142157:[9,6,5,4,4,3,1,1]\rightarrow[9,6,5,4,4,3,1]$, $45854:[7,6,6,4,4,1,1,1,1]\rightarrow[7,6,6,4,4,1,1]$, $142493:[9,6,5,5,3,1$, $1,1]\rightarrow[8,6,5,5,3,1,1,1]$,
$136621:[9,5,4,3,3,2,1,1]\rightarrow[8,5,4,3,3,2,1,1]$, $35293:[7,4,2,2,2,1,1,1]\rightarrow[7,4,2,2,2,1,1]$ and
$18107:[6,3,3,2,1,1,1]\rightarrow[6,3,3,2,1,1]$. For $q\in\{2,3\}$ we have
\begin{eqnarray*}
  \overline{M}_q(18,6;9) &\le& q^{63}+q^{54}+q^{49}+q^{47}+7q^{46}+22q^{45}+46q^{44}+90q^{43}+149q^{42}+106q^{41}
\end{eqnarray*}
using $max\_dive=5$, $ub=424$ and
\begin{eqnarray*}
  \overline{M}_q(18,6;9) &\ge& q^{63}+q^{54}+q^{49}+q^{47}+4q^{46}+6q^{45}+5q^{44}+q^{43}+3q^{42}+5q^{41}+11q^{40}\\
  && +q^{39}+7q^{38}+3q^{37}+7q^{36}+5q^{35}+6q^{34}+7q^{33}+4q^{32}+5q^{31}+7q^{30}\\
  && +2q^{29}+7q^{28}+12q^{27}+5q^{26}+8q^{25}+6q^{24}+12q^{23}+6q^{22}+5q^{21}\\
  && +6q^{20}+5q^{19}+5q^{18}+10q^{17}+6q^{16}+4q^{15}+5q^{14}+3q^{13}+4q^{12}\\
  && +2q^{11}+q^{10}+3q^9+q^7+q^6+2q^5+q^3+q^1+q^0
\end{eqnarray*}
with
\begin{eqnarray*}
  \mathcal{U}_{18,6,9} &=& \Big\{261632,258496,249248,256280,240480,242992,252016,254640,236432,\\
  && 240848,243880,247632,250664,255080,227620,236872,239496,248520,\\
  && 251032,235232,163216,224578,256134,193700,219012,224129,242264,\\
  && 255237,64416,97504,114050,129169,161632,187204,187586,193802,\\
  && 226066,226388,252163,211652,97096,122132,124706,127585,190145,\\
  && 194633,219810,112452,127628,191538,125524,208660,216676,233016,\\
  && 236677,242214,242755,159400,171810,177810,183956,250389,110244,\\
  && 121898,178514,184073,218698,254475,60872,63249,89889,186906,191013,\\
  && 211505,211994,93842,95800,120451,175690,95625,106066,155217,177705,\\
  && 215508,102337,175404,177029,184754,206308,221610,246118,154088,\\
  && 221553,32518,81233,104932,112909,146314,209308,232206,60227,63590,\\
  && 87024,93709,117849,160409,169428,173684,174513,188636,209138,\\
  && 229880,107474,110954,123934,208035,245971,48248,54753,115956,\\
  && 202521,214332,218163,233678,245933,115609,146533,156614,217421,\\
  && 230307,230989,40844,58232,77240,88474,108854,117299,135152,167322,\\
  && 168817,173401,181806,206972,31173,54218,79509,161815,202969,205545,\\
  && 40625,54876,84908,89159,167145,55609,61611,108747,152293,166630,\\
  && 168636,16073,28020,137155,213662,215147,71594,94299,150326,150919,\\
  && 199507,26533,27377,29902,30908,72583,79143,90734,100149,141114,\\
  && 142157,27931,44567,45854,54055,76587,138422,43706,138767,142493,\\
  && 150622,20330,23254,69340,75379,90519,39531,42439,45301,51855,72765,\\
  && 102743,136621,13292,180511,36211,13235,38107,172143,18263,35293,\\
  && 18107,135550,67831,8927,16879\Big\}.
\end{eqnarray*}

\bigskip

\begin{eqnarray*}
  M_q(19,6;3) &=& q^{16}+q^{13}+q^{10}+q^7+q^4+q^0
\end{eqnarray*}
\begin{eqnarray*}
  \mathcal{U}_{19,6,3} &=& \Big\{458752,57344,7168,896,112,7\Big\}
\end{eqnarray*}

\bigskip

\begin{eqnarray*}
  M_q(19,6;4) &=& q^{30}+q^{24}+q^{20}+q^{19}+q^{18}+2q^{16}+2q^{15}+2q^{14}+q^{13}\\
  && +q^{12}+q^6+q^2+q^1+q^0
\end{eqnarray*}
\begin{eqnarray*}
  \mathcal{U}_{19,6,4} &=& \Big\{491520,61440,72704,141824,280832,34560,148608,266880,271424,\\
  && 74112,82496,135488,35008,{\color{blue}312},{\color{green}102},149,523\Big\}
\end{eqnarray*}

\bigskip

For $q\ge 3$ we have
\begin{eqnarray*}
  M_q(19,6;5) &=& q^{42}+q^{33}+q^{32}+q^{30}+3q^{28}+2q^{27}+2q^{26}+q^{23}+2q^{22}+5q^{20}\\
  && +2q^{18}+4q^{17}+5q^{16}+5q^{15}+4q^{14}+4q^{13}+3q^{12}+4q^{11}+2q^{10}\\
  && +2q^9+q^8+2q^7+q^6+q^5+q^4+3q^2
\end{eqnarray*}
with
\begin{eqnarray*}
  \mathcal{U}_{19,6,5} &=& \Big\{507904,63488,209920,338432,284160,304384,399616,156416,167424,\\
  && 85248,103168,106688,151744,286880,198816,265408,299104,331920,\\
  && 401488,50320,172072,5536,19040,71752,90136,2960,9864,11312,67152,\\
  && 86052,25668,135728,296984,328744,393860,49480,82562,149524,274444,\\
  && 12866,148514,164226,328004,35108,41492,132376,10561,37898,74018,\\
  && 167941,20754,262945,100355,197129,1798,6278,18569,66693,\\
  && 248,278595,8339,{\color{green}131150,262198}\Big\}.
\end{eqnarray*}
For $q=2$ we have
\begin{eqnarray*}
  \overline{M}_2(19,6;5) &\le& 2^{42}+2^{33}+2^{32}+2^{31}+4\cdot 2^{30}+4\cdot 2^{29}+6\cdot 2^{28}+16\cdot 2^{27}+26\cdot 2^{26}\\
  && +23\cdot 2^{25}= 4\,425\,796\,026\,368
\end{eqnarray*}
using $max\_dive=5$, $ub=83$ and
\begin{eqnarray*}
  \overline{M}_2(19,6;5) &\ge & 2^{42}+2^{33}+2^{32}+3^{30}+3\cdot 2^{28}+2\cdot 2^{27}+2\cdot 2^{26}+2^{23}+2\cdot 2^{22}+5\cdot 2^{20}\\
  && +2\cdot 2^{18}+4\cdot 2^{17}+5\cdot 2^{16}+5\cdot 2^{15}+4\cdot 2^{14}+4\cdot 2^{13}+3\cdot 2^{12}+4\cdot 2^{11}\\
  && +2\cdot 2^{10}+2\cdot 2^9+2^8+2\cdot 2^7+2^6+2^5+2^4+3\cdot 2^2 \\
  &=& 4\,413\,236\,797\,052
\end{eqnarray*}
with
\begin{eqnarray*}
  \mathcal{U}_{19,6,5} &=& \Big\{507904,63488,209920,338432,284160,304384,399616,156416,167424,\\
  && 85248,103168,106688,151744,286880,198816,265408,299104,331920,\\
  && 401488,50320,172072,5536,19040,71752,90136,2960,9864,11312,67152,\\
  && 86052,25668,135728,296984,328744,393860,49480,82562,149524,274444,\\
  && 12866,148514,164226,328004,35108,41492,132376,10561,37898,74018,\\
  && 167941,20754,262945,100355,197129,1798,6278,18569,66693,\\
  && 248,278595,8339,131150,262198\Big\}.
\end{eqnarray*}

\bigskip

For $q\ge 4$ we have
\begin{eqnarray*}
  \overline{M}_q(19,6;6) &=& q^{52}+q^{43}+q^{40}+3q^{38}+5q^{37}+8q^{35}+2q^{32}+q^{31}+2q^{30}\\
  && +3q^{29}+4q^{28}+6q^{27}+6q^{26}+q^{25}+2q^{24}+5q^{23}+9q^{22}+13q^{21}+8q^{20}\\
  && +5q^{19}+5q^{18}+8q^{17}+3q^{16}+3q^{15}+3q^{14}+q^{13}+5q^{12}+2q^{11}+2q^{10}\\
  && +4q^9+q^7+q^5+2q^4+q^3+2q^2+q^0
\end{eqnarray*}
with
\begin{eqnarray*}
  \mathcal{U}_{19,6,6} &=& \Big\{516096,465920,121344,161280,293120,430848,111872,178304,183552,\\
  && 306688,316544,93312,207616,217472,231040,340608,345856,362880,\\
  && 412288,364640,402816,58240,215136,428112,184368,208976,413768,\\
  && 8064,115792,287840,459312,61508,94248,108592,172640,337992,427048,\\
  && 233484,307224,313380,348180,405540,459076,157720,40008,311874,\\
  && 136544,173076,205858,267856,410642,11600,20016,22736,90689,114978,\\
  && 137768,155970,265056,361478,14914,58378,69188,106690,180417,214025,\\
  && 229649,287244,303425,336913,360969,409889,458890,13104,15393,38436,\\
  && 68904,70472,70832,213510,346115,9928,21784,39188,{\color{cyan}132048},153605,\\
  && 134338,148644,198804,265368,268450,10916,34244,37288,82584,143625,\\
  && 168963,397457,401923,34578,132876,295572,19210,71817,263841,13446,\\
  && 25861,74124,263476,{\color{cyan}66017},133426,263498,272406,278918,21123,24754,\\
  && 10627,43021,{\color{red}16748},35363,264389,267277,2649,4435,{\color{cyan}4638},{\color{green}65658,8309},\\
  && {\color{green}32923},69671,131151\Big\}.
\end{eqnarray*}
For $q=3$ we have
\begin{eqnarray*}
  \overline{M}_3(19,6;6) &\le& 3^{52}+3^{43}+3^{40}+3\cdot 3^{38}+7\cdot 3^{37}+16\cdot 3^{36}+31\cdot 3^{35}+50\cdot 3^{34}\\
  && +71\cdot 3^{33}+47\cdot 3^{32}\\
  &=& 6\,461\,434\,776\,538\,418\,054\,680\,440
\end{eqnarray*}
using $max\_dive=5$, $ub=228$ and
\begin{eqnarray*}
  \overline{M}_3(19,6;6) &\ge & 3^{52}+3^{43}+3^{40}+3\cdot 3^{38}+5\cdot 3^{37}+8\cdot 3^{35}+2\cdot 3^{32}+3^{31}+2\cdot 3^{30}\\
  && +3\cdot 3^{29}+4\cdot 3^{28}+6\cdot 3^{27}+6\cdot 3^{26}+3^{25}+2\cdot 3^{24}+5\cdot 3^{23}+9\cdot 3^{22}\\
  && +13\cdot 3^{21}+8\cdot 3^{20}+5\cdot 3^{19}+5\cdot 3^{18}+8\cdot 3^{17}+3\cdot 3^{16}+3\cdot 3^{15}\\
  && +3\cdot 3^{14}+3^{13}+5\cdot 3^{12}+2\cdot 3^{11}+2\cdot 3^{10}+4\cdot 3^9+3^7+3^5+2\cdot 3^4\\
  && +3^3+2\cdot 3^2+3^0\\
  &=& 6\,461\,429\,013\,182\,807\,423\,722\,756
\end{eqnarray*}
with
\begin{eqnarray*}
  \mathcal{U}_{19,6,6} &=& \Big\{516096,465920,121344,161280,293120,430848,111872,178304,183552,\\
  && 306688,316544,93312,207616,217472,231040,340608,345856,362880,\\
  && 412288,364640,402816,58240,215136,428112,184368,208976,413768,\\
  && 8064,115792,287840,459312,61508,94248,108592,172640,337992,427048,\\
  && 233484,307224,313380,348180,405540,459076,157720,40008,311874,\\
  && 136544,173076,205858,267856,410642,11600,20016,22736,90689,114978,\\
  && 137768,155970,265056,361478,14914,58378,69188,106690,180417,214025,\\
  && 229649,287244,303425,336913,360969,409889,458890,13104,15393,38436,\\
  && 68904,70472,70832,213510,346115,9928,21784,39188,132048,153605,\\
  && 134338,148644,198804,265368,268450,10916,34244,37288,82584,143625,\\
  && 168963,397457,401923,34578,132876,295572,19210,71817,263841,13446,\\
  && 25861,74124,263476,66017,133426,263498,272406,278918,21123,24754,\\
  && 10627,43021,16748,35363,264389,267277,2649,4435,4638,65658,8309,\\
  && 32923,69671,131151\Big\}.
\end{eqnarray*}
For $q=2$ we have
\begin{eqnarray*}
  \overline{M}_2(19,6;6) &\le&  2^{52}+2^{43}+9\cdot 2^{38}+11\cdot 2^{37}+21\cdot 2^{36}+33\cdot 2^{35}+54\cdot 2^{34}+78\cdot 2^{33}\\\
  && +20\cdot 2^{32}=4\,520\,642\,057\,601\,024
\end{eqnarray*}
using $max\_dive=5$, $ub=228$ and
\begin{eqnarray*}
  \overline{M}_2(19,6;6) &\ge & 2^{52}+2^{43}+2^{40}+3\cdot 2^{38}+5\cdot 2^{37}+8\cdot 2^{35}+2\cdot 2^{32}+2^{31}+2\cdot 2^{30}\\
  && +3\cdot 2^{29}+4\cdot 2^{28}+6\cdot 2^{27}+6\cdot 2^{26}+2^{25}+2\cdot 2^{24}+5\cdot 2^{23}+9\cdot 2^{22}\\
  && +13\cdot 2^{21}+8\cdot 2^{20}+5\cdot 2^{19}+5\cdot 2^{18}+8\cdot 2^{17}+3\cdot 2^{16}+3\cdot 2^{15}\\
  && +3\cdot 2^{14}+2^{13}+5\cdot 2^{12}+2\cdot 2^{11}+2\cdot 2^{10}+4\cdot 2^9+2^7+2^5+2\cdot 2^4\\
  && +2^3+2\cdot 2^2+2^0\\
  &=& 4\,515\,298\,903\,445\,713
\end{eqnarray*}
with
\begin{eqnarray*}
  \mathcal{U}_{19,6,6} &=& \Big\{516096,465920,121344,161280,293120,430848,111872,178304,183552,\\
  && 306688,316544,93312,207616,217472,231040,340608,345856,362880,\\
  && 412288,364640,402816,58240,215136,428112,184368,208976,413768,\\
  && 8064,115792,287840,459312,61508,94248,108592,172640,337992,427048,\\
  && 233484,307224,313380,348180,405540,459076,157720,40008,311874,\\
  && 136544,173076,205858,267856,410642,11600,20016,22736,90689,114978,\\
  && 137768,155970,265056,361478,14914,58378,69188,106690,180417,214025,\\
  && 229649,287244,303425,336913,360969,409889,458890,13104,15393,38436,\\
  && 68904,70472,70832,213510,346115,9928,21784,39188,132048,153605,\\
  && 134338,148644,198804,265368,268450,10916,34244,37288,82584,143625,\\
  && 168963,397457,401923,34578,132876,295572,19210,71817,263841,13446,\\
  && 25861,74124,263476,66017,133426,263498,272406,278918,21123,24754,\\
  && 10627,43021,16748,35363,264389,267277,2649,4435,4638,65658,8309,\\
  && 32923,69671,131151\Big\}.
\end{eqnarray*}

\bigskip

For $q\ge 5$ we have
\begin{eqnarray*}
  \overline{M}_q(19,6;7) &=& q^{60}+q^{51}+2q^{46}+q^{45}+q^{44}+4q^{43}+6q^{42}+q^{41}+2q^{40}+3q^{39}\\
  && +4q^{38}+q^{37}+3q^{36}+8q^{35}+q^{34}+9q^{33}+7q^{32}+10q^{31}+6q^{30}+7q^{29}\\
  && +6q^{28}+7q^{27}+7q^{26}+11q^{25}+7q^{24}+15q^{23}+4q^{22}+7q^{21}+5q^{20}\\
  && +5q^{19}+5q^{18}+5q^{17}+q^{16}+4q^{15}+7q^{14}+5q^{13}+5q^{12}+6q^{11}\\
  && +3q^{10}+5q^9+q^8+2q^7+q^6+2q^5+q^4
\end{eqnarray*}
with
\begin{eqnarray*}
  \mathcal{U}_{19,6,7} &=& \Big\{520192,495104,126208,477568,163328,322176,318528,343168,\\
  && 349952,468288,235840,238464,246976,431488,436512,443200,121376,\\
  && 112192,463552,223328,370768,418976,186512,293136,377904,464944,\\
  && 475688,160192,238632,491800,62600,94864,188976,291424,307976,\\
  && 407624,444428,491684,467476,119172,175236,234516,246050,352456,\\
  && 369186,414744,435218,476166,40720,80644,110896,308260,346692,\\
  && 352548,366602,23968,47328,95266,123404,222225,315924,400130,\\
  && 430625,434369,468995,85640,174658,183329,184424,219145,{\color{cyan}266176},\\
  && 63494,77360,123041,150864,215570,315554,418058,137056,207000,\\
  && 209418,269864,364689,463109,23368,90946,188677,289801,312842,\\
  && 348242,360898,15682,141008,149380,199972,361548,430150,459913,\\
  && 26408,61521,71120,115473,144036,144652,166568,201128,275985,\\
  && 287828,298136,101538,138433,290949,303508,313603,331026,410257,\\
  && 22210,27794,76481,101012,103689,167178,202886,213324,229961,\\
  && 267184,273926,280816,330529,340009,396386,{\color{cyan}36452},137812,142097,\\
  && 268644,27020,42328,71329,84280,86369,151860,152835,11170,37825,\\
  && 74980,100102,164336,49880,75146,268681,296289,328588,53530,85059,\\
  && 143474,149706,296581,23573,27173,77916,274586,409701,51397,{\color{cyan}10053},\\
  && 39437,140835,272444,{\color{red}9649},{\color{cyan}17865,18033},{\color{red}34220},{\color{cyan}68202},86535,197722,\\
  && {\color{red}66420},165942,168075,286771,344093,5788,10601,41318,264793,393436,\\
  && 6803,19502,106523,\emph{131994},135513,393635,25227,37939,267339,20665,\\
  && {\color{cyan}\emph{34006}},82102,\textbf{262886},295034,3463,{\color{cyan}4907},67757,\emph{4558},12567,393743,{\color{green}132157}\Big\},
\end{eqnarray*}
where $34006:[9,5,3,3,2,1,1]\rightarrow[9,5,3,3,2,1]$, $4588:[6,3,3,3,1,1,1]\rightarrow[6,3,3,3,1,1]$ and  $131994:[11,4,4,4,2,2,1] \rightarrow [7,4,4,4,2,2,1]$, and for $q=4$ we have
\begin{eqnarray*}
  M_4(19,6;7) &=& 4^{60}+4^{51}+2\cdot 4^{46}+4^{45}+4^{44}+4\cdot 4^{43}+6\cdot 4^{42}+4^{41}+2\cdot 4^{40}\\
  && +3\cdot 4^{39}+4\cdot 4^{38}+4^{37}+3\cdot 4^{36}+7\cdot 4^{35}+6\cdot 4^{34}+6\cdot 4^{33}+6\cdot 4^{32}\\
  && +11\cdot 4^{31}+7\cdot 4^{30}+7\cdot 4^{29}+3\cdot 4^{28}+11\cdot 4^{27}+3\cdot 4^{26}+13\cdot 4^{25}+8\cdot 4^{24}\\
  && +8\cdot 4^{23}+4\cdot 4^{22}+7\cdot 4^{21}+7\cdot 4^{20}+7\cdot 4^{19}+7\cdot 4^{18}+7\cdot 4^{17}+6\cdot 4^{16}\\
  && +5\cdot 4^{15}+4\cdot 4^{14}+5\cdot 4^{13}+2\cdot 4^{12}+2\cdot 4^{11}+3\cdot 4^{10}+2\cdot 4^9\\
  && +4\cdot 4^8+2\cdot 4^7+4^5+2\cdot 4^4+3\cdot 4^3+4^0\\
  &=& 1\,329\,233\,078\,272\,310\,224\,711\,325\,051\,281\,770\,177
\end{eqnarray*}
with
\begin{eqnarray*}
  \mathcal{U}_{19,6,7} &=& \Big\{520192,495104,126208,477568,163328,322176,318528,343168,349952,\\
  && 468288,235840,238464,246976,431488,436512,443200,121376,112192,463552,\\
  && 223328,370768,418976,186512,293136,377904,464944,475688,160192,238632,\\
  && 491800,62600,94864,188976,291424,406280,444428,491684,62212,304912,\\
  && 407620,434376,467476,469002,175236,185380,246050,369186,414744,476166,\\
  && 110896,234514,315688,346692,352548,365580,23968,40712,47328,95266,\\
  && 96268,119169,222225,338689,352450,378883,430625,85640,135104,174658,\\
  && 217364,246277,307522,315922,77360,90952,118872,215570,273992,309257,\\
  && 320517,27856,188681,307348,72580,144496,181770,209057,217226,269860,\\
  && 296672,331048,418307,427092,463107,52756,142616,231561,14210,23362,\\
  && 71120,141992,157958,203781,268744,275745,288010,298136,361105,411729,\\
  && 413829,22212,51652,101476,122902,141092,156748,199842,360801,59434,\\
  && 137816,150305,168588,184387,267184,345161,401810,35696,271300,298246,\\
  && 336056,30851,42433,72778,100754,136977,148376,279956,11618,75148,166961,\\
  && 198241,198868,329298,395910,50514,51785,99974,137516,197096,201254,\\
  && 267913,13852,22644,78149,148194,280754,311398,328586,26153,42162,\\
  && 43557,99116,136518,229454,397418,84249,139985,144395,148867,264988,\\
  && 273427,10676,39203,84133,102923,265413,7318,37993,99605,164630,\\
  && 21773,24945,41308,68291,401453,12697,25230,37461,49340,132314,279093,\\
  && 295309,5434,12851,9799,131915,264747,266451,74027,263342,65910,69693,\\
  && 133223,935,2297,147515,34847\Big\}.
\end{eqnarray*}
For $q=3$ we have
\begin{eqnarray*}
  \overline{M}_3(19,6;7) &\le& q^{60}+q^{51}+2q^{46}+6q^{44}+27q^{43}+60q^{42}+89q^{41}+133q^{40}+144q^{39}\\
  &=& 42\,393\,356\,478\,392\,524\,679\,477\,084\,409
\end{eqnarray*}
using $max\_dive=5$, $ub=463$ and
\begin{eqnarray*}
  \overline{M}_3(19,6;7) &\ge& 3^{60}+3^{51}+2\cdot 3^{46}+3^{45}+3^{44}+4\cdot 3^{43}+6\cdot 3^{42}+3^{41}+2\cdot 3^{40}\\
  && +3\cdot 3^{39}+4\cdot 3^{38}+3^{37}+3\cdot 3^{36}+7\cdot 3^{35}+6\cdot 3^{34}+6\cdot 3^{33}+6\cdot 3^{32}\\
  && +11\cdot 3^{31}+7\cdot 3^{30}+7\cdot 3^{29}+3\cdot 3^{28}+11\cdot 3^{27}+3\cdot 3^{26}+13\cdot 3^{25}+8\cdot 3^{24}\\
  && +8\cdot 3^{23}+4\cdot 3^{22}+7\cdot 3^{21}+7\cdot 3^{20}+7\cdot 3^{19}+7\cdot 3^{18}+7\cdot 3^{17}+6\cdot 3^{16}\\
  && +5\cdot 3^{15}+4\cdot 3^{14}+5\cdot 3^{13}+2\cdot 3^{12}+2\cdot 3^{11}+3\cdot 3^{10}+2\cdot 3^9+4\cdot 3^8\\
  && +2\cdot 3^7+3^5+2\cdot 3^4+3\cdot 3^3+3^0\\
  &=& 42\,393\,335\,683\,434\,545\,764\,040\,230\,261
\end{eqnarray*}
with
\begin{eqnarray*}
  \mathcal{U}_{19,6,7} &=& \Big\{520192,495104,126208,477568,163328,322176,318528,343168,349952,\\
  && 468288,235840,238464,246976,431488,436512,443200,121376,112192,463552,\\
  && 223328,370768,418976,186512,293136,377904,464944,475688,160192,238632,\\
  && 491800,62600,94864,188976,291424,406280,444428,491684,62212,304912,\\
  && 407620,434376,467476,469002,175236,185380,246050,369186,414744,476166,\\
  && 110896,234514,315688,346692,352548,365580,23968,40712,47328,95266,96268,\\
  && 119169,222225,338689,352450,378883,430625,85640,{\color{cyan}135104},174658,217364,\\
  && 246277,307522,315922,77360,90952,118872,215570,273992,309257,320517,\\
  && 27856,188681,307348,72580,144496,181770,209057,217226,269860,296672,\\
  && 331048,418307,427092,463107,52756,142616,231561,14210,23362,71120,\\
  && 141992,157958,203781,268744,275745,288010,298136,361105,411729,413829,\\
  && 22212,51652,101476,122902,141092,156748,199842,360801,59434,137816,\\
  && 150305,168588,184387,267184,345161,401810,{\color{red}35696},271300,298246,336056,\\
  && 30851,42433,72778,100754,136977,148376,279956,11618,75148,166961,\\
  && 198241,198868,329298,395910,50514,51785,99974,137516,197096,201254,\\
  && 267913,13852,22644,78149,148194,280754,311398,328586,26153,42162,\\
  && 43557,99116,136518,229454,397418,84249,139985, 144395,148867,{\color{cyan}264988},\\
  && 273427,10676,39203,84133,102923,265413,{\color{red}7318},37993,99605,164630,\\
  &&21773,24945,41308,68291,401453,12697,25230,37461,49340,\emph{132314},\\
  && 279093, 295309,\emph{5434},12851,9799,{\color{cyan}131915},264747,266451,74027,{\color{green}263342},\\
  && {\color{green}65910},69693,133223,935,2297,147515,34847\Big\},
\end{eqnarray*}
where $132314:[11,5,3,3,2,2,1]\rightarrow[7,5,3,3,2,2,1]$ and $5434:[6,5,4,2,2,2,1]\rightarrow[6,5,4,2,2,2]$. For $q=2$ we have
\begin{eqnarray*}
  \overline{M}_2(19,6;7) &\le & 2^{60}+2^{51}+2^{45}+31\cdot 2^{44}+33\cdot 2^{43}+59\cdot 2^{42}+109\cdot 2^{41}+144\cdot 2^{40}\\
  && +84\cdot 2^{39}=1\,156\,747\,805\,071\,507\,456
\end{eqnarray*}
using $max\_dive=5$, $ub=463$ and
\begin{eqnarray*}
  \overline{M}_2(19,6;7) &\ge& 2^{60}+2^{51}+2^{46}+3\cdot 2^{45}+3\cdot 2^{44}+3\cdot 2^{43}+3\cdot 2^{42}+3\cdot 2^{41}+4\cdot 2^{40}\\
  && +2^{39}+2\cdot 2^{38}+5\cdot 2^{37}+4\cdot 2^{36}+3\cdot 2^{35}+4\cdot 2^{34}+8\cdot 2^{33}+6\cdot 2^{32}\\
  && +7\cdot 2^{31}+7\cdot 2^{30}+10\cdot 2^{29}+9\cdot 2^{28}+5\cdot 2^{27}+4\cdot 2^{26}+11\cdot 2^{25}\\
  && +11\cdot 2^{24}+9\cdot 2^{23}+3\cdot 2^{22}+7\cdot 2^{21}+8\cdot 2^{20}+5\cdot 2^{19}+5\cdot 2^{18}\\
  && +4\cdot 2^{17}+5\cdot 2^{16}+5\cdot 2^{15}+4\cdot 2^{14}+3\cdot 2^{13}+7\cdot 2^{12}+5\cdot 2^{11}\\
  &&+2\cdot 2^{10}+3\cdot 2^9+4\cdot 2^8+2\cdot 2^5+\cdot 2^3+\cdot 2^0\\
  &=& 1\,155\,454\,940\,188\,871\,241
\end{eqnarray*}
with
\begin{eqnarray*}
  \mathcal{U}_{19,6,7} &=& \Big\{520192,495104,477568,122112,191232,322688,224320,354048,408832,\\
  && 238976,371008,443712,113280,318016,415360,219680,246464,463680,\\
  && 177728,365760,432288,467616,378928,235600,476200,95392,161936,\\
  && 239656,308000,435248,61888,418384,469012,491800,32264,246804,\\
  && 442916,221488,340368,419874,434376,63524,123416,203912,234530,\\
  && 291908,309272,369676,444425,184616,312208,350220,365076,463889,\\
  && 491617,85648,94786,189450,274016,316434,342049,491654,{\color{red}36800},\\
  && 209092,348450,352449,413972,430602,475651,40496,79176,86920,\\
  && 109586,142980,184849,188577,238085,376906,377093,15184,22368,\\
  && 73284,116257,153922,168836,176402,273288,396496,72160,117254,\\
  && 276674,298280,396080,77092,92426,150808,289041,14210,43440,\\
  && 46228,51880,103016,141096,154629,217161,231683,287080,405548,\\
  && 84801,118915,136624,167012,198256,268056,282864,301318,305667,\\
  && 330914,361097,58449,83396,137921,156044,165528,199436,201362,\\
  && 330328,402182,50956,106722,275593,29445,39689,40010,43596,50594,\\
  && 51540,143713,26164,101509,140628,180338,{\color{red}264100},299345,394828,\\
  && 402499,281221,283177,300069,329042,394634,12984,19681,140177,\\
  && 141850,286874,99122,151718,198833,395589,13426,91143,{\color{cyan}265750},\\
  && 266700,303190,\emph{18122},71957,98732,\emph{132066},134691,21084,22586,70716,\\
  && 204939,70309,71179,266854,{\color{red}6798},7315,11321,18195,{\color{green}33524},{\color{cyan}66969},\\
  && {\color{cyan}133526},9452,45099,168007,{\color{cyan}264380},270755,35033,74286,147669,{\color{cyan}9035},\\
  && 164149,279581,18733,{\color{cyan}67702,262867},73821,264299,1383,33311\Big\},
\end{eqnarray*}
where $18122:[8,5,5,4,4,2,1]\rightarrow[7,5,5,4,4,2,1]$ and $132066:[11,4,4,4,4,4,1]\rightarrow[7,4,4,4,4,4,1]$.

\bigskip

For $q\ge 5$ we have
\begin{eqnarray*}
  \overline{M}_q(19,6;8) &=& q^{66}+q^{57}+q^{52}+2q^{50}+4q^{49}+6q^{48}+7q^{47}+3q^{46}+2q^{45}+7q^{44}\\
  && +5q^{43}+2q^{42}+3q^{41}+6q^{40}+10q^{39}+13q^{38}+4q^{37}+4q^{36}+8q^{35}\\
  && +9q^{34}+13q^{33}+11q^{32}+8q^{31}+7q^{30}+11q^{29}+7q^{28}+17q^{27}+9q^{26}\\
  && +9q^{25}+9q^{24}+4q^{23}+7q^{22}+6q^{21}+13q^{20}+7q^{19}+8q^{18}+7q^{17}+6q^{16}\\
  && +6q^{15}+6q^{14}+2q^{13}+2q^{12}+7q^{11}+q^{10}+q^9+2q^8+2q^7+q^6+q^5\\
  && +q^4+2q^3+q^1
\end{eqnarray*}
with
\begin{eqnarray*}
  \mathcal{U}_{19,6,8} &=& \Big\{522240,509696,500928,128640,472608,423296,433792,469888,479936,\\
  && 421440,439104,451232,466240,478368,496032,126272,241168,374032,\\
  && 382048,437536,483680,494176,254352,385616,444864,351760,484376,\\
  && 193616,250416,255012,322704,357128,379528,416544,113440,191240,\\
  && 250184,318728,496652,493848,500234,162448,343176,479508,235728,\\
  && 323784,356528,379172,444948,508038,121412,162856,211528,239784,\\
  && 248842,293200,310340,316304,407776,483845,63876,232836,235074,\\
  && 236577,323874,350594,354500,354850,438420,446985,451587,470022,\\
  && 499761,147204,215888,309784,471122,193030,225420,321036,479274,\\
  && 53152,186642,190562,202640,241925,362160,408593,477257,91952,181960,\\
  && 185506,317601,339056,345552,418578,419978,465411,{\color{red}79328},97285,154820,\\
  && 174788,230768,{\color{cyan}276176},298832,305026,337732,366642,369036,372867,\\
  && 432202,124956,127017,169768,210186,210996,223491,245985,291428,\\
  && 364037,365318,442488,108136,{\color{red}142800},213928,284328,284482,331560,\\
  && 334500,491843,115650,120067,177289,395232,405816,412933,460369,\\
  && 56372,88296,{\color{red}138864},170340,205552,{\color{red}269744},331458,332656,418001,427409,\\
  && 436237,48643,{\color{cyan}143010},{\color{red}269256},303600,314115,319813,403209,29633,40386,\\
  && 59953,80273,94534,152048,173388,281328,287970,291862,298216,338250,\\
  && 368742,410444,426706,427558,463001,55850,85601,85766,91529,118938,\\
  && 153993,172852,214659,462949,45794,47433,{\color{cyan}79388},87340,116249,181125,\\
  && 230044,289065,402604,58650,94739,103305,142177,153114,284828,305430,\\
  && 333189,361562,153125,199138,215189,336585,107670,199788,213606,\\
  && 315475,337187,344636,401862,107173,{\color{cyan}144486},174227,275561,299717,\\
  && 398531,27058,27308,50802,{\color{cyan}76442},79971,\emph{134986},159819,172633,{\color{cyan}\emph{269838}},\\
  && 275221,299356,395945,410019,27737,54350,99754,103509,148892,166998,\\
  && 281658,23699,{\color{cyan}27222},39574,52365,57995,76557,168334,345159,44078,\\
  && {\color{cyan}72249},90234,140949,148841,155950,295786,{\color{cyan}21414},49644,100947,{\color{red}135916},\\
  && 330069,331982,11685,201011,{\color{cyan}\emph{265574}},273031,344347,395446,13651,{\color{red}36153},\\
  && 104463,{\color{cyan}\emph{136022}},{\color{cyan}265611},267827,\emph{34661},279733,13498,{\color{red}68028},13099,35452,\\
  && {\color{cyan}37301,137309},150031,180279,197069,9934,18887,{\color{red}66809},204831,6522,\\
  && 68663,{\color{green}66455},295099,4827,{\color{green}1854},82095,{\color{green}8573}\Big\},
\end{eqnarray*}
where $134986:[10,5,5,5,5,4,2,1]\rightarrow[8,5,5,5,5,4,2,1]$, $269838:[11,6,6,6,6,1,1,1]\rightarrow[11,6,6,6,6,1,1]$,
$265574:[11,5,5,4,3,3,1,1]\rightarrow[11,5,5,4,3,3]$, $136022:[10,6,4,4,3,2$, $1,1]\rightarrow[10,6,4,4,3,2,1]$ and  $34661:[8,4,4,4,3,3,1]\rightarrow[7,4,4,4,3,3,1]$. For $q\in\{2,3,4\}$ we have
\begin{eqnarray*}
  \overline{M}_q(19,6;8) &\le& q^{66}+q^{57}+q^{52}+q^{50}+20q^{49}+51q^{48}+93q^{47}+140q^{46}+210q^{45}+175q^{44}
\end{eqnarray*}
using $max\_dive=5$, $ub=693$ and
\begin{eqnarray*}
  \overline{M}_q(19,6;8) &\ge& q^{66}+q^{57}+q^{52}+2q^{50}+4q^{49}+6q^{48}+7q^{47}+3q^{46}+2q^{45}+7q^{44}\\
  && +5q^{43}+2q^{42}+3q^{41}+6q^{40}+10q^{39}+13q^{38}+4q^{37}+4q^{36}+8q^{35}\\
  && +9q^{34}+13q^{33}+11q^{32}+8q^{31}+7q^{30}+11q^{29}+7q^{28}+17q^{27}+9q^{26}\\
  && +9q^{25}+9q^{24}+4q^{23}+7q^{22}+6q^{21}+13q^{20}+7q^{19}+8q^{18}+7q^{17}+6q^{16}\\
  && +6q^{15}+6q^{14}+2q^{13}+2q^{12}+7q^{11}+q^{10}+q^9+2q^8+2q^7+q^6+q^5\\
  && +q^4+2q^3+q^1
\end{eqnarray*}
with
\begin{eqnarray*}
  \mathcal{U}_{19,6,8} &=& \Big\{522240,509696,500928,128640,472608,423296,433792,469888,479936,\\
  && 421440,439104,451232,466240,478368,496032,126272,241168,374032,\\
  && 382048,437536,483680,494176,254352,385616,444864,351760,484376,\\
  && 193616,250416,255012,322704,357128,379528,416544,113440,191240,\\
  && 250184,318728,496652,493848,500234,162448,343176,479508,235728,\\
  && 323784,356528,379172,444948,508038,121412,162856,211528,239784,\\
  && 248842,293200,310340,316304,407776,483845,63876,232836,235074,\\
  && 236577,323874,350594,354500,354850,438420,446985,451587,470022,\\
  && 499761,147204,215888,309784,471122,193030,225420,321036,479274,\\
  && 53152,186642,190562,202640,241925,362160,408593,477257,91952,181960,\\
  && 185506,317601,339056,345552,418578,419978,465411,79328,97285,154820,\\
  && 174788,230768,276176,298832,305026,337732,366642,369036,372867,\\
  && 432202,124956,127017,169768,210186,210996,223491,245985,291428,\\
  && 364037,365318,442488,108136,142800,213928,284328,284482,331560,\\
  && 334500,491843,115650,120067,177289,395232,405816,412933,460369,\\
  && 56372,88296,138864,170340,205552,269744,331458,332656,418001,427409,\\
  && 436237,48643,143010,269256,303600,314115,319813,403209,29633,40386,\\
  && 59953,80273,94534,152048,173388,281328,287970,291862,298216,338250,\\
  && 368742,410444,426706,427558,463001,55850,85601,85766,91529,118938,\\
  && 153993,172852,214659,462949,45794,47433,79388,87340,116249,181125,\\
  && 230044,289065,402604,58650,94739,103305,142177,153114,284828,305430,\\
  && 333189,361562,153125,199138,215189,336585,107670,199788,213606,\\
  && 315475,337187,344636,401862,107173,144486,174227,275561,299717,\\
  && 398531,27058,27308,50802,76442,79971,134986,159819,172633,269838,\\
  && 275221,299356,395945,410019,27737,54350,99754,103509,148892,166998,\\
  && 281658,23699,27222,39574,52365,57995,76557,168334,345159,44078,\\
  && 72249,90234,140949,148841,155950,295786,21414,49644,100947,135916,\\
  && 330069,331982,11685,201011,265574,273031,344347,395446,13651,36153,\\
  && 104463,136022,265611,267827,34661,279733,13498,68028,13099,35452,\\
  && 37301,137309,150031,180279,197069,9934,18887,66809,204831,6522,\\
  && 68663,66455,295099,4827,1854,82095,8573\Big\}.
\end{eqnarray*}

\bigskip

For $q\ge 5$ we have
\begin{eqnarray*}
  \overline{M}_q(19,6;9) &=& q^{70}+q^{61}+q^{56}+q^{54}+4q^{53}+7q^{52}+5q^{51}+q^{50}+7q^{49}+5q^{48}\\
  && +6q^{47}+4q^{46}+8q^{45}+5q^{44}+6q^{43}+7q^{42}+11q^{41}+7q^{40}+4q^{39}+8q^{38}\\
  && +10q^{37}+10q^{36}+10q^{35}+11q^{34}+11q^{33}+9q^{32}+10q^{31}+6q^{30}+17q^{29}\\
  && +13q^{28}+9q^{27}+12q^{26}+9q^{25}+17q^{24}+6q^{23}+9q^{22}+8q^{21}+8q^{20}+5q^{19}\\
  && +8q^{18}+7q^{17}+6q^{16}+6q^{15}+3q^{14}+4q^{13}+4q^{12}+2q^{11}+3q^{10}+2q^9\\
  && +3q^8+q^7+2q^6+2q^5+q^4
\end{eqnarray*}
with
\begin{eqnarray*}
  \mathcal{U}_{19,6,9} &=& \Big\{523264,516992,512608,498112,473888,479040,497424,502352,{\color{red}326464},\\
  && 472768,481936,487760,495264,501088,510256,252680,480672,485600,\\
  && 503984,509136,484912,195232,255560,387624,389320,448136,454952,\\
  && 504332,128800,193984,256324,438024,470416,{\color{red}323216},375426,380296,\\
  && 423684,455186,510470,122560,256152,358788,512261,224900,227906,\\
  && {\color{red}310688},375064,382274,386324,451748,485642,243816,433732,448600,\\
  && 500874,516163,258182,{\color{red}324720},371908,383137,439841,466032,129556,\\
  && 236836,242497,249361,250936,351780,480332,194833,212504,242772,\\
  && 248930,424457,440582,444194,445121,472329,480771,501797,355416,\\
  && 357522,416962,421153,423988,494620,494851,220434,226401,240034,\\
  && 368137,350568,355089,365480,371762,377688,433297,451609,472102,\\
  && 97448,110465,171906,189296,218052,239366,246248,353729,408716,\\
  && 434920,236082,{\color{red}286593,291760},{\color{cyan}317842},324229,338888,353124,412584,\\
  && 433194,473107,65034,191619,234200,236165,246452,{\color{cyan}314724},363412,\\
  && 378437,475850,483477,120209,{\color{cyan}161425},175344,183700,214936,222508,\\
  && 363362,405090,422275,434644,459760,81732,123617,128013,146384,\\
  && 155240,{\color{cyan}293957},{\color{red}301808},{\color{cyan}308556,319970},373027,430922,111842,112849,\\
  && 177458,209802,\emph{278290},337826,380982,462025,491929,183625,217905,\\
  && 240139,306513,340600,353806,413210,460677,461420,491750,88994,\\
  && 109898,202594,{\color{cyan}321622},321805,418396,28640,60728,185614,186469,\\
  && 203192,215505,231209,331576,333272,341525,366670,377491,403832,\\
  && 414108,414374,430449,463446,60294,91986,93731,108216,117541,181958,\\
  && 190510,207537,217330,{\color{red}284372},404293,406618,442554,81201,{\color{red}146755},\\
  && 176742,200402,202412,219275,377899,442925,475257,46996,56611,87881,\\
  && 105626,116134,157354,166884,176549,234571,{\color{cyan}298698},365703,404118,\\
  && 104049,123275,143156,{\color{cyan}156617},159980,237773,291427,306695,350237,\\
  && 40792,47820,47913,54962,62041,94983,123004{\color{cyan},282225},\emph{285454},296929,\\
  && {\color{cyan}300604,304026,305340},340197,345276,410966,427603,55724,87412,92749,\\
  && 94522,{\color{red}274089},396710,31178,47476,54211,100300,{\color{red}117022},170198,{\color{cyan}\emph{275398}},\\
  && {\color{cyan}283881},459955,52877,58933,84728,{\color{cyan}107222},203085,206966,346439,410085,\\
  && 40389,58586,85221,\emph{101934},{\color{cyan}\emph{144973}},151303,288947,402222,58695,{\color{cyan}150713},\\
  && 176283,398009,402549,16038,107293,156086,181019,208957,338027,\\
  && 360813,428175,{\color{red}72531},115059,{\color{red}137955},157787,166521,283931,396083,\\
  && \emph{86686},213607,{\color{cyan}\emph{269613}},{\color{cyan}272203,296314},336222,29141,{\color{cyan}44141},76957,172395,\\
  && 299799,311517,23703,140750,330407,101463,201111,214047,{\color{cyan}\emph{280814}},\\
  && 23158,{\color{cyan}50030},136060,{\color{cyan}\emph{136619}},7666,271767,{\color{cyan}13883},74233,{\color{cyan}\emph{268507}},11870,\\
  && {\color{cyan}\emph{134045}},14623,{\color{cyan}\emph{70126}},397423,
  {\color{red}67307},37551,{\color{cyan}\emph{67383}},10727,35259,{\color{green}18813}\Big\},\end{eqnarray*}
  where $278290:[10,6,6,6,6,6,6,3,1]\rightarrow[9,6,6,6,6,6,6,3,1]$, $285454:[10,7,6,6,5,5,1,1,$ $1]\rightarrow[9,7,6,6,5,5,1,1,1]$,
$275398:[10,6,6,4,4,4,4,1,1]\rightarrow[10,6,6,4,4,4,4]$,
$101934:[8,8,5,5,5,2,1,1,1]\rightarrow[8,8,5,5,5,2,1,1]$,
$144973:[9,6,6,5,5,3,1,1]\rightarrow[9,6,6,5,5,3,1]$,
$86686:[8,7,6,4,3,1,1,1,1]\rightarrow[8,7,6,4,3,1,1,1]$,
$269613:[10,5,5,5,4,2,1,1]\rightarrow[10,5,5,5,4,2]$,
$280814:[10,7,5,2,2,2,1,1,1]\rightarrow[7,7,5,2,2,2,1,1,1]$,
$136619:[9,5,4,3$, $3,2,1]\rightarrow[9,5,4,3,3,2]$,
$268507:[10,5,5,2,2,1,1]\rightarrow[10,5,5,2,2,1]$,
$134045:[9,4,3$, $3,3,1,1,1]\rightarrow[9,4,3,3,3]$,
$70126:[8,5,2,2,2,2,1,1,1]\rightarrow[8,5,2,2,2,2]$ and
$67383:[8,3,3,3,1,1]\rightarrow[8,3,3,3]$. For $q\in\{2,3,4\}$ we have
\begin{eqnarray*}
  \overline{M}_q(19,6;9) &\le& q^{70}+q^{61}+q^{56}+q^{54}+7q^{53}+26q^{52}+69q^{51}+138q^{50}+206q^{49}\\
  &&+285q^{48}+54q^{47}
\end{eqnarray*}
using $max\_dive=5$, $ub=789$ and
\begin{eqnarray*}
  \overline{M}_q(19,6;9) &\ge& q^{70}+q^{61}+q^{56}+q^{54}+4q^{53}+7q^{52}+5q^{51}+q^{50}+7q^{49}+5q^{48}\\
  && +6q^{47}+4q^{46}+8q^{45}+5q^{44}+6q^{43}+7q^{42}+11q^{41}+7q^{40}+4q^{39}+8q^{38}\\
  && +10q^{37}+10q^{36}+10q^{35}+11q^{34}+11q^{33}+9q^{32}+10q^{31}+6q^{30}+17q^{29}\\
  && +13q^{28}+9q^{27}+12q^{26}+9q^{25}+17q^{24}+6q^{23}+9q^{22}+8q^{21}+8q^{20}+5q^{19}\\
  && +8q^{18}+7q^{17}+6q^{16}+6q^{15}+3q^{14}+4q^{13}+4q^{12}+2q^{11}+3q^{10}+2q^9\\
  && +3q^8+q^7+2q^6+2q^5+q^4
\end{eqnarray*}
with
\begin{eqnarray*}
  \mathcal{U}_{19,6,9} &=& \Big\{523264,516992,512608,498112,473888,479040,497424,502352,326464,\\
  && 472768,481936,487760,495264,501088,510256,252680,480672,485600,\\
  && 503984,509136,484912,195232,255560,387624,389320,448136,454952,\\
  && 504332,128800,193984,256324,438024,470416,323216,375426,380296,\\
  && 423684,455186,510470,122560,256152,358788,512261,224900,227906,\\
  && 310688,375064,382274,386324,451748,485642,243816,433732,448600,\\
  && 500874,516163,258182,324720,371908,383137,439841,466032,129556,\\
  && 236836,242497,249361,250936,351780,480332,194833,212504,242772,\\
  && 248930,424457,440582,444194,445121,472329,480771,501797,355416,\\
  && 357522,416962,421153,423988,494620,494851,220434,226401,240034,\\
  && 368137,350568,355089,365480,371762,377688,433297,451609,472102,\\
  && 97448,110465,171906,189296,218052,239366,246248,353729,408716,\\
  && 434920,236082,286593,291760,317842,324229,338888,353124,412584,\\
  && 433194,473107,65034,191619,234200,236165,246452,314724,363412,\\
  && 378437,475850,483477,120209,161425,175344,183700,214936,222508,\\
  && 363362,405090,422275,434644,459760,81732,123617,128013,146384,\\
  && 155240,293957,301808,308556,319970,373027,430922,111842,112849,\\
  && 177458,209802,278290,337826,380982,462025,491929,183625,217905,\\
  && 240139,306513,340600,353806,413210,460677,461420,491750,88994,\\
  && 109898,202594,321622,321805,418396,28640,60728,185614,186469,\\
  && 203192,215505,231209,331576,333272,341525,366670,377491,403832,\\
  && 414108,414374,430449,463446,60294,91986,93731,108216,117541,181958,\\
  && 190510,207537,217330,284372,404293,406618,442554,81201,146755,\\
  && 176742,200402,202412,219275,377899,442925,475257,46996,56611,87881,\\
  && 105626,116134,157354,166884,176549,234571,298698,365703,404118,\\
  && 104049,123275,143156,156617,159980,237773,291427,306695,350237,\\
  && 40792,47820,47913,54962,62041,94983,123004,282225,285454,296929,\\
  && 300604,304026,305340,340197,345276,410966,427603,55724,87412,92749,\\
  && 94522,274089,396710,31178,47476,54211,100300,117022,170198,275398,\\
  && 283881,459955,52877,58933,84728,107222,203085,206966,346439,410085,\\
  && 40389,58586,85221,101934,144973,151303,288947,402222,58695,150713,\\
  && 176283,398009,402549,16038,107293,156086,181019,208957,338027,\\
  && 360813,428175,72531,115059,137955,157787,166521,283931,396083,\\
  && 86686,213607,269613,272203,296314,336222,29141,44141,76957,172395,\\
  && 299799,311517,23703,140750,330407,101463,201111,214047,280814,\\
  && 23158,50030,136060,136619,7666,271767,13883,74233,268507,11870,\\
  && 134045,14623,70126,397423,67307,37551,67383,10727,35259,18813\Big\}.
\end{eqnarray*}

\subsection{Minimum subspace distance $8$}
\label{subsec_mindist_8}

\begin{eqnarray*}
  M_q(8,8;4) &=& q^4+q^0
\end{eqnarray*}
\begin{eqnarray*}
  \mathcal{U}_{8,8,4} &=& \Big\{240,15\Big\}
\end{eqnarray*}

\bigskip

\begin{eqnarray*}
  M_q(9,8;4) &=& q^5+q^0
\end{eqnarray*}
\begin{eqnarray*}
  \mathcal{U}_{9,8,4} &=& \Big\{480,15\Big\}
\end{eqnarray*}

\bigskip

\begin{eqnarray*}
  M_q(10,8;4) &=& q^6+q^0
\end{eqnarray*}
\begin{eqnarray*}
  \mathcal{U}_{10,8,4} &=& \Big\{960,15\Big\}
\end{eqnarray*}

\bigskip

\begin{eqnarray*}
  M_q(10,8;5) &=& q^{10}+q^0
\end{eqnarray*}
\begin{eqnarray*}
  \mathcal{U}_{10,8,5} &=& \Big\{992,31\Big\}
\end{eqnarray*}

\bigskip

\begin{eqnarray*}
  M_q(11,8;4) &=& q^7+q^0
\end{eqnarray*}
\begin{eqnarray*}
  \mathcal{U}_{11,8,4} &=& \Big\{1920,15\Big\}
\end{eqnarray*}

\bigskip

\begin{eqnarray*}
  M_q(11,8;5) &=& q^{12}+q^0
\end{eqnarray*}
\begin{eqnarray*}
  \mathcal{U}_{11,8,5} &=& \Big\{1984,31\Big\}
\end{eqnarray*}

\bigskip

\begin{eqnarray*}
  M_q(12,8;4) &=& q^8+q^4+q^0
\end{eqnarray*}
\begin{eqnarray*}
  \mathcal{U}_{12,8,4} &=& \Big\{3840,240,15\Big\}
\end{eqnarray*}

\bigskip

\begin{eqnarray*}
  M_q(12,8;5) &=& q^{14}+q^4+q^0
\end{eqnarray*}
\begin{eqnarray*}
  \mathcal{U}_{12,8,5} &=& \Big\{3968,{\color{green}2168},143\Big\}
\end{eqnarray*}

\bigskip

\begin{eqnarray*}
  M_q(12,8;6) &=& q^{18}+q^4+q^2+q^0
\end{eqnarray*}
\begin{eqnarray*}
  \mathcal{U}_{12,8,6} &=& \Big\{4032,{\color{green}3132,819},207\Big\}
\end{eqnarray*}

\bigskip

\begin{eqnarray*}
  M_q(13,8;4) &=& q^9+q^5+q^0
\end{eqnarray*}
\begin{eqnarray*}
  \mathcal{U}_{13,8,4} &=& \Big\{7680,480,15\Big\}
\end{eqnarray*}

\bigskip

\begin{eqnarray*}
  \overline{M}_q(13,8;5) &=& q^{16}+q^8+q^0
\end{eqnarray*}
\begin{eqnarray*}
  \mathcal{U}_{13,8,5} &=& \Big\{7936,{\color{red}2288},31\Big\}
\end{eqnarray*}

\bigskip

\begin{eqnarray*}
  \overline{M}_q(13,8;6) &=& q^{21}+q^8+q^4+q^0
\end{eqnarray*}
\begin{eqnarray*}
  \mathcal{U}_{13,8,6} &=& \Big\{8064,{\color{red}6264,1637},159\Big\}
\end{eqnarray*}

\bigskip

\begin{eqnarray*}
  M_q(14,8;4) &=& q^{10}+q^6+q^0
\end{eqnarray*}
\begin{eqnarray*}
  \mathcal{U}_{14,8,4} &=& \Big\{15360,960,15\Big\}
\end{eqnarray*}

\bigskip

\begin{eqnarray*}
  M_q(14,8;5) &=& q^{18}+q^{10}+q^3+q^0
\end{eqnarray*}
\begin{eqnarray*}
  \mathcal{U}_{14,8,5} &=& \Big\{15872,992,{\color{green}1308},2087\Big\}
\end{eqnarray*}

\bigskip

\begin{eqnarray*}
  \overline{M}_q(14,8;6) &=& q^{24}+q^{12}+q^8+q^4+q^3+q^2+q^0
\end{eqnarray*}
\begin{eqnarray*}
  \mathcal{U}_{14,8,6} &=& \Big\{16128,{\color{red}6384},{\color{magenta}9420_{r=2}},963,{\color{green}1593,8502},6159\Big\}
\end{eqnarray*}

\bigskip

\begin{eqnarray*}
  \overline{M}_q(14,8;7) &=& q^{28}+q^{12}+q^8+q^6+3q^4+q^2
\end{eqnarray*}
\begin{eqnarray*}
  \mathcal{U}_{14,8,7} &=& \Big\{16256,14456,{\color{red}9830,5461,2861,3251,4811},{\color{green}8606}\Big\}
\end{eqnarray*}

\bigskip

\begin{eqnarray*}
  M_q(15,8;4) &=& q^{11}+q^7+q^0
\end{eqnarray*}
\begin{eqnarray*}
  \mathcal{U}_{15,8,4} &=& \Big\{30720,1920,15\Big\}
\end{eqnarray*}

\bigskip

\begin{eqnarray*}
  \overline{M}_q(15,8;5) &=& q^{20}+q^{12}+q^6+q^2+q^1+q^0
\end{eqnarray*}
\begin{eqnarray*}
  \mathcal{U}_{15,8,5} &=& \Big\{31744,1984,{\color{red}2360},{\color{green}4198,8341},16907\Big\}
\end{eqnarray*}

\bigskip

\begin{eqnarray*}
  \overline{M}_q(15,8;6) &=& q^{27}+q^{15}+q^{11}+q^9+q^6+q^4+q^0
\end{eqnarray*}
\begin{eqnarray*}
  \mathcal{U}_{15,8,6} &=& \Big\{32256,{\color{red}5600},{\color{cyan}9112},{\color{magenta}18644_{r=2}},{\color{cyan}18730},{\color{red}8805},1055\Big\}
\end{eqnarray*}

\bigskip

\begin{eqnarray*}
  \overline{M}_q(15,8;7) &=& q^{32}+q^{16}+q^{12}+q^{11}+q^{10}+q^8+2q^7+3q^6+2q^4+q^3+q^2
\end{eqnarray*}
\begin{eqnarray*}
  \mathcal{U}_{15,8,7} &=& \Big\{32512,{\color{red}21744},{\color{magenta}13004_{r=2}},{\color{red}10721},{\color{magenta}19114_{r=2}},{\color{red}11324,6550},{\color{magenta}\emph{24922}_{r=2}},\\
  && 9875,{\color{cyan}\emph{17805},19029,\emph{4921}},7243,1894,28711\Big\},
\end{eqnarray*}
where $24922:[8,8,4,3,2,2,1]\rightarrow [8,8,4,3,2,2]$, $17805:[8,5,4,4,1,1]\rightarrow[5,5,4,4,1,1]$ and $4921:[6,4,4,2,2,2]\rightarrow[6,4,4,2,2]$.

\bigskip

\begin{eqnarray*}
  M_q(16,8;4) &=& q^{12}+q^8+q^4+q^0
\end{eqnarray*}
\begin{eqnarray*}
  \mathcal{U}_{16,8,4} &=& \Big\{61440,3840,240,15\Big\}
\end{eqnarray*}

\bigskip

\begin{eqnarray*}
  M_q(16,8;5) &=& q^{22}+q^{14}+q^8+q^4+q^3+q^2
\end{eqnarray*}
\begin{eqnarray*}
  \mathcal{U}_{16,8,5} &=& \Big\{63488,3968,{\color{cyan}4336},{\color{green}8492},{\color{cyan}16970},{\color{green}33817}\Big\}
\end{eqnarray*}

\bigskip

\begin{eqnarray*}
  M_q(16,8;6) &=& q^{30}+q^{18}+q^{14}+q^{12}+q^{10}+q^8+q^5+2q^4
\end{eqnarray*}
\begin{eqnarray*}
  \mathcal{U}_{16,8,6} &=& \Big\{64512,4032,13104,{\color{yellow}49392_{2,4,8}},49932,{\color{yellow}12492_{2,4,8}},20867,{\color{green}3132},41539\Big\}
\end{eqnarray*}

\bigskip

\begin{eqnarray*}
  \overline{M}_q(16,8;7) &=& q^{36}+q^{20}+q^{19}+2q^{14}+2q^{12}+5q^{10}+2q^9+q^6
\end{eqnarray*}
\begin{eqnarray*}
  \mathcal{U}_{16,8,7} &=& \Big\{65024,36320,{\color{red}25552,13708},22840,{\color{red}37556},{\color{magenta}37706_{2}},{\color{red}10922},{\color{cyan}13426},\\
  && {\color{red}18214},22726,57452,43286,{\color{cyan}50330},{\color{red}3669}\Big\}
\end{eqnarray*}

\bigskip

\begin{eqnarray*}
  \overline{M}_q(16,8;8) &=& q^{40}+q^{24}+q^{18}+q^{15}+q^{14}+3q^{13}+9q^{12}+4q^{11}+4q^{10}\\
  && +3q^8+q^4+q^0
\end{eqnarray*}
\begin{eqnarray*}
  \mathcal{U}_{16,8,8} &=& \Big\{65280,61680,{\color{red}52428,43690},15555,{\color{red}23190,26265,38565},{\color{yellow}15420_{2,4,8}},{\color{cyan}22953},\\
  && {\color{cyan}26022},{\color{red}26970},{\color{cyan}27237},{\color{red}39270},{\color{cyan}39513},
  {\color{red}42345},{\color{yellow}50115_{2,4,8}},{\color{cyan}22122},{\color{magenta}\emph{38298}_2,
  \emph{42582}_2},\\
  && {\color{cyan}43413},13260,{\color{red}21845},{\color{yellow}49980_{2,4,8}},52275,4080,{\color{blue}13107},
  61455,3855,255\Big\},\end{eqnarray*}
where $38298:[8,6,5,4,4,2,2,1]\rightarrow [8,6,5,4,4,2,1]$ and $42582:[8,7,5,5,3,2,1,1]\rightarrow [8,7,5,5,3,2,1]$.
\bigskip

\begin{eqnarray*}
  M_q(17,8;4) &=& q^{13}+q^9+q^5+q^0
\end{eqnarray*}
\begin{eqnarray*}
  \mathcal{U}_{17,8,4} &=& \Big\{122880,7680,480,15\Big\}
\end{eqnarray*}

\bigskip

\begin{eqnarray*}
  M_q(17,8;5) &=& q^{24}+q^{16}+q^{10}+q^6+q^5+q^4+q^0
\end{eqnarray*}
\begin{eqnarray*}
  \mathcal{U}_{17,8,5} &=& \Big\{126976,7936,8672,{\color{yellow}16952_{2,3,6}},{\color{cyan}33876},67724,4135\Big\}
\end{eqnarray*}

\bigskip

\begin{eqnarray*}
  \overline{M}_q(17,8;6) &=& q^{33}+q^{21}+q^{17}+q^{15}+q^{13}+q^{11}+2q^8+q^7+q^6\\
  && +2q^4+q^3
\end{eqnarray*}
\begin{eqnarray*}
  \mathcal{U}_{17,8,6} &=& \Big\{129024,8064,26208,{\color{cyan}98784},99864,{\color{cyan}24984},{\color{red}6264},41734,83205,\\
  && {\color{cyan}\emph{18630}},{\color{cyan}12453},37955,68131\Big\},
\end{eqnarray*}
where $18630:[9,7,4,4,1,1]\rightarrow[5,5,4,4,1,1]$.

\bigskip

\begin{eqnarray*}
  \overline{M}_q(17,8;7) &=& q^{40}+q^{24}+q^{23}+2q^{18}+2q^{16}+5q^{14}+3q^{13}+q^{12}+q^{10}\\
  && +q^7+q^2+2q^0
\end{eqnarray*}
\begin{eqnarray*}
  \mathcal{U}_{17,8,7} &=& \Big\{130048,15296,{\color{red}51104},{\color{cyan}69232},86808,106856,107156,{\color{cyan}25940,38104},\\
  && 51788,{\color{cyan}69004},86244,{\color{red}13868},26808,39220,101123,22083,45219,74779,\\
  && 2263,16687\Big\}
\end{eqnarray*}

\bigskip

For $q\ge 3$ we have
\begin{eqnarray*}
  \overline{M}_q(17,8;8) &=& q^{45}+q^{29}+q^{23}+q^{20}+6q^{19}+q^{18}+3q^{17}+4q^{16}+5q^{15}\\
  && +q^{13}+q^{12}+q^8+q^0
\end{eqnarray*}
with
\begin{eqnarray*}
  \mathcal{U}_{17,8,8} &=& \Big\{130560,123360,{\color{red}88472},47442,{\color{red}40353,45964,54484,71618},{\color{cyan}79156},{\color{red}101708},\\
  && {\color{red}28038},30025,{\color{red}76500,103032},23396,{\color{red}50994,51914},76586,29361,{\color{magenta}59452_{2}},\\
  && {\color{red}83628},{\color{cyan}85105},107674,119047,{\color{red}42597,7709},255\Big\}
\end{eqnarray*}
and for $q=2$ we have
\begin{eqnarray*}
  \overline{M}_2(17,8;8) &=& 2^{45}+2^{29}+2^{23}+2\cdot 2^{20}+2\cdot 2^{19}+6\cdot 2^{18}+2^{17}+4\cdot 2^{16}+2\cdot 2^{15}\\
  && +2^{13}+2^{12}+2^8+2^2+2^0=35\,184\,922\,538\,245
\end{eqnarray*}
with
\begin{eqnarray*}
  \mathcal{U}_{17,8,8} &=& \Big\{130560,123360,88472,{\color{red}40385},{\color{cyan}79188},45964,{\color{cyan}47410},28038,{\color{red}52052,54452},\\
  && {\color{red}71586,100056,101676},{\color{magenta}76618_{2}},29394,{\color{red}30828,76468,85106,51882},58649,\\
  && 119047,{\color{red}42597},{\color{yellow}7710_{2,4,8}},{\color{green}4985},65775\Big\}.
\end{eqnarray*}

\bigskip

\begin{eqnarray*}
  M_q(18,8;4) &=& q^{14}+q^{10}+q^6+q^0
\end{eqnarray*}
\begin{eqnarray*}
  \mathcal{U}_{18,8,4} &=& \Big\{245760,15360,{\color{green}960},15\Big\}
\end{eqnarray*}

\bigskip

\begin{eqnarray*}
  M_q(18,8;5) &=& q^{26}+q^{18}+q^{12}+q^8+q^7+q^6+q^2+q^1+q^0
\end{eqnarray*}
\begin{eqnarray*}
  \mathcal{U}_{18,8,5} &=& \Big\{253952,15872,17344,{\color{cyan}33904},67752,135448,{\color{green}8486,18453},33291\Big\}
\end{eqnarray*}

\bigskip

\begin{eqnarray*}
  M_q(18,8;6) &=& q^{36}+q^{24}+q^{20}+q^{18}+q^{16}+q^{14}+q^{12}+q^{11}+q^{10}\\
  && +q^9+4q^7+3q^6+2q^5+q^0
\end{eqnarray*}
\begin{eqnarray*}
  \mathcal{U}_{18,8,6} &=& \Big\{258048,16128,52416,197568,199728,49968,12528,83468,166410,{\color{cyan}37260},\\
  && 24906,75910,140425,149765,71753,99587,136262,21123,41541,63\Big\}
\end{eqnarray*}

\bigskip

\begin{eqnarray*}
  M_q(18,8;7) &=& q^{44}+q^{28}+q^{27}+2q^{22}+2q^{20}+5q^{18}+3q^{17}+q^{16}+2q^{14}\\
  && +q^{12}+2q^{11}+q^8+2q^5+q^4+q^2
\end{eqnarray*}
\begin{eqnarray*}
  \mathcal{U}_{18,8,7} &=& \Big\{260096,16256,116544,150752,169520,205520,206120,58520,86696,\\
  && 100784,150296,168392,29040,43624,72792,157190,103558,167173,\\
  && 201475,23109,76102,143525,82997,180307,{\color{green}49454,1739}\Big\}
\end{eqnarray*}

\bigskip

\begin{eqnarray*}
  \overline{M}_q(18,8;8) &=& q^{50}+q^{34}+q^{30}+2q^{26}+q^{25}+3q^{24}+4q^{23}+7q^{22}+3q^{21}+4q^{20}+q^{19}\\
  && +q^{18}+2q^{16}+q^{15}+q^{12}+q^{11}+q^{10}+q^7+3q^6+q^5+2q^4+2q^3\\
  && +q^2+q^0
\end{eqnarray*}
\begin{eqnarray*}
  \mathcal{U}_{18,8,8} &=& \Big\{261120,246720,{\color{red}146336},62232,{\color{red}102192},{\color{cyan}79684},{\color{red}87720},110832,157296,\\
  && {\color{red}55908},92520,92820,{\color{cyan}203352},44744,58788,{\color{magenta}151308_2},{\color{cyan}169320,169620},\\
  && 206232,238124,{\color{cyan}104844},174420,217396,24016,159948,182456,{\color{red}199908},\\
  && 88835,176707,167299,221347,115804,{\color{yellow}15420_{2,4,8}},76883,54323,59407,{\color{magenta}10027_2},\\
  && {\color{cyan}18119},201743,{\color{cyan}98971},{\color{yellow}12695_{2,3,6}},{\color{green}98663,6379,147803,133687},1020\Big\}
\end{eqnarray*}

\bigskip

For $q\ge 3$ we have
\begin{eqnarray*}
  \overline{M}_q(18,8;9) &=& q^{54}+q^{38}+q^{30}+q^{27}+q^{26}+q^{25}+10q^{24}+8q^{23}+q^{22}+2q^{21}\\
  && +q^{18}+2q^{14}+q^{12}+q^{10}+q^6+2q^2+q^0
\end{eqnarray*}
with
\begin{eqnarray*}
  \mathcal{U}_{18,8,9} &=& \Big\{261632,254432,235928,{\color{red}218452,162182,177482},{\color{cyan}95625,112965,182162},\\
  && {\color{red}183116,185516},{\color{cyan}186482},{\color{red}207698,208076,209708,210098},117969,{\color{red}118058},\\
  && 119601,{\color{red}123548},{\color{cyan}175412},{\color{red}176849,200609},{\color{cyan}217802},231032,{\color{red}174762},{\color{cyan}223289},\\
  && {\color{red}157286},24387,62487,{\color{red}138781},72436,{\color{red}37775},{\color{green}10043,70767},16887\Big\}.
\end{eqnarray*}
For $q=2$ we have
\begin{eqnarray*}
  \overline{M}_2(18,8;9) &\le & 2^{54}+2^{38}+2^{30}+2^{27}+2^{26}+2^{25}+52\cdot 2^{24}\\
  &=& 18\,014\,675\,568\,427\,008
\end{eqnarray*}
using $max\_dive=15$, $ub=58$ and 
\begin{eqnarray*}
  \overline{M}_2(18,8;9) &\ge& 2^{54}+2^{38}+2^{30}+2^{27}+2^{26}+2^{25}+10\cdot 2^{24}+8\cdot 2^{23}+2^{22}+2\cdot 2^{21}\\
  && +2^{18}+2\cdot 2^{14}+2^{12}+2^{10}+3^6+2\cdot 2^2+2^0\\
  &=& 18\,014\,674\,939\,581\,513
\end{eqnarray*}
with
\begin{eqnarray*}
  \mathcal{U}_{18,8,9} &=& \Big\{261632,254432,235928,218452,162182,177482,95625,112965,182162,\\
  && 183116,185516,186482,207698,208076,209708,210098,117969,118058,\\
  && 119601,123548,175412,176849,200609,217802,231032,174762,223289,\\
  && 157286,24387,62487,138781,72436,37775,10043,70767,16887\Big\}.
\end{eqnarray*}

\bigskip

\begin{eqnarray*}
  M_q(19,8;4) &=& q^{15}+q^{11}+q^7+q^0
\end{eqnarray*}
\begin{eqnarray*}
  \mathcal{U}_{19,8,4} &=& \Big\{491520,30720,1920,15\Big\}
\end{eqnarray*}

\bigskip

\begin{eqnarray*}
  M_q(19,8;5) &=& q^{28}+q^{20}+q^{14}+q^{10}+q^9+q^8+q^5+q^3+2q^2\\
  && +q^1+q^0
\end{eqnarray*}
\begin{eqnarray*}
  \mathcal{U}_{19,8,5} &=& \Big\{507904,31744,34688,67808,135504,270920,{\color{yellow}16948_{2,3,6}},{\color{green}36906,66585},\\
  && {\color{green}73990,266373},133635\Big\}
\end{eqnarray*}

\bigskip

\begin{eqnarray*}
  \overline{M}_q(19,8;6) &=& q^{39}+q^{27}+q^{23}+q^{21}+q^{19}+q^{17}+q^{15}+q^{14}+q^{13}\\
  && +q^{12}+4q^{10}+3q^9+2q^8+q^7+2q^6+2q^5+2q^4+q^3+q^0
\end{eqnarray*}
\begin{eqnarray*}
  \mathcal{U}_{19,8,6} &=& \Big\{516096,32256,104832,395136,399456,99936,{\color{red}25056},166936,332820,\\
  && 74520,49812,151826,280844,299530,143500,199174,272530,42246,\\
  && 83082,45137,{\color{red}2761},84017,5417,148549,2708
\end{eqnarray*}

\bigskip

For $q\ge 3$ we have
\begin{eqnarray*}
  M_q(19,8;7) &=& q^{48}+q^{32}+q^{31}+2q^{26}+2q^{24}+5q^{22}+3q^{21}+q^{20}+2q^{18}+q^{16}\\
  && +2q^{15}+2q^{14}+2q^{13}+q^{12}+q^{11}+2q^9+q^8+q^7+2q^6+q^0\\
\end{eqnarray*}
with
\begin{eqnarray*}
  \mathcal{U}_{19,8,7} &=& \Big\{520192,32512,233088,301504,339040,411040,412240,117040,173392,\\
  && 201568,300592,336784,58080,87248,145584,314380,207116,334346,\\
  && 402950,46218,152204,103685,426761,92291,138313,287050,83497,\\
  && {\color{magenta}98908_{2}},360614,22630,151827,{\color{cyan}3478},274517,251\Big\}.
\end{eqnarray*}
For $q=2$ we have
\begin{eqnarray*}
  \overline{M}_2(19,8;7) &\le & 2^{48}+2^{32}+2^{31}+2^{25}+5\cdot 2^{24}+43\cdot 2^{23}\\
  &=& 281\,481\,897\,312\,256
\end{eqnarray*}
using $max\_dive=10$, $ub=52$ and 
\begin{eqnarray*}
  \overline{M}_2(19,8;7) &\ge& 2^{48}+2^{32}+2^{31}+2\cdot 2^{26}+2\cdot 2^{24}+5\cdot 2^{22}+3\cdot 2^{21}+2^{20}+2^{18}\\
  && +5\cdot 2^{16}+3\cdot 2^{15}+2^{14}+2^{12}+2\cdot 2^{10}+2\cdot 2^9+2\cdot 2^8+2\cdot 2^7\\
  && +4\cdot 2^6+2\cdot 2^5+2^3 \\
  &=& 281\,481\,615\,958\,088
\end{eqnarray*}
with
\begin{eqnarray*}
  \mathcal{U}_{19,8,7} &=& \Big\{520192,32512,233088,301504,339040,411040,412240,116048,172896,\\
  && 203056,300592,336784,59568,86752,144592,314380,334346,108806,\\
  && 169226,213772,402950,403721,45708,91274,153734,333061,312067,\\
  && {\color{cyan}263881},397420,{\color{yellow}25029_{2,4,8},98729_{2,4,8}},{\color{red}18794},303194,274595,426133,{\color{red}3475,34022},\\
  && {\color{red}67804},344118,{\color{cyan}\emph{4950}},{\color{magenta}\emph{9532}_{2}},131770\Big\},
\end{eqnarray*}
where $4950:[6,4,4,3,2,1,1]\rightarrow[6,4,4,3,2]$ and $9532:[7,5,4,2,2,2,2]\rightarrow[7,5,4,2,2,1]$.

\bigskip

For $q\ge 3$ we have
\begin{eqnarray*}
  \overline{M}_q(19,8;8) &=& q^{55}+q^{39}+q^{35}+q^{32}+q^{31}+2q^{30}+2q^{29}+6q^{28}+9q^{27}+2q^{25}\\
  && +q^{24}+q^{20}+2q^{17}+q^{16}+q^{14}+3q^8+q^6+q^4+q^2+2q^0\\
\end{eqnarray*}
with
\begin{eqnarray*}
  \mathcal{U}_{19,8,8} &=& \Big\{522240,493440,63296,{\color{red}89760},419376,{\color{red}146832},339296,222408,{\color{cyan}400968},\\
  && 174816,215888,289672,314576,340688,414176,110104,123312,183592,\\
  && 209704,234608,{\color{cyan}301872},308392,349464,377448,{\color{red}104904},184984,434520,\\
  && 460984,{\color{red}31758},476167,30833,176647,{\color{red}18326},76167,{\color{red}132963},{\color{cyan}\emph{34509}},\\
  && 297063,{\color{green}9533},147582,266399\Big\},
\end{eqnarray*}
where $34509:[8,4,4,3,3,1,1]\rightarrow[8,4,4,3,3]$. For $q=2$ we have
\begin{eqnarray*}
  \overline{M}_2(19,8;8) &\le & 2^{55}+2^{39}+2^{35}+2^{32}+4\cdot 2^{30}+2\cdot 2^{29}+68\cdot 2^{28}\\
  &=& 36\,029\,409\,051\,803\,648
\end{eqnarray*}
using $max\_dive=15$, $ub=78$ and 
\begin{eqnarray*}
  \overline{M}_2(19,8;8) &\ge& 2^{55}+2^{39}+2^{35}+2^{32}+2^{31}+2\cdot 2^{30}+2\cdot 2^{29}+6\cdot 2^{28}+9\cdot 2^{27}\\
  && +2\cdot 2^{25}+2^{24}+2^{20}+2\cdot 2^{17}+2^{16}+2^{14}+3\cdot 2^8+2^6+2^4+2^2\\
  &=& 36\,029\,393\,702\,044\,502
\end{eqnarray*}
with
\begin{eqnarray*}
  \mathcal{U}_{19,8,8} &=& \Big\{522240,493440,63296,89760,419376,146832,339296,222408,400968,\\
  && 174816,215888,289672,314576,340688,414176,110104,123312,183592,\\
  && 209704,234608,301872,308392,349464,377448,104904,184984,434520,\\
  && 460984,31758,476167,30833,176647,18326,76167,132963,34509,\\
  && 297063,9533,147582,266399\Big\}.
\end{eqnarray*}

\bigskip

For $q\ge 4$ we have 
\begin{eqnarray*}
  \overline{M}_q(19,8;9) &=& q^{60}+q^{44}+q^{37}+2q^{34}+3q^{32}+4q^{31}+5q^{30}+3q^{29}+4q^{28}+4q^{26}\\
  && +2q^{25}+2q^{22}+q^{21}+q^{20}+q^{19}+q^{18}+q^{17}+q^{16}+q^{15}+q^{14}\\
  && +2q^{13}+q^{12}+q^{11}+q^8+2q^7+q^4+q^2+q^1
\end{eqnarray*}
with
\begin{eqnarray*}
  \mathcal{U}_{19,8,9} &=& \Big\{523264,508864,{\color{red}408480},242288,{\color{red}354096,226060},{\color{cyan}324244},367384,239016,\\
  && {\color{cyan}\emph{323944}},{\color{red}341704,420440},125601,186146,{\color{red}307010},{\color{cyan}350820},479416,{\color{red}204609},\\
  &&{\color{cyan} 223586},444724,113092,{\color{red}216724},363760,{\color{cyan}419044,186578},189777,356739,\\
  && 493093,{\color{cyan}413066},499790,{\color{red}119900},464150,178699,{\color{cyan}286227},339213,{\color{red}60474},\\
  && 430477,{\color{red}154669},434355,{\color{red}166636},91719,{\color{red}100243},{\color{cyan}\emph{79029}},{\color{red}84441},{\color{magenta}\emph{140246}_{2}},\\
  && {\color{cyan}102635},{\color{magenta}\emph{328622}_{2}},{\color{cyan}\emph{41789}},{\color{green}132731},6519\Big\},
\end{eqnarray*}
where $323944:[10,8,8,8,8,5,4,4,3]\rightarrow[10,8,8,8,8,5,4,4,2]$, $79029:[8,6,6,5,3,2,2,1]$ $\rightarrow[8,6,6,5,3,2,2]$,
$140246:[9,6,3,3,3,3,2,1,1]\rightarrow[8,6,3,3,3,3,2,1,1]$, $328622:[10,9,3,3,3,2,1,1,1]\rightarrow[7,6,3,3,3,2,1,1,1]$ and
$41789:[7,6,3,3,1,1,1,1]\rightarrow[5,5,3,3,1$, $1,1,1]$. For $q\in\{2,3\}$ we have
\begin{eqnarray*}
  \overline{M}_q(19,8;9) &\le & q^{60}+q^{44}+q^{37}+2q^{34}+98q^{32}
\end{eqnarray*}
and
\begin{eqnarray*}
  \overline{M}_q(19,8;9) &\ge& q^{60}+q^{44}+q^{37}+2q^{34}+3q^{32}+4q^{31}+5q^{30}+3q^{29}+4q^{28}+4q^{26}\\
  && +2q^{25}+2q^{22}+q^{21}+q^{20}+q^{19}+q^{18}+q^{17}+q^{16}+q^{15}+q^{14}\\
  && +2q^{13}+q^{12}+q^{11}+q^8+2q^7+q^4+q^2+q^1
\end{eqnarray*}
with
\begin{eqnarray*}
  \mathcal{U}_{19,8,9} &=& \Big\{523264,508864,408480,242288,354096,226060,324244,367384,239016,\\
  && 323944,341704,420440,125601,186146,307010,350820,479416,204609,\\
  && 223586,444724,113092,216724,363760,419044,186578,189777,356739,\\
  && 493093,413066,499790,119900,464150,178699,286227,339213,60474,\\
  && 430477,154669,434355,166636,91719,100243,79029,84441,140246,\\
  && 102635,328622,41789,132731,6519\Big\}.
\end{eqnarray*}

\subsection{Minimum subspace distance $10$}
\label{subsec_mindist_10}

\begin{eqnarray*}
  M_q(10,10;5) &=& q^5+q^0
\end{eqnarray*}
\begin{eqnarray*}
  \mathcal{U}_{10,10,5} &=& \Big\{992,31\Big\}
\end{eqnarray*}

\bigskip

\begin{eqnarray*}
  M_q(11,10;5) &=& q^6+q^0
\end{eqnarray*}
\begin{eqnarray*}
  \mathcal{U}_{11,10,5} &=& \Big\{1984,31\Big\}
\end{eqnarray*}

\bigskip

\begin{eqnarray*}
  M_q(12,10;5) &=& q^7+q^0
\end{eqnarray*}
\begin{eqnarray*}
  \mathcal{U}_{12,10,5} &=& \Big\{3968,31\Big\}
\end{eqnarray*}

\bigskip

\begin{eqnarray*}
  M_q(12,10;6) &=& q^{12}+q^0
\end{eqnarray*}
\begin{eqnarray*}
  \mathcal{U}_{12,10,6} &=& \Big\{4032,63\Big\}
\end{eqnarray*}

\bigskip

\begin{eqnarray*}
  M_q(13,10;5) &=& q^8+q^0
\end{eqnarray*}
\begin{eqnarray*}
  \mathcal{U}_{13,10,5} &=& \Big\{7936,31\Big\}
\end{eqnarray*}

\bigskip

\begin{eqnarray*}
  M_q(13,10;6) &=& q^{14}+q^0
\end{eqnarray*}
\begin{eqnarray*}
  \mathcal{U}_{13,10,6} &=& \Big\{8064,63\Big\}
\end{eqnarray*}

\bigskip

\begin{eqnarray*}
  M_q(14,10;5) &=& q^9+q^0
\end{eqnarray*}
\begin{eqnarray*}
  \mathcal{U}_{14,10,5} &=& \Big\{15872,31\Big\}
\end{eqnarray*}

\bigskip

\begin{eqnarray*}
  M_q(14,10;6) &=& q^{16}+q^0
\end{eqnarray*}
\begin{eqnarray*}
  \mathcal{U}_{14,10,6} &=& \Big\{16128,63\Big\}
\end{eqnarray*}

\bigskip

\begin{eqnarray*}
  M_q(14,10;7) &=& q^{21}+q^0
\end{eqnarray*}
\begin{eqnarray*}
  \mathcal{U}_{14,10,7} &=& \Big\{16256,127\Big\}
\end{eqnarray*}

\bigskip

\begin{eqnarray*}
  M_q(15,10;5) &=& q^{10}+q^5+q^0
\end{eqnarray*}
\begin{eqnarray*}
  \mathcal{U}_{15,10,5} &=& \Big\{31744,992,31\Big\}
\end{eqnarray*}

\bigskip

\begin{eqnarray*}
  M_q(15,10;6) &=& q^{18}+q^5+q^0
\end{eqnarray*}
\begin{eqnarray*}
  \mathcal{U}_{15,10,6} &=& \Big\{32256,16880,543\Big\}
\end{eqnarray*}

\bigskip

\begin{eqnarray*}
  M_q(15,10;7) &=& q^{24}+q^5+q^0
\end{eqnarray*}
\begin{eqnarray*}
  \mathcal{U}_{15,10,7} &=& \Big\{32512,{\color{green}24824},799\Big\}
\end{eqnarray*}

\bigskip

\begin{eqnarray*}
  M_q(16,10;5) &=& q^{11}+q^6+q^0
\end{eqnarray*}
\begin{eqnarray*}
  \mathcal{U}_{16,10,5} &=& \Big\{63488,1984,31\Big\}
\end{eqnarray*}

\bigskip

\begin{eqnarray*}
  \overline{M}_q(16,10;6) &=& q^{20}+q^{10}+q^0
\end{eqnarray*}
\begin{eqnarray*}
  \mathcal{U}_{16,10,6} &=& \Big\{64512,{\color{red}17376},63\Big\}
\end{eqnarray*}

\bigskip

\begin{eqnarray*}
  \overline{M}_q(16,10;7) &=& q^{27}+q^{10}+q^3+q^0
\end{eqnarray*}
\begin{eqnarray*}
  \mathcal{U}_{16,10,7} &=& \Big\{65024,{\color{red}49648,5518},2615\Big\}
\end{eqnarray*}

\bigskip

\begin{eqnarray*}
  \overline{M}_q(16,10;8) &=& q^{32}+q^{10}+q^3+q^1
\end{eqnarray*}
\begin{eqnarray*}
  \mathcal{U}_{16,10,8} &=& \Big\{65280,{\color{red}57592,7339},9559\Big\}
\end{eqnarray*}

\bigskip

\begin{eqnarray*}
  M_q(17,10;5) &=& q^{12}+q^7+q^0
\end{eqnarray*}
\begin{eqnarray*}
  \mathcal{U}_{17,10,5} &=& \Big\{126976,3968,31\Big\}
\end{eqnarray*}

\bigskip

\begin{eqnarray*}
  M_q(17,10;6) &=& q^{22}+q^{12}+q^0
\end{eqnarray*}
\begin{eqnarray*}
  \mathcal{U}_{17,10,6} &=& \Big\{129024,4032,63\Big\}
\end{eqnarray*}

\bigskip

\begin{eqnarray*}
  \overline{M}_q(17,10;7) &=& q^{30}+q^{15}+q^6+2q^2
\end{eqnarray*}
\begin{eqnarray*}
  \mathcal{U}_{17,10,7} &=& \Big\{130048,{\color{red}50144,6940},{\color{green}9421},{\color{cyan}66867}\Big\}
\end{eqnarray*}

\bigskip

\begin{eqnarray*}
  \overline{M}_q(17,10;8) &=& q^{36}+q^{15}+2q^6+q^5+q^0
\end{eqnarray*}
\begin{eqnarray*}
  \mathcal{U}_{17,10,8} &=& \Big\{130560,{\color{red}115184},{\color{yellow}14563_{2,3,6}},{\color{red}21901,41806},67127\Big\}
\end{eqnarray*}

\bigskip

\begin{eqnarray*}
  M_q(18,10;5) &=& q^{13}+q^8+q^0
\end{eqnarray*}
\begin{eqnarray*}
  \mathcal{U}_{18,10,5} &=& \Big\{253952,7936,31\Big\}
\end{eqnarray*}

\bigskip

\begin{eqnarray*}
  M_q(18,10;6) &=& q^{24}+q^{14}+q^4+q^0
\end{eqnarray*}
\begin{eqnarray*}
  \mathcal{U}_{18,10,6} &=& \Big\{258048,8064,{\color{green}10360},16527\Big\}
\end{eqnarray*}

\bigskip

\begin{eqnarray*}
  \overline{M}_q(18,10;7) &=& q^{33}+q^{18}+q^9+q^8+q^3+q^0
\end{eqnarray*}
\begin{eqnarray*}
  \mathcal{U}_{18,10,7} &=& \Big\{260096,{\color{red}26560},{\color{yellow}99128_{2,5,10}},{\color{red}136372},{\color{green}6499},24607\Big\}
\end{eqnarray*}

\bigskip

\begin{eqnarray*}
  \overline{M}_q(18,10;8) &=& q^{40}+q^{20}+q^{15}+q^9+q^8+2q^5+q^3+q^0
\end{eqnarray*}
\begin{eqnarray*}
  \mathcal{U}_{18,10,8} &=& \Big\{261120,{\color{red}173024,86932,84586,36633,13646},{\color{yellow}{137427_{2,3,6}}},
  {\color{cyan}\emph{148653}},106551\Big\}
\end{eqnarray*}
where $148653:[10,8,5,3,2,1,1]\rightarrow[10,8,5,3,2,1]$.

\bigskip

\begin{eqnarray*}
  \overline{M}_q(18,10;9) &=& q^{45}+q^{20}+q^{11}+2q^{10}+2q^7+q^6+q^5+q^3
\end{eqnarray*}
\begin{eqnarray*}
  \mathcal{U}_{18,10,9} &=& \Big\{261632,246256,{\color{cyan}59790},{\color{red}80213,144809,86731},{\color{yellow}165703_{2,4,8}},{\color{red}40058,75446},{\color{green}150077}\Big\}
\end{eqnarray*}

\bigskip

\begin{eqnarray*}
  M_q(19,10;5) &=& q^{14}+q^9+q^0
\end{eqnarray*}
\begin{eqnarray*}
  \mathcal{U}_{19,10,5} &=& \Big\{507904,15872,31\Big\}
\end{eqnarray*}

\bigskip

\begin{eqnarray*}
  \overline{M}_q(19,10;6) &=& q^{26}+q^{16}+q^8+q^0
\end{eqnarray*}
\begin{eqnarray*}
  \mathcal{U}_{19,10,6} &=& \Big\{516096,16128,{\color{red}18672},287\Big\}
\end{eqnarray*}

\bigskip

\begin{eqnarray*}
  \overline{M}_q(19,10;7) &=& q^{36}+q^{21}+q^{12}+q^{11}+q^6+q^4+2q^0
\end{eqnarray*}
\begin{eqnarray*}
  \mathcal{U}_{19,10,7} &=& \Big\{520192,16256,{\color{red}50032,197864,280732},298051,{\color{green}12347},197383\Big\}
\end{eqnarray*}

\bigskip

\begin{eqnarray*}
  \overline{M}_q(19,10;8) &=& q^{44}+q^{24}+q^{20}+q^{15}+q^{12}+q^9+q^8+2q^7+q^6+q^2
\end{eqnarray*}
\begin{eqnarray*}
  \mathcal{U}_{19,10,8} &=& \Big\{522240,{\color{red}108480,284464},{\color{magenta}151180_{2}},{\color{red}395745,40042},{\color{yellow}459354_{2,5,10}},{\color{red}72086,172348},\\
  && {\color{red}28889},90663\Big\}
\end{eqnarray*}

\bigskip

For $q\ge 3$ we have
\begin{eqnarray*}
  \overline{M}_q(19,10;9) &=& q^{50}+q^{25}+q^{21}+2q^{16}+2q^{12}+3q^{11}+q^9+q^8+q^1
\end{eqnarray*}
with
\begin{eqnarray*}
  \mathcal{U}_{19,10,9} &=& \Big\{523264,{\color{red}418784,363416,119412,202629,59155,170345,107754,207190},\\
  && {\color{red}398938},31374,{\color{magenta}281813_{2}},344367\Big\}
\end{eqnarray*}
and for $q=2$ we have
\begin{eqnarray*}
  \overline{M}_2(19,10;9) &=& 2^{50}+2^{25}+2^{21}+2\cdot2^{16}+2\cdot 2^{12}+2^{11}+4\cdot 2^{10}+3\cdot 2^9+2^8\\
  && +2^7+2^5+2^4=1\,125\,899\,942\,641\,584
\end{eqnarray*}
with
\begin{eqnarray*}
  \mathcal{U}_{19,10,9} &=& \Big\{523264,{\color{red}418784,363416,119412,202630,59149,176474,230633},{\color{cyan}47779},\\
  && {\color{red}298342},{\color{magenta}337491_2,399949_2},{\color{red}92366},{\color{magenta}151098_2},{\color{red}207157,275644},348459,24017,442519\Big\}.
\end{eqnarray*}

\subsection{Minimum subspace distance $12$}
\label{subsec_mindist_12}

\begin{eqnarray*}
  M_q(12,12;6) &=& q^6+q^0
\end{eqnarray*}
\begin{eqnarray*}
  \mathcal{U}_{12,12,6} &=& \Big\{4032,63\Big\}
\end{eqnarray*}

\bigskip

\begin{eqnarray*}
  M_q(13,12;6) &=& q^7+q^0
\end{eqnarray*}
\begin{eqnarray*}
  \mathcal{U}_{13,12,6} &=& \Big\{8064,63\Big\}
\end{eqnarray*}

\bigskip

\begin{eqnarray*}
  M_q(14,12;6) &=& q^8+q^0
\end{eqnarray*}
\begin{eqnarray*}
  \mathcal{U}_{14,12,6} &=& \Big\{16128,63\Big\}
\end{eqnarray*}

\bigskip

\begin{eqnarray*}
  M_q(14,12;7) &=& q^{14}+q^0
\end{eqnarray*}
\begin{eqnarray*}
  \mathcal{U}_{14,12,7} &=& \Big\{16256,127\Big\}
\end{eqnarray*}

\bigskip

\begin{eqnarray*}
  M_q(15,12;6) &=& q^9+q^0
\end{eqnarray*}
\begin{eqnarray*}
  \mathcal{U}_{15,12,6} &=& \Big\{32256,63\Big\}
\end{eqnarray*}

\bigskip

\begin{eqnarray*}
  M_q(15,12;7) &=& q^{16}+q^0
\end{eqnarray*}
\begin{eqnarray*}
  \mathcal{U}_{15,12,7} &=& \Big\{32512,127\Big\}
\end{eqnarray*}

\bigskip

\begin{eqnarray*}
  M_q(16,12;6) &=& q^{10}+q^0
\end{eqnarray*}
\begin{eqnarray*}
  \mathcal{U}_{16,12,6} &=& \Big\{64512,63\Big\}
\end{eqnarray*}

\bigskip

\begin{eqnarray*}
  M_q(16,12;7) &=& q^{18}+q^0
\end{eqnarray*}
\begin{eqnarray*}
  \mathcal{U}_{16,12,7} &=& \Big\{65024,127\Big\}
\end{eqnarray*}

\bigskip

\begin{eqnarray*}
  M_q(16,12;8) &=& q^{24}+q^0
\end{eqnarray*}
\begin{eqnarray*}
  \mathcal{U}_{16,12,8} &=& \Big\{65280,255\Big\}
\end{eqnarray*}

\bigskip

\begin{eqnarray*}
  M_q(17,12;6) &=& q^{11}+q^0
\end{eqnarray*}
\begin{eqnarray*}
  \mathcal{U}_{17,12,6} &=& \Big\{129024,63\Big\}
\end{eqnarray*}

\bigskip

\begin{eqnarray*}
  M_q(17,12;7) &=& q^{20}+q^0
\end{eqnarray*}
\begin{eqnarray*}
  \mathcal{U}_{17,12,7} &=& \Big\{130048,127\Big\}
\end{eqnarray*}

\bigskip

\begin{eqnarray*}
  M_q(17,12;8) &=& q^{27}+q^0
\end{eqnarray*}
\begin{eqnarray*}
  \mathcal{U}_{17,12,8} &=& \Big\{130560,255\Big\}
\end{eqnarray*}

\bigskip

\begin{eqnarray*}
  M_q(18,12;6) &=& q^{12}+q^6+q^0
\end{eqnarray*}
\begin{eqnarray*}
  \mathcal{U}_{18,12,6} &=& \Big\{258048,4032,63\Big\}
\end{eqnarray*}

\bigskip

\begin{eqnarray*}
  M_q(18,12;7) &=& q^{22}+q^6+q^0
\end{eqnarray*}
\begin{eqnarray*}
  \mathcal{U}_{18,12,7} &=& \Big\{260096,{\color{green}133088},2111\Big\}
\end{eqnarray*}

\bigskip

\begin{eqnarray*}
  M_q(18,12;8) &=& q^{30}+q^6+q^0
\end{eqnarray*}
\begin{eqnarray*}
  \mathcal{U}_{18,12,8} &=& \Big\{261120,{\color{green}197616},3135\Big\}
\end{eqnarray*}

\bigskip

\begin{eqnarray*}
  M_q(18,12;9) &=& q^{36}+q^6+q^3+q^0
\end{eqnarray*}
\begin{eqnarray*}
  \mathcal{U}_{18,12,9} &=& \Big\{261632,{\color{green}229880,29127},3647\Big\}
\end{eqnarray*}

\bigskip

\begin{eqnarray*}
  M_q(19,12;6) &=& q^{13}+q^7+q^0
\end{eqnarray*}
\begin{eqnarray*}
  \mathcal{U}_{19,12,6} &=& \Big\{516096,8064,63\Big\}
\end{eqnarray*}

\bigskip

\begin{eqnarray*}
  \overline{M}_q(19,12;7) &=& q^{24}+q^{12}+q^0
\end{eqnarray*}
\begin{eqnarray*}
  \mathcal{U}_{19,12,7} &=& \Big\{520192,{\color{red}135104},127\Big\}
\end{eqnarray*}

\bigskip

\begin{eqnarray*}
  \overline{M}_q(19,12;8) &=& q^{33}+q^{12}+q^0
\end{eqnarray*}
\begin{eqnarray*}
  \mathcal{U}_{19,12,8} &=& \Big\{522240,{\color{red}395232},2175\Big\}
\end{eqnarray*}

\bigskip

\begin{eqnarray*}
  \overline{M}_q(19,12;9) &=& q^{40}+q^{12}+q^6+q^0
\end{eqnarray*}
\begin{eqnarray*}
  \mathcal{U}_{19,12,9} &=& \Big\{523264,{\color{red}459760,58253},3199\Big\}
\end{eqnarray*}

\subsection{Minimum subspace distance $14$}
\label{subsec_mindist_14}

\begin{eqnarray*}
  M_q(14,14;7) &=& q^7+q^0
\end{eqnarray*}
\begin{eqnarray*}
  \mathcal{U}_{14,14,7} &=& \Big\{16256,127\Big\}
\end{eqnarray*}

\bigskip

\begin{eqnarray*}
  M_q(15,14;7) &=& q^8+q^0
\end{eqnarray*}
\begin{eqnarray*}
  \mathcal{U}_{15,14,7} &=& \Big\{32512,127\Big\}
\end{eqnarray*}

\bigskip

\begin{eqnarray*}
  M_q(16,14;7) &=& q^9+q^0
\end{eqnarray*}
\begin{eqnarray*}
  \mathcal{U}_{16,14,7} &=& \Big\{65024,127\Big\}
\end{eqnarray*}

\bigskip

\begin{eqnarray*}
  M_q(16,14;8) &=& q^{16}+q^0
\end{eqnarray*}
\begin{eqnarray*}
  \mathcal{U}_{16,14,8} &=& \Big\{65280,255\Big\}
\end{eqnarray*}

\bigskip

\begin{eqnarray*}
  M_q(17,14;7) &=& q^{10}+q^0
\end{eqnarray*}
\begin{eqnarray*}
  \mathcal{U}_{17,14,7} &=& \Big\{130048,127\Big\}
\end{eqnarray*}

\bigskip

\begin{eqnarray*}
  M_q(17,14;8) &=& q^{18}+q^0
\end{eqnarray*}
\begin{eqnarray*}
  \mathcal{U}_{17,14,8} &=& \Big\{130560,255\Big\}
\end{eqnarray*}

\bigskip

\begin{eqnarray*}
  M_q(18,14;7) &=& q^{11}+q^0
\end{eqnarray*}
\begin{eqnarray*}
  \mathcal{U}_{18,14,7} &=& \Big\{260096,127\Big\}
\end{eqnarray*}

\bigskip

\begin{eqnarray*}
  M_q(18,14;8) &=& q^{20}+q^0
\end{eqnarray*}
\begin{eqnarray*}
  \mathcal{U}_{18,14,8} &=& \Big\{261120,255\Big\}
\end{eqnarray*}

\bigskip

\begin{eqnarray*}
  M_q(18,14;9) &=& q^{27}+q^0
\end{eqnarray*}
\begin{eqnarray*}
  \mathcal{U}_{18,14,9} &=& \Big\{261632,511\Big\}
\end{eqnarray*}

\bigskip

\begin{eqnarray*}
  M_q(19,14;7) &=& q^{12}+q^0
\end{eqnarray*}
\begin{eqnarray*}
  \mathcal{U}_{19,14,7} &=& \Big\{520192,127\Big\}
\end{eqnarray*}

\bigskip

\begin{eqnarray*}
  M_q(19,14;8) &=& q^{22}+q^0
\end{eqnarray*}
\begin{eqnarray*}
  \mathcal{U}_{19,14,8} &=& \Big\{522240,255\Big\}
\end{eqnarray*}

\bigskip

\begin{eqnarray*}
  M_q(19,14;9) &=& q^{30}+q^0
\end{eqnarray*}
\begin{eqnarray*}
  \mathcal{U}_{19,14,9} &=& \Big\{523264,511\Big\}
\end{eqnarray*}

\subsection{Minimum subspace distance $16$}
\label{subsec_mindist_16}

\begin{eqnarray*}
  M_q(16,16;8) &=& q^8+q^0
\end{eqnarray*}
\begin{eqnarray*}
  \mathcal{U}_{16,16,8} &=& \Big\{65280,255\Big\}
\end{eqnarray*}

\bigskip

\begin{eqnarray*}
  M_q(17,16;8) &=& q^9+q^0
\end{eqnarray*}
\begin{eqnarray*}
  \mathcal{U}_{17,16,8} &=& \Big\{130560,255\Big\}
\end{eqnarray*}

\bigskip

\begin{eqnarray*}
  M_q(18,16;8) &=& q^{10}+q^0
\end{eqnarray*}
\begin{eqnarray*}
  \mathcal{U}_{18,16,8} &=& \Big\{261120,255\Big\}
\end{eqnarray*}

\bigskip

\begin{eqnarray*}
  M_q(18,16;9) &=& q^{18}+q^0
\end{eqnarray*}
\begin{eqnarray*}
  \mathcal{U}_{18,16,9} &=& \Big\{261632,511\Big\}
\end{eqnarray*}

\bigskip

\begin{eqnarray*}
  M_q(19,16;8) &=& q^{11}+q^0
\end{eqnarray*}
\begin{eqnarray*}
  \mathcal{U}_{19,16,8} &=& \Big\{522240,255\Big\}
\end{eqnarray*}

\bigskip

\begin{eqnarray*}
  M_q(19,16;9) &=& q^{20}+q^0
\end{eqnarray*}
\begin{eqnarray*}
  \mathcal{U}_{19,16,9} &=& \Big\{523264,511\Big\}
\end{eqnarray*}

\subsection{Minimum subspace distance $18$}
\label{subsec_mindist_18}

\begin{eqnarray*}
  M_q(18,18;9) &=& q^9+q^0
\end{eqnarray*}
\begin{eqnarray*}
  \mathcal{U}_{18,18,9} &=& \Big\{261632,511\Big\}
\end{eqnarray*}

\bigskip

\begin{eqnarray*}
  M_q(19,18;9) &=& q^{10}+q^0
\end{eqnarray*}
\begin{eqnarray*}
  \mathcal{U}_{19,18,9} &=& \Big\{523264,511\Big\}
\end{eqnarray*}

\newpage

\section{The best known dimension of FDRM codes marked in red}
\label{the_best_known_dimension_of_given_FDRMCs}

\subsection{Minimum subspace distance 6}\label{AppB:Minimum subspace distance 6}
\begin{center}
$\begin{array}{|c|cc|c|c|c|c}\hline
(12,6,5)_q & 1256   & [6,4,4,4,3]& [5,4,4,4,3]^{\rm Thm \ref{thm:shortening}\ or\ \ref{cor:delta=3}} & 10\rightarrow 11 \\\hline
(12,6,6)_q & 1830   & [5,5,5,3,1,1] & [5,5,5,3,1]^{\rm Thm \ref{thm:shortening}\ or\ \ref{cor:delta=3}}& 9\rightarrow 10 \\
  & 2265   & [6,3,3,2,2] & [5,3,3,2,2]^{\rm Thm \ref{thm:shortening}\ or\ \ref{cor:delta=3}}& 5\rightarrow 6 \\\hline
(13,6,6)_q & 4968   & [7,5,5,4,4,3] & [6,5,5,4,4,3]^{\rm Thm \ref{thm:shortening}\ or\ \ref{cor:delta=3}}& 15\rightarrow 16 \\\hline
(14,6,6)_q & 5040   & [7,5,5,5,4,4] & [6,5,5,5,4,4]^{\rm Thm \ref{thm:shortening}\ or\ \ref{cor:delta=3}}& 17 \rightarrow 18 \\\hline
(14,6,7)_q & 3725   & [5,5,5,4,1,1] & [5,5,5,4,1]^{\rm Thm \ref{thm:shortening}\ or\ \ref{cor:delta=3}}& 10\rightarrow11 \\\hline
(15,6,7)_{q\geq 4} & 20176  & [8,6,6,6,5,5,4] & [7,6,6,6,5,5,4]^{\rm Thm \ref{thm:shortening}\ or\ \ref{cor:delta=3}}& 25\rightarrow26 \\
 & 19880  & [8,6,6,5,5,4,3] & [7,6,6,5,5,4,3]^{\rm Thm \ref{thm:shortening}\ or\ \ref{cor:delta=3}}& 22\rightarrow23 \\
 & 18204  & [8,5,5,5,2,2,2] & [7,5,5,5,2,2,2]^{\rm Thm \ref{thm:shortening}\ or\ \ref{cor:delta=3}} & 14\rightarrow15 \\
 & 3229  & [5,5,3,1,1,1] & [5,5,3,1,1]^{\rm Thm \ref{thm:shortening}\ or\ \ref{cor:delta=3}}& 5\rightarrow6 \\
(15,6,7)_{3} & 7462  & [6,6,6,5,3,1,1] & [6,6,6,5,3,1]^{\rm Thm \ref{thm:shortening}\ or\ \ref{cor:delta=3}}& 15\rightarrow16 \\
& 9457  & [7,5,3,3,3,3] & [6,5,3,3,3,3]^{\rm Thm \ref{thm:shortening}\ or\ \ref{cor:delta=3}}& 11\rightarrow12 \\\hline
(16,6,6)_{q\geq 3} & 4508  & [7,4,4,2,2,2] & [6,4,4,2,2,2]^{\rm Thm \ref{thm:shortening}\ or\ \ref{cor:delta=3}} & 8\rightarrow9 \\
& 33132  & [10,4,3,3,2,2] & [6,4,3,3,2,2]^{\rm Thm \ref{thm:shortening}\ or\ \ref{cor:delta=3}}& 8\rightarrow9 \\
(16,6,6)_{2} & 2233  & [6,3,2,2,2] & [5,3,2,2,2]^{\rm Thm \ref{thm:shortening}\ or\ \ref{cor:delta=3}} & 4\rightarrow5  \\\hline
(16,6,7)_{q\geq 4} & 20416  & [8,6,6,6,6,6,6]  & [7,6,6,6,6,6,6]^{\rm Thm \ref{thm:shortening}\ or\ \ref{cor:delta=3}}& 29\rightarrow30 \\
 & 20024  & [8,6,6,6,3,3,3] & [7,6,6,6,3,3,3]^{\rm Thm \ref{thm:shortening}\ or\ \ref{cor:delta=3}} & 20\rightarrow21 \\
 & 17268 & [8,4,4,3,3,3,2] & [7,4,4,3,3,3,2] ^{\rm Thm \ref{thm:shortening}\ or\ \ref{cor:delta=3}}& 12\rightarrow13 \\
 & 17129  & [8,4,3,3,3,2] & [6,4,3,3,3,2]^{\rm Thm \ref{thm:shortening}\ or\ \ref{cor:delta=3}}& 9\rightarrow10 \\
(16,6,7)_{3} & 7942  & [6,6,6,6,6,1,1] & [6,6,6,6,6,1]^{\rm Thm \ref{thm:shortening}\ or\ \ref{cor:delta=3}}& 19\rightarrow20 \\
 & 34412  & [9,5,5,3,3,2,2] & [7,5,5,3,3,2,2]^{\rm Thm \ref{thm:shortening}\ or\ \ref{cor:delta=3}}& 13\rightarrow14 \\
 & 33272  & [9,3,3,3,3,3,3] & [7,3,3,3,3,3,3]^{\rm Thm \ref{thm:shortening}\ or\ \ref{cor:delta=3}} & 11\rightarrow12 \\
 (16,6,7)_{2} & 7694  & [5,5,5,5,1,1,1] & [5,5,5,4,1,1,1]^{\rm Thm \ref{thm:shortening}\ or\ \ref{cor:delta=3}}& 11\rightarrow12 \\
 & 34252  & [9,5,4,4,4,2,2] & [7,5,4,4,4,2,2]^{\rm Thm \ref{thm:shortening}\ or\ \ref{cor:delta=3}}& 14\rightarrow15 \\
 & 33649  & [9,4,4,3,3,3] & [6,4,4,3,3,3]^{\rm Thm \ref{thm:shortening}\ or\ \ref{cor:delta=3}}& 11\rightarrow12 \\
 & 17212  & [8,4,4,2,2,2,2] & [[7,4,4,2,2,2,2]]^{\rm Thm \ref{thm:shortening}\ or\ \ref{cor:delta=3}} & 9\rightarrow10 \\\hline
(16,6,8)_{q\geq 4} & 31110  & [7,7,7,7,5,5,1,1] & [7,7,7,7,5,5,1]^{\rm Thm \ref{thm:shortening}\ or\ \ref{cor:delta=3}}& 25\rightarrow26 \\\hline
(17,6,7)_{q\geq 4} & 66544  & [10,4,4,4,4,4,4] & [7,4,4,4,4,4,4]^{\rm Thm \ref{thm:shortening}\ or\ \ref{cor:delta=3}}& 17\rightarrow18 \\
 & 34264  & [9,5,4,4,4,3,3] & [7,5,4,4,4,3,3]^{\rm Thm \ref{thm:shortening}\ or\ \ref{cor:delta=3}}& 16\rightarrow17 \\
 & 66988  & [10,5,4,4,3,2,2] & [7,5,4,4,3,2,2]^{\rm Thm \ref{thm:shortening}\ or\ \ref{cor:delta=3}}& 13\rightarrow14 \\
(17,6,7)_{3} & 18316  & [8,5,5,5,5,2,2]& [7,5,5,5,5,2,2]^{\rm Thm \ref{thm:shortening}\ or\ \ref{cor:delta=3}} & 17\rightarrow18 \\
 & 33776  & [9,4,4,4,4,4,4] & [7,4,4,4,4,4,4]^{\rm Thm \ref{thm:shortening}\ or\ \ref{cor:delta=3}}& 17\rightarrow18 \\
 & 66236  & [10,4,3,2,2,2,2] & [7,4,3,2,2,2,2]^{\rm Thm \ref{thm:shortening}\ or\ \ref{cor:delta=3}}& 8\rightarrow9 \\
 (17,6,7)_{2} & 17392 & [8,4,4,4,4,4,4]& [7,4,4,4,4,4,4]^{\rm Thm \ref{thm:shortening}\ or\ \ref{cor:delta=3}} & 17\rightarrow18 \\\hline
(17,6,8)_{q\geq 4} & 81312  & [9,7,7,7,7,6,6,5]& [8,7,7,7,7,6,6,5]^{\rm Thm \ref{thm:shortening}\ or\ \ref{cor:delta=3}} & 37\rightarrow38 \\
 & 80720  & [9,7,7,7,6,6,5,4] & [8,7,7,7,6,6,5,4]^{\rm Thm \ref{thm:shortening}\ or\ \ref{cor:delta=3}}& 34\rightarrow35 \\
 & 77512  & [9,7,6,6,6,5,5,3] & [8,7,6,6,6,5,5,3]^{\rm Thm \ref{thm:shortening}\ or\ \ref{cor:delta=3}}& 30\rightarrow31 \\
 & 79416  & [9,7,7,6,6,3,3,3] & [8,7,7,6,6,3,3,3]^{\rm Thm \ref{thm:shortening}\ or\ \ref{cor:delta=3}}& 27\rightarrow28 \\
 & 71128  & [9,6,5,4,4,4,3,3] & [8,6,5,4,4,4,3,3]^{\rm Thm \ref{thm:shortening}\ or\ \ref{cor:delta=3}}& 21\rightarrow22 \\
 & 29462  & [7,7,7,5,5,2,1,1] & [7,7,7,5,5,2,1]^{\rm Thm \ref{thm:shortening}\ or\ \ref{cor:delta=3}}& 20\rightarrow21 \\\hline
 (18,6,7)_{q\geq5} & 133088  & [11,5,5,5,5,5,5] & [7,5,5,5,5,5,5]^{\rm Thm \ref{thm:shortening}\ or\ \ref{cor:delta=3}}& 23\rightarrow24 \\
& 17876  & [8,5,4,4,4,3,2] & [7,5,4,4,4,3,2]^{\rm Thm \ref{thm:shortening}\ or\ \ref{cor:delta=3}}& 15\rightarrow16 \\

 \end{array}$
\end{center}

\begin{center}
$\begin{array}{|c|cc|c|c|c}
& 17241  & [8,4,4,3,2,2] & [6,4,4,3,2,2]^{\rm Thm \ref{thm:shortening}\ or\ \ref{cor:delta=3}} & 9\rightarrow10 \\\hline
(18,6,7)_{2} & 18400  & [8,5,5,5,5,5,5] & [7,5,5,5,5,5,5]^{\rm Thm \ref{thm:shortening}\ or\ \ref{cor:delta=3}}& 23\rightarrow24 \\
& 20248  & [8,6,6,6,6,3,3] & [7,6,6,6,6,3,3]^{\rm Thm \ref{thm:shortening}\ or\ \ref{cor:delta=3}}& 23\rightarrow24 \\
& 34252  & [9,5,4,4,4,2,2] & [7,5,4,4,4,2,2]^{\rm Thm \ref{thm:shortening}\ or\ \ref{cor:delta=3}} & 15\rightarrow17 \\
& 66505  & [10,4,4,4,4,2] & [6,4,4,4,4,2]^{\rm Thm \ref{thm:shortening}\ or\ \ref{cor:delta=3}} & 12\rightarrow13 \\\hline
(18,6,8)_{q\geq 4} & 81224  & [9,7,7,7,7,6,5,3] & [8,7,7,7,7,6,5,3]^{\rm Thm \ref{thm:shortening}\ or\ \ref{cor:delta=3}} & 34\rightarrow35 \\
& 73601  & [9,6,6,6,6,6,6] & [7,6,6,6,6,6,6]^{\rm Thm \ref{thm:shortening}\ or\ \ref{cor:delta=3}} & 29\rightarrow30 \\
& 138088  & [10,6,6,5,5,4,4,3] & [8,6,6,5,5,4,4,3]^{\rm Thm \ref{thm:shortening}\ or\ \ref{cor:delta=3}}  & 26\rightarrow27 \\
& 71396  & [9,6,5,5,4,4,4,2] & [8,6,5,5,4,4,4,2]^{\rm Thm \ref{thm:shortening}\ or\ \ref{cor:delta=3}} & 22\rightarrow23 \\
& 69432  & [9,5,5,5,5,3,3,3] & [8,5,5,5,5,3,3,3]^{\rm Thm \ref{thm:shortening}\ or\ \ref{cor:delta=3}}  & 21\rightarrow22 \\
& 134872  & [10,5,5,5,4,4,3,3] & [8,5,5,5,4,4,3,3]^{\rm Thm \ref{thm:shortening}\ or\ \ref{cor:delta=3}} & 22\rightarrow23 \\
& 76024  & [9,7,6,3,3,3,3,3] & [8,7,6,3,3,3,3,3]^{\rm Thm \ref{thm:shortening}\ or\ \ref{cor:delta=3}}  & 20\rightarrow21 \\
& 134628  & [10,5,5,4,4,4,4,2] & [8,5,5,4,4,4,4,2]^{\rm Thm \ref{thm:shortening}\ or\ \ref{cor:delta=3}}  & 20\rightarrow21 \\\hline
(18,6,9)_{q\geq 4} & 155217  & [9,7,6,6,6,6,4,3] & [8,7,6,6,6,6,4,3]^{\rm Thm \ref{thm:shortening}\ or\ \ref{cor:delta=3}} & 30\rightarrow31 \\
& 123934  & [8,8,8,8,6,1,1,1,1] & [8,8,8,8,6,1,1,1]^{\rm Thm \ref{thm:shortening}\ or\ \ref{cor:delta=3}} & 25\rightarrow26 \\
& 137155  & [9,5,4,4,4,4,4] & [7,5,4,4,4,4,4]^{\rm Thm \ref{thm:shortening}\ or\ \ref{cor:delta=3}}  & 18\rightarrow19 \\
& 72583  & [8,5,5,4,4,4] & [6,5,5,4,4,4]^{\rm Thm \ref{thm:shortening}\ or\ \ref{cor:delta=3}}  & 16\rightarrow17 \\\hline
(19,6,6)_{q\geq 4} & 16748 &  [9,4,3,3,2,2] & [6,4,3,3,2,2]^{\rm Thm \ref{thm:shortening}\ or\ \ref{cor:delta=3}} & 8\rightarrow9 \\\hline
(19,6,7)_{q\geq 5} & 9649  & [7,5,4,4,3,3] & [7,5,4,4,3,2]^{\rm Thm \ref{thm:shortening}\ or\ \ref{cor:delta=3}} & 13\rightarrow14 \\
& 34220  & [9,5,4,4,3,2,2] & [7,5,4,4,3,2,2]^{\rm Thm \ref{thm:shortening}\ or\ \ref{cor:delta=3}} & 13\rightarrow14 \\
& 66420  & [10,4,4,3,3,3,2] & [7,4,4,3,3,3,2]^{\rm Thm \ref{thm:shortening}\ or\ \ref{cor:delta=3}}  & 12\rightarrow13 \\
(19,6,7)_{3} & 35696  & [9,6,5,5,4,4,4] & [7,6,5,5,4,4,4]^{\rm Thm \ref{thm:shortening}\ or\ \ref{cor:delta=3}} & 21\rightarrow22 \\
& 7318  & [6,6,6,4,2,1,1] & [6,6,5,4,2,1,1]^{\rm Thm \ref{thm:from subcodes}}& 13\rightarrow14 \\
(19,6,7)_{2} & 36800  & [9,6,6,6,6,6,6] & [7,6,6,6,6,6,6]^{\rm Thm \ref{thm:shortening}\ or\ \ref{cor:delta=3}} & 29\rightarrow30 \\
& 264100  & [12,5,5,5,5,4,2] & [7,5,5,5,5,4,2]^{\rm Thm \ref{thm:shortening}\ or\ \ref{cor:delta=3}} & 19\rightarrow20 \\
& 6798  & [6,6,5,4,1,1,1] & [6,6,5,4,1,1]^{\rm Thm \ref{thm:shortening}\ or\ \ref{cor:delta=3}}  & 11\rightarrow12 \\\hline
(19,6,8)_{q\geq 5} & 79328  & [9,7,7,6,5,5,5,5] & [8,7,7,6,5,5,5,5]^{\rm Thm \ref{thm:shortening}\ or\ \ref{cor:delta=3}} & 32\rightarrow33 \\
& 142800  & [10,7,6,6,5,5,5,4] & [10,7,6,6,5,5,5,3]^{\rm Thm \ref{thm:from subcodes}}& 30\rightarrow31 \\
& 138864  & [10,6,6,6,6,4,4,4] & [8,6,6,6,6,4,4,4]^{\rm Thm \ref{thm:shortening}\ or\ \ref{cor:delta=3}} & 28\rightarrow30 \\
& 269744  & [11,6,6,6,5,5,4,4] & [8,6,6,6,5,5,4,4]^{\rm Thm \ref{thm:shortening}\ or\ \ref{cor:delta=3}} & 28\rightarrow30 \\
& 269256  & [11,6,6,5,5,5,5,3] & [8,6,6,5,5,5,5,3]^{\rm Thm \ref{thm:shortening}\ or\ \ref{cor:delta=3}}  & 27\rightarrow28 \\
& 135916  & [10,6,4,3,3,3,2,2] & [10,6,4,3,3,3,2]^{\rm Thm \ref{thm:from subcodes}}& 15\rightarrow16 \\
& 36153  & [8,5,5,4,2,2,2] & [7,5,5,4,2,2,2]^{\rm Thm \ref{thm:shortening}\ or\ \ref{cor:delta=3}} & 13\rightarrow14 \\
& 68028  & [9,5,4,4,2,2,2,2] & [8,5,4,4,2,2,2,2]^{\rm Thm \ref{thm:shortening}\ or\ \ref{cor:delta=3}}  & 11\rightarrow12 \\
& 66809  & [9,4,2,2,2,2,2] & [7,4,2,2,2,2,2]^{\rm Thm \ref{thm:shortening}\ or\ \ref{cor:delta=3}}  & 7\rightarrow8 \\
\hline
(19,6,9)_{q\geq 5} & 326464  & [10,8,8,8,8,8,7,7,6]  & [9,8,8,8,8,8,7,7,6]^{\rm Thm \ref{thm:shortening}\ or\ \ref{cor:delta=3}}& 51\rightarrow52 \\
& 323216  & [10,8,8,8,7,7,7,6,4] & [9,8,8,8,7,7,7,6,4]^{\rm Thm \ref{thm:shortening}\ or\ \ref{cor:delta=3}} & 46\rightarrow47 \\
& 310688  & [10,8,7,7,7,7,6,6,5] & [9,8,7,7,7,7,6,6,5]^{\rm Thm \ref{thm:shortening}\ or\ \ref{cor:delta=3}} & 44\rightarrow45 \\
& 324720  & [10,8,8,8,8,7,4,4,4] & [9,8,8,8,8,7,4,4,4]^{\rm Thm \ref{thm:shortening}\ or\ \ref{cor:delta=3}} & 42\rightarrow43 \\
& 286593  & [10,7,6,6,6,6,6,6] & [8,7,6,6,6,6,6,6]^{\rm Thm \ref{thm:shortening}\ or\ \ref{cor:delta=3}}  & 35\rightarrow36 \\
& 291760  & [10,7,7,7,5,5,5,4,4] & [9,7,7,7,5,5,5,4,4]^{\rm Thm \ref{thm:shortening}\ or\ \ref{cor:delta=3}} & 35\rightarrow36 \\
& 301808  & [10,8,6,6,5,4,4,4,4] & [9,8,6,6,5,4,4,4,4]^{\rm Thm \ref{thm:shortening}\ or\ \ref{cor:delta=3}} & 32\rightarrow33 \\
& 284372  & [10,7,6,5,5,4,4,3,2] & [9,7,6,5,5,4,4,3,2]^{\rm Thm \ref{thm:shortening}\ or\ \ref{cor:delta=3}}  & 27\rightarrow28 \\
& 146755  & [9,6,6,6,6,5,4] & [7,6,6,6,6,5,4]^{\rm Thm \ref{thm:shortening}\ or\ \ref{cor:delta=3}}  & 26\rightarrow27 \\
& 274089  & [10,6,5,5,5,4,3,2] & [8,6,5,5,5,4,3,2]^{\rm Thm \ref{thm:shortening}\ or\ \ref{cor:delta=3}} & 23\rightarrow24 \\
& 117022  & [8,8,8,6,4,1,1,1,1] & [8,8,8,6,4,1,1,1]^{\rm Thm \ref{thm:shortening}\ or\ \ref{cor:delta=3}}  & 21\rightarrow22 \\
& 72531  & [8,5,5,4,4,3,2] & [7,5,5,4,4,3,2]^{\rm Thm \ref{thm:shortening}\ or\ \ref{cor:delta=3}}  & 16\rightarrow17 \\
& 137955  & [9,5,5,4,3,3,3] & [7,5,5,4,3,3,3]^{\rm Thm \ref{thm:shortening}\ or\ \ref{cor:delta=3}} & 16\rightarrow17 \\
& 67307 & [8,3,3,2,2,2,1] & [7,3,3,2,2,2,1]^{\rm Thm \ref{thm:shortening}\ or\ \ref{cor:delta=3}} & 6\rightarrow7 \\
\hline
\end{array}$
\end{center}

\subsection{Minimum subspace distance 8}
\label{AppB:Minimum subspace distance 8}

\begin{center}
$\begin{array}{|c|cc|c|c|c}\hline
(13,8,5)_q & 2288 & [7,4,4,4,4]& [7,4,4,4,3]^{\rm Thm \ref{thm:from subcodes}}&  7\rightarrow8\\\hline
(13,8,6)_q & 6264 & [7,7,3,3,3,3] & [6,6,3,3,3,3]^{\rm Thm \ref{thm:shortening}}&  6\rightarrow8\\
& 1637  & [5,5,3,3,1] & [5,4,3,3,1]^{\rm Thm \ref{thm:from subcodes}}&  3\rightarrow4\\\hline
(14,8,6)_q & 6384  & [7,7,4,4,4,4] & [6,6,4,4,4,4]^{\rm Thm \ref{thm:shortening}}&  10\rightarrow12\\\hline
(14,8,7)_q & 9830  & [7,5,5,3,3,1,1] & [7,5,5,3,3,1]^{\rm Thm \ref{thm:from subcodes}}&  7\rightarrow8\\
& 5461  & [6,5,4,3,2,1] & [6,5,4,3,1,1]^{\rm Thm \ref{thm:subcodes from Gab}}_{(r=2)}&   q<5,\ 5\rightarrow6\\
&   &  & [6,5,4,3,2,1]^{\rm Thm \ref{thm:from MDS}}&   q\geq5,\ 6\rightarrow6\\
& 2861  & [5,4,4,2,1,1] & [5,4,4,2,1]^{\rm Thm \ref{thm:from subcodes}}&   3\rightarrow4\\
& 3251  & [5,5,3,2,2] & [5,5,2,2,2]^{\rm Thm \ref{thm:combine with same dim}}&  3\rightarrow4\\
& 4811  & [6,4,3,3,1] & [5,4,3,3,1]^{\rm Thm \ref{thm:from subcodes}}  &  3\rightarrow4\\\hline
(15,8,5)_q &  2360  &  [7,5,3,3,3] &  [5,5,3,3,3]^{\rm Thm \ref{thm:shortening}}&  4\rightarrow6\\\hline
(15,8,6)_q &  5600 &  [7,6,5,5,5,5] &  [6,6,5,5,5,5]^{\rm Thm \ref{thm:shortening}}&  14\rightarrow15\\
&  8805  &  [8,5,3,3,1] &  [8,5,3,3]^{\rm Thm \ref{thm:from subcodes}}&  3\rightarrow4\\\hline
(15,8,7)_q &  21744  &  [8,7,6,4,4,4,4] &  [7,7,6,4,4,4,4]^{\rm Thm \ref{thm:shortening}}&  15\rightarrow16  \\
& 10721  &  [7,6,4,4,4,4]&  [6,6,4,4,4,4]^{\rm Thm \ref{thm:shortening}} &  10\rightarrow11  \\
& 11324  &  [7,6,6,2,2,2,2] &  [7,6,6,2,2,2,1]^{\rm Thm \ref{thm:from subcodes}}&  7\rightarrow8  \\
& 6550  &  [6,6,4,4,2,1,1] &  [6,6,4,4,2]^{\rm Thm \ref{thm:from subcodes}}&  6\rightarrow7\\\hline
(16,8,7)_q &  25552  &  [8,8,5,5,5,5,4] &  [7,7,5,5,5,5,4]^{\rm Thm \ref{thm:shortening}}&  17\rightarrow19  \\
& 13708  &  [7,7,6,5,5,2,2] &  [7,7,6,5,5,2,1]^{\rm Thm \ref{thm:from subcodes}}&  13\rightarrow14  \\
& 37556  &  [9,7,5,4,3,3,2] &  [7,7,5,4,3,3,2]^{\rm Thm \ref{thm:from subcodes}}&  11\rightarrow12  \\
& 10922  &  [7,6,5,4,3,2,1] &  [7,6,5,4,3,2]^{\rm Thm \ref{thm:from subcodes}} &   q<6,\ 9\rightarrow10  \\
&    &    &  [7,6,5,4,3,2,1]^{\rm Thm \ref{thm:from MDS}} &   q\geq 6,\ 10\rightarrow10  \\
& 18214  &  [8,5,5,5,3,1,1] &  [6,5,5,5,3,1,1]^{\rm Thm \ref{thm:from subcodes}} &  9\rightarrow10  \\
& 3669  &  [5,5,5,3,2,1] &  [5,5,5,3,2]^{\rm Thm \ref{thm:shortening}}&  5\rightarrow6\\\hline
(16,8,8)_{q} & 52428  & [8,8,6,6,4,4,2,2] & [8,8,6,6,4,4,2]^{\rm Thm \ref{thm:from subcodes}} & 16\rightarrow18 \\
& 43690  & [8,7,6,5,4,3,2,1] & [8,7,6,5,4,3,2]^{\rm Thm \ref{thm:from subcodes}}& q<7, \ 14\rightarrow15 \\
&    &  & [8,7,6,5,4,3,2,1]^{\rm Thm \ref{thm:from MDS}}& q\geq7, \ 15\rightarrow15 \\
& 23190  & [7,6,6,5,4,2,1,1] & [7,6,6,5,4,2,1]^{\rm Thm \ref{thm:from subcodes}}&  \ 12\rightarrow13 \\
& 26265  & [7,7,5,5,4,2,2] & [7,7,5,5,4,2]^{\rm Thm \ref{thm:from subcodes}}&  \ 11\rightarrow13 \\
& 38565  & [8,6,5,5,4,3,1] & [8,6,5,5,4,3]^{\rm Thm \ref{thm:from subcodes}} &  \ 12\rightarrow13 \\
& 26970  & [7,7,6,4,3,2,2,1] & [7,7,6,4,3,2,1]^{\rm Thm \ref{cfdrm}}&  \ 11\rightarrow12 \\
& 39270  & [8,6,6,4,3,3,1,1] & [8,6,6,4,3,3,1]^{\rm Thm \ref{thm:from subcodes}}&  \ 11\rightarrow12 \\
& 42345  & [8,7,5,4,3,3,2] & [7,7,5,4,3,3,2]^{\rm Thm \ref{cfdrm}}&  \ 11\rightarrow12 \\
& 21845  & [7,6,5,4,3,2,1] & [7,6,5,4,3,2]^{\rm Thm \ref{thm:from subcodes}}& q<6, \ 9\rightarrow10 \\
&    &    & [7,6,5,4,3,2,1]^{\rm Thm \ref{thm:from MDS}}& q\geq 6, \ 10\rightarrow10 \\\hline
(17,8,6)_q &  6264  &  [7,7,3,3,3,3] &  [6,6,3,3,3,3]^{\rm Thm \ref{thm:shortening}}&  6\rightarrow8\\\hline
(17,8,7)_q &  51104  &  [9,9,6,6,6,6,5] &  [9,9,6,6,6,6,3]^{\rm Thm \ref{thm:from subcodes}}&  22\rightarrow23  \\
& 13868  &  [7,7,6,6,3,2,2] &  [7,7,6,6,3,2,1]^{\rm Thm \ref{thm:from subcodes}}&  12\rightarrow13\\\hline
(17,8,8)_{q\geq 3} &  88472  &  [9,8,7,7,5,5,3,3] &  [8,8,7,7,5,5,3,3]^{\rm Thm \ref{thm:shortening}}&  22\rightarrow23  \\
& 40353  &  [8,6,6,6,5,5,4] &  [7,6,6,6,5,5,4]^{\rm Thm \ref{thm:shortening}}&  18\rightarrow19  \\
& 45964  &  [8,7,7,5,5,5,2,2] &  [8,7,7,5,5,5,2,1]^{\rm Thm \ref{thm:from subcodes}}&  18\rightarrow19  \\
& 54484  &  [8,8,7,6,4,4,3,2]&  [8,8,7,6,4,4,3,1]^{\rm Thm \ref{thm:from subcodes}} &  18\rightarrow19  \\
\end{array}$
\end{center}

\begin{center}
$\begin{array}{|c|cc|c|c|c}
& 71618  &  [9,6,5,5,5,5,5,1] &  [8,6,5,5,5,5,5,1]^{\rm Thm \ref{thm:from subcodes}}&  18\rightarrow19  \\
& 101708  &  [9,9,6,6,5,4,2,2]&  [9,9,6,6,5,4,2,1]^{\rm Thm \ref{thm:subcodes from Gab}} _{(r=2)}&  18\rightarrow19  \\
& 28038  &  [7,7,6,6,5,5,1,1] &  [7,7,6,6,5,5,1]^{\rm Thm \ref{thm:from subcodes}}&  17\rightarrow18  \\
& 76500  &  [9,7,6,5,4,4,3,2] &  [8,7,6,5,4,4,3,2]^{\rm Thm \ref{thm:from subcodes}}&  16\rightarrow17  \\
& 103032  &  [9,9,7,5,3,3,3,3]  &  [9,9,7,5,3,3,3,2]^{\rm Thm \ref{thm:from subcodes}}&  16\rightarrow17  \\
& 50994  &  [8,8,5,5,5,3,3,1] &  [8,7,5,5,5,3,3,1]^{\rm Thm \ref{thm:from subcodes}} &  15\rightarrow16  \\
& 51914  &  [8,8,6,5,4,4,2,1] &  [8,8,6,5,4,4,2]^{\rm Thm \ref{thm:from subcodes}}&  15\rightarrow16  \\
& 83628  &  [9,8,5,5,4,3,2,2] &  [8,7,5,5,4,3,2,2]^{\rm Thm \ref{cfdrm}}&  14\rightarrow16  \\
& 42597  &  [8,7,5,5,3,3,1] &  [8,7,5,5,3,3]^{\rm Thm \ref{thm:from subcodes}} &  11\rightarrow12  \\
& 7709  &  [5,5,5,5,1,1,1] &  [5,5,5,4,1,1,1]^{\rm Thm \ref{thm:from subcodes}}&  7\rightarrow8 \\
(17,8,8)_{2} & 40385   & [8,6,6,6,5,5,5]  & [7,6,6,6,5,5,5]^{\rm Thm \ref{thm:shortening}}& 19\rightarrow20\\
 &52052   & [8,8,6,5,5,4,3,2] & [8,7,6,5,5,4,3,2]^{\rm Thm \ref{thm:from subcodes}}& 17\rightarrow18\\
 &54452   & [8,8,7,6,4,3,3,2] & [8,8,7,6,4,3,3,1]^{\rm Thm \ref{thm:from subcodes}}& 17\rightarrow18\\
&71586   & [9,6,5,5,5,5,4,1] & [8,6,5,5,5,5,4,1]^{\rm Thm \ref{thm:from subcodes}}& 17\rightarrow18\\
&100056   & [9,9,5,5,4,4,3,3] & [8,8,5,5,4,4,3,3]^{\rm Thm \ref{thm:shortening}}& 16\rightarrow18\\
&101617   & [9,9,6,6,5,3,2,2] & [9,9,6,6,5,3,2,1]^{\rm Thm \ref{thm:from subcodes}} & 17\rightarrow18\\
&30828   & [7,7,7,7,3,3,2,2] & [7,7,7,7,3,3,2]^{\rm Thm \ref{thm:shortening}}& 15\rightarrow16\\
&76468   & [9,7,6,5,4,3,3,2] & [8,7,6,5,4,3,3,2]^{\rm Thm \ref{thm:from subcodes}} & 15\rightarrow16\\
&85106   & [9,8,6,6,3,3,3,1] & [9,8,6,6,3,3,3]^{\rm Thm \ref{thm:from subcodes}}& 15\rightarrow16\\
&51882   & [8,8,6,5,4,3,2,1] & [8,8,6,5,4,3,2]^{\rm Thm \ref{thm:from subcodes}} & 14\rightarrow15\\
&42597   & [8,7,5,5,3,3,1] & [8,7,5,5,3,3]^{\rm Thm \ref{thm:from subcodes}}& 11\rightarrow12\\\hline
(18,8,8)_{q} & 146336  & [10,7,7,7,6,6,6,5] & [8,7,7,7,6,6,6,5]^{\rm Thm \ref{thm:shortening}}& \ 28\rightarrow30 \\
& 102192  & [9,9,6,6,6,6,4,4] & [8,8,6,6,6,6,4,4]^{\rm Thm \ref{thm:shortening}}&  \ 24\rightarrow26 \\
& 87720  & [9,8,7,6,6,5,4,3] & [8,8,7,6,6,5,4,3]^{\rm Thm \ref{thm:shortening}}&  \ 23\rightarrow24 \\
& 55908  & [8,8,7,7,6,4,4,2] & [8,8,7,7,6,4,4,1]^{\rm Thm \ref{thm:from subcodes}}&  \ 22\rightarrow23 \\
& 199908 & [10,10,6,6,4,4,4,2] & [10,10,6,6,4,4,4]^{\rm Thm \ref{thm:from subcodes}}&  \ 18\rightarrow20 \\
\hline
(18,8,9)_{q} & 218452 & [9,9,8,7,6,5,4,3,2]  & [9,9,8,7,6,5,4,3,1]^{\rm Thm \ref{thm:from subcodes}}& 26\rightarrow27\\
&162182   & [9,7,7,7,7,5,5,1,1]  & [9,7,7,7,7,5,5,1]^{\rm Thm \ref{thm:from subcodes}} & 25\rightarrow26\\
&177482   & [9,8,7,7,6,5,4,2,1] & [9,8,7,7,6,5,4,2]^{\rm Thm \ref{thm:from subcodes}} & 24\rightarrow25 \\
&183116   & [9,8,8,6,5,5,4,2,2] & [9,8,8,6,5,5,4,2,1]^{\rm Thm \ref{thm:from subcodes}}& 23\rightarrow24\\
&185516   & [9,8,8,7,6,4,3,2,2] & [9,8,8,7,6,4,3,2,1]^{\rm Thm \ref{thm:from subcodes}}& 23\rightarrow24\\
&207698   & [9,9,7,6,5,5,4,3,1] & [9,8,7,6,5,5,4,3,1]^{\rm Thm \ref{thm:from subcodes}}& 23\rightarrow24\\
&208076   & [9,9,7,6,6,4,4,2,2] & [9,9,7,6,6,4,4,2]^{\rm Thm \ref{thm:from subcodes}}& 22\rightarrow24\\
&209708   & [9,9,7,7,5,5,3,2,2] & [9,9,7,7,5,5,3,2]^{\rm Thm \ref{thm:from subcodes}} & 22\rightarrow24\\
&210098   & [9,9,7,7,6,4,3,3,1] & [9,8,7,7,6,4,3,3,1]^{\rm Thm \ref{thm:from subcodes}}& 23\rightarrow24\\
&118058   & [8,8,8,6,6,5,3,2,1] & [8,8,8,6,6,5,3,2]^{\rm Thm \ref{thm:shortening}}& 22\rightarrow23\\
&123548   & [8,8,8,8,5,4,2,2,2]  & [8,8,8,8,5,4,2,2]^{\rm Thm \ref{thm:shortening}}& 21\rightarrow23\\
&176849  & [9,8,7,7,5,4,4,3] & [8,8,7,7,5,4,4,3]^{\rm Thm \ref{thm:shortening}}& 22\rightarrow23\\
&200609   & [9,9,5,5,5,5,5,4] & [8,8,5,5,5,5,5,4]^{\rm Thm \ref{thm:shortening}}& 21\rightarrow23\\
&174762   & [9,8,7,6,5,4,3,2,1] & [9,8,7,6,5,4,3,2]^{\rm Thm \ref{thm:from subcodes}}&  q<8,\ 20\rightarrow21\\
&    &  & [9,8,7,6,5,4,3,2,1]^{\rm Thm \ref{thm:from MDS}}&  q\geq 8,\ 21\rightarrow21\\
&157286   & [9,7,7,5,5,3,3,1,1] & [9,7,7,5,5,3,3,1]^{\rm Thm \ref{thm:from subcodes}}& 17\rightarrow18\\
&138781   & [9,5,5,5,5,1,1,1] & [9,5,5,5,5,1]^{\rm Thm \ref{thm:from subcodes}} & 11\rightarrow12\\
&37775   & [7,5,3,3,3] & [5,5,3,3,3]^{\rm Thm \ref{thm:shortening}}& 4\rightarrow6\\\hline
\end{array}$
\end{center}

\begin{center}
$\begin{array}{|c|cc|c|c|c}\hline
(19,8,6)_q & 25056   & [9,9,5,5,5,5]& [9,9,5,5,5,4]^{\rm Thm \ref{thm:from subcodes}} & 14\rightarrow15\\
\hline
(19,8,7)_{2} & 18794  & [8,6,4,3,3,2,1]  & [7,5,4,3,3,2,1]^{\rm Thm \ref{thm:from subcodes}}  & 7\rightarrow8\\
&3475  & [5,5,4,4,2] & [5,5,4,4,1]^{\rm Thm \ref{thm:from subcodes}} & 5\rightarrow6\\
&34022   & [9,5,3,3,3,1,1] & [6,5,3,3,3,1,1]^{\rm Thm \ref{thm:from subcodes}}& 5\rightarrow6\\
&67804   & [10,6,3,3,2,2,2] & [6,5,3,3,2,2,2]^{\rm Thm \ref{thm:from subcodes}} & 5\rightarrow6\\\hline
(19,8,8)_{q} & 89760   & [9,8,7,7,7,7,6,5] & [8,8,7,7,7,7,6,5]^{\rm Thm \ref{thm:shortening}}& 31\rightarrow32\\
&146832   & [10,7,7,7,7,6,6,4] & [10,7,7,7,7,6,6,3]^{\rm Thm \ref{thm:from subcodes}}& 29\rightarrow30\\
&104904   & [9,9,7,7,5,5,5,3] & [9,9,7,7,5,5,5,2]^{\rm Thm \ref{thm:from subcodes}}& 24\rightarrow25\\
&31758   & [6,6,6,6,6,1,1,1] & [6,6,6,5,5,1,1,1]^{\rm Thm \ref{thm:from subcodes}}& 13\rightarrow15\\
&132963   & [10,4,4,4,3,3] & [10,4,4,4,3]^{\rm Thm \ref{thm:from subcodes}}& 7\rightarrow8\\\hline
(19,8,9)_{q\geq5} & 408480   & [10,10,7,7,7,6,6,6,5]  & [9,9,7,7,7,6,6,6,5]^{\rm Thm \ref{thm:shortening}}& 35\rightarrow37\\
&354096   & [10,9,8,8,6,6,6,4,4] & [9,9,8,8,6,6,6,4,4]^{\rm Thm \ref{thm:shortening}}& 33\rightarrow34\\
&226060   & [9,9,8,8,8,6,6,2,2] & [9,9,8,8,8,6,6,2,1]^{\rm Thm \ref{thm:from subcodes}}& 31\rightarrow32\\
&341704   & [10,9,7,7,6,6,5,5,3] & [9,9,7,7,6,6,5,5,3]^{\rm Thm \ref{thm:shortening}}& 30\rightarrow31\\
&420440   & [10,10,8,8,7,6,4,3,3] & [9,9,8,8,7,6,4,3,3]^{\rm Thm \ref{thm:shortening}}& 30\rightarrow31\\
&307010   & [10,8,7,6,6,6,6,5,1]  & [9,8,7,6,6,6,6,5,1]^{\rm Thm \ref{thm:from subcodes}}& 29\rightarrow30\\
&204609   & [9,9,6,6,6,6,6,5] & [8,8,6,6,6,6,6,5]^{\rm Thm \ref{thm:shortening}} & 27\rightarrow29\\
&216724   & [9,9,8,6,6,6,5,3,2] & [9,9,8,6,6,6,5,3,1]^{\rm Thm \ref{thm:from subcodes}}& 27\rightarrow28\\
&119900   & [8,8,8,7,6,3,2,2,2] & [8,8,8,7,6,3,2,2]^{\rm Thm \ref{thm:shortening}}& 20\rightarrow22\\
&60474   & [7,7,7,6,6,2,2,2,1] & [7,7,7,6,6,2,2]^{\rm Thm \ref{thm:shortening}}& 16\rightarrow18\\
&166636   & [9,8,5,4,3,3,3,2,2] & [9,7,5,4,3,3,3,2,2]^{\rm Thm \ref{thm:from subcodes}}& 13\rightarrow14\\
&100243   & [8,8,4,4,4,4,2] & [8,8,4,4,4,4]^{\rm Thm \ref{the:from sys MRD-new1}}_{l=2}& 12\rightarrow13\\
&84441  & [8,7,5,3,3,3,2,2] & [8,6,5,3,3,3,2,2]^{\rm Thm \ref{thm:from subcodes}}& 10\rightarrow11\\\hline
\end{array}$
\end{center}

\subsection{Minimum subspace distance 10}

\begin{center}
$\begin{array}{|c|cc|c|c|c}\hline
(16,10,6)_q & 17376  &  [9,5,5,5,5,5] &  [9,5,5,5,5,4]^{\rm Thm \ref{thm:from subcodes}} & 9\rightarrow10\\\hline
(16,10,7)_q & 49648  &  [9,9,4,4,4,4,4] &  [8,8,4,4,4,4]^{\rm Thm \ref{the:from sys MRD-new1}}_{l=2}& 8\rightarrow10\\
& 5518  &  [6,5,4,4,1,1,1] &  [5,5,4,4,1,1,1]^{\rm Thm \ref{thm:from subcodes}}& 5\rightarrow6\\\hline
(16,10,8)_q & 57592  &  [8,8,8,3,3,3,3,3] &  [6,6,6,3,3,3]^{\rm Thm \ref{the:from sys MRD-new1}}_{l=2}& 6\rightarrow10\\
& 7339  &  [5,5,5,3,2,1] &  [5,5,4,3,2,1]^{\rm Thm \ref{thm:from MDS}}& 2\rightarrow3\\\hline
(17,10,7)_q & 50144  &  [9,9,5,5,5,5,5]&  [7,7,5,5,5,5,5]^{\rm Thm \ref{thm:shortening}} & 11\rightarrow15\\
& 6940  &  [6,6,5,5,2,2,2] &  [6,6,5,5,2,1]^{\rm Thm \ref{thm:from subcodes}}& 3\rightarrow6\\\hline
(17,10,8)_q & 115184  &  [9,9,9,4,4,4,4,4] &  [8,8,8,4,4,4,4,4]^{\rm Thm \ref{thm:shortening}} & 12\rightarrow15\\
& 21901  &  [7,6,5,4,4,1,1] &  [6,6,5,4,4,1,1]^{\rm Thm \ref{thm:from subcodes}} & 5\rightarrow6\\
& 41806  &  [8,7,4,4,3,1,1,1] &  [8,7,4,4,3,1]^{\rm Thm \ref{thm:from subcodes}}& 4\rightarrow5\\\hline
(18,10,7)_q & 26560  &  [8,8,6,6,6,6,6] &  [7,7,6,6,6,6,6]^{\rm Thm \ref{thm:shortening}} & 16\rightarrow18\\
& 136372  &  [11,7,6,4,3,3,2] & [8,7,6,4,3,3,2]^{\rm Thm \ref{thm:from MDS}}   & q\geq 7, 7\rightarrow8\\
& & &  [7,7,6,4,3,3,2]^{\rm Thm \ref{thm:combine with same dim}} &  q<7,\ 4\rightarrow8 \\\hline
(18,10,8)_q & 173024  &  [10,9,8,5,5,5,5,5] &  [8,8,8,5,5,5,5,5]^{\rm Thm \ref{thm:shortening}}& 17\rightarrow20\\
& 86932  &  [9,8,7,5,5,5,3,2]&  [7,6,6,5,5,5,3,2]^{\rm Thm \ref{thm:from subcodes}} & 11\rightarrow15\\
& 84586  &  [9,8,6,5,3,3,2,1]&  [9,8,6,5,3,3,1]^{\rm Thm \ref{thm:subcodes from Gab}} _{(r=2)} & 7\rightarrow9\\
& 36633  &  [8,5,5,5,5,2,2] &  [7,5,5,5,5,2,2]^{\rm Thm \ref{thm:subcodes from Gab}} _{(r=2)} & 7\rightarrow8\\
& 13646  &  [6,6,5,4,3,1,1,1] &  [5,5,5,4,3,1,1,1]^{\rm Thm \ref{the:from sys MRD-new1}}_{l=2}& 4\rightarrow5\\\hline
\end{array}$
\end{center}

\begin{center}
$\begin{array}{|c|cc|c|c|c}\hline
(18,10,9)_q & 80213  &  [8,6,6,6,4,3,2,1] &  [8,6,6,6,4,3,2]^{\rm Thm \ref{thm:from subcodes}} & 9\rightarrow10\\
& 144809  &  [9,6,6,5,4,4,3,2] &  [8,6,6,5,4,4,3,2]^{\rm Thm \ref{thm:from subcodes}}& 9\rightarrow10\\
& 86731  &  [8,7,6,4,3,3,1] & [7,6,5,4,3,2,1]^{\rm Thm \ref{thm:from MDS}}&  q\geq 7,\ 6\rightarrow7\\
& & &  [8,7,6,4,3,1]^{\rm Thm \ref{thm:subcodes from Gab}} _{(r=2)}&  q<7,\ 4\rightarrow7 \\
& 40058  &  [7,5,5,5,2,2,2,2,1] &  [6,5,5,5,2,2,2,2]^{\rm Thm \ref{the:from sys MRD-new1}}_{l=2}& 5\rightarrow6\\\hline
(19,10,7)_q & 50032  &  [9,9,5,5,4,4,4] &  [9,9,5,5,4,4]^{\rm Thm \ref{thm:from subcodes}} & 8\rightarrow12\\
& 197864  &  [11,11,6,4,4,4,3] &  [8,8,4,4,4,4]^{\rm Thm \ref{the:from sys MRD-new1}}_{l=2} & 8\rightarrow11\\\hline
(19,10,8)_q & 108480  &  [9,9,8,6,6,6,6,6] &  [8,8,8,6,6,6,6,6]^{\rm Thm \ref{thm:shortening}}& 22\rightarrow24\\
& 284464  &  [11,8,7,6,6,6,4,4] &  [8,8,7,6,6,6,4,4]^{\rm Thm \ref{thm:shortening}}& 17\rightarrow20\\
& 395745  &  [11,11,6,4,4,4,4] &[8,8,4,4,4,4]^{\rm Thm \ref{the:from sys MRD-new1}}_{l=2}& 8\rightarrow12\\
& 40042  &  [8,6,6,6,3,3,2,1]&  [8,6,6,6,3,3,2]^{\rm Thm \ref{thm:from subcodes}} & 8\rightarrow9\\
& 72086  &  [9,6,6,4,4,2,1,1] &  [7,6,5,4,4,2,1,1]^{\rm Thm \ref{thm:from subcodes}} & 6\rightarrow7\\
& 172348  &  [10,9,8,4,2,2,2,2] &  [8,8,8,4,2,2,2]^{\rm Thm \ref{the:from sys MRD-new1}}_{l=2}& 6\rightarrow7\\
& 28889  &  [7,7,7,3,3,2,2] &  [7,7,7,3,3,2,2]^{\rm Thm \ref{thm:combine with same dim}}& 2\rightarrow4\\\hline
(19,10,9)_{q\geq 3} & 418784  &  [10,10,8,8,5,5,5,5,5]  &  [9,9,8,8,5,5,5,5,5]^{\rm Thm \ref{thm:shortening}}& 23\rightarrow25\\
& 363416  &  [10,9,9,6,5,5,5,3,3] &  [8,8,7,6,5,5,5,3,3]^{\rm Thm \ref{thm:from subcodes}}& 16\rightarrow21\\
& 119412  &  [8,8,8,7,5,3,3,3,2] &  [8,8,8,7,5,3,3,1]^{\rm Thm \ref{thm:from subcodes}} & 12\rightarrow16\\
& 202629  &  [9,9,6,5,5,5,5,1]  &  [8,7,6,5,5,5,5,1]^{\rm Thm \ref{thm:from subcodes}} & 13\rightarrow16\\
& 59155  &  [7,7,7,5,5,5,2] &  [7,6,6,5,5,5,2]^{\rm Thm \ref{thm:from subcodes}}& 10\rightarrow12\\
& 170345  &  [9,8,6,6,4,3,3,2] &  [9,8,6,6,4,3,2,1]^{\rm Thm \ref{thm:subcodes from Gab}}_{r=2}& 10\rightarrow12\\
& 107754  &  [8,8,7,5,3,3,3,2,1] &  [8,8,7,5,3,3]^{\rm Thm \ref{thm:from subcodes}}& 6\rightarrow11\\
& 207190  &  [9,9,7,6,4,3,2,1,1] &  [9,9,7,6,4,3,2,1]^{\rm Thm \ref{thm:subcodes from Gab}}_{r=2}& 10\rightarrow11\\
& 398938  &  [10,10,6,5,5,3,2,2,1] &  [10,10,6,5,5,3,2]^{\rm Thm \ref{thm:subcodes from Gab}}_{r=2}& 10\rightarrow11\\
(19,10,9)_{2} & 202630  &  [9,9,6,5,5,5,5,1,1]  &  [8,8,6,5,5,5,5,1,1]^{\rm Thm \ref{thm:from subcodes}} & 14\rightarrow16\\
& 59149  &  [7,7,7,5,5,5,1,1] &  [7,7,6,5,5,5,1,1]^{\rm Thm \ref{thm:from subcodes}}& 11\rightarrow12\\
& 176473  &  [9,8,7,7,4,3,2,2,1] &  [9,8,7,7,4,3,2,2]^{\rm Thm \ref{thm:from subcodes}}& 11\rightarrow12\\
& 230633  &  [9,9,9,5,3,3,3,2] &  [9,9,9,5,3,3,1]^{\rm Thm \ref{thm:subcodes from Gab}}_{r=2}& 7\rightarrow11\\
& 298342  &  [10,8,5,5,4,3,3,1,1] &  [10,8,5,5,4,3,2]^{\rm Thm \ref{thm:subcodes from Gab}}_{r=2}& 8\rightarrow10\\
& 92366  &  [8,7,7,6,3,3,1,1,1] &  [8,7,7,6,3,3,1,1]^{\rm Thm \ref{thm:subcodes from Gab}}_{r=2}& 8\rightarrow9\\
& 275644  &  [10,6,6,5,3,2,2,2,2] &[10,6,6,5,3,2,1]^{\rm Thm \ref{thm:subcodes from Gab}}_{r=2} & 6\rightarrow8\\\hline
\end{array}$
\end{center}

\subsection{Minimum subspace distance 12}

\begin{center}
$\begin{array}{|c|cc|c|c|c}\hline
(19,12,7)_q & 135104  & [11,6,6,6,6,6,6] & [11,6,6,6,6,6,5]^{\rm Thm \ref{thm:from subcodes}}& 11\rightarrow12 \\\hline
(19,12,8)_q & 395232  & [11,11,5,5,5,5,5,5]& [10,10,5,5,5,5,5]^{\rm Thm \ref{the:from sys MRD-new1}}_{l=2} & 10\rightarrow12 \\\hline
(19,12,9)_q & 459760  & [10,10,4,4,4,4,4,4] &[10,10,4,4,4,4,4,4]^{\rm Thm\ref{thm:combine with same dim}}& 6\rightarrow12 \\
 & 58253  & [7,7,7,4,4,4,1,1] & [7,7,7,4,4,4,1,1]^{\rm Thm\ref{thm:combine with same dim}} & 3\rightarrow6 \\\hline
\end{array}$
\end{center}


\end{document}